\documentstyle[epsf,psfig]{mn}

\def\K{{\rm\thinspace K}}

\def\km{{\rm\thinspace km}}
\def\kpc{{\rm\thinspace kpc}}

\def\Mpc{{\rm\thinspace Mpc}}
\def\Msun{\hbox{$\rm\thinspace M_{\odot}$}}

\def\s{{\rm\thinspace s}}

\def\yrs{{\rm\thinspace yrs}}
\def\h50{\hbox{$\rm\thinspace h_{50}$}}
\def\h50m1{\hbox{$\rm\thinspace h_{50}^{-1}$}}

\def\kmpspMpc{\hbox{$\km\s^{-1}\Mpc^{-1}\,$}}

\def\Msun{\mathop{\rm M_{\odot}}\nolimits}

\def\Mpc{\mathop{\rm Mpc}\nolimits}
\def\kpc{\mathop{\rm kpc}\nolimits}

\def\s{\mathop{\rm s}\nolimits}
\def\K{\mathop{\rm K}\nolimits}
\def\Ga{\mathop{\times 10^9 \yrs}\nolimits}
\def\yrs{\mathop{\rm yrs}\nolimits}

\def\km{\mathop{\rm km}\nolimits}

\def\kps{\mathop{\rm km/s}\nolimits}

\def\etal{{\it et al.\thinspace}}

\def\fig{figure}

\def\figs{figures}

\newcommand{\expd}[1]{
 \times 10^{#1}
}

\def\Hydra{{\sc hydra}}
\def\<{\left<}
\def\>{\right>}
\def\fig{Fig.}
\def\figs{Figs.}

\title[Comparative study of SPH implementations]{
Smoothed Particle Hydrodynamics in cosmology: a comparative study of implementations}
\author[R. J. Thacker \etal]
  {R.~J.~Thacker,$^{1,2}$\thanks{Current Address: Department of Physics
and
Astronomy, University of Western Ontario, London, Ontario, N6A 3K7, 
Canada.} E.~R.~Tittley,$^2$
F.~R.~Pearce,$^3$ 
H.~M.~P.~Couchman,$^2$ 
  \newauthor 
  and P.~A.~Thomas$^4$\\
  $^1$Theoretical Physics Institute, Department of Physics, University of
Alberta, Edmonton, Alberta, T6G 2J1, Canada \\
  $^2$Department of Physics and Astronomy,
University
of Western Ontario, London, Ontario, N6A 3K7, Canada \\
  $^3$Department of Physics, University of Durham, Durham, DH1 3LE, United
Kingdom\\
  $^4$Astronomy Centre, CPES, University of Sussex, Falmer, Brighton,
Sussex, BN1 9QJ, United Kingdom     
}
\date{Accepted 1998 January 00.
      Received 1998 January 00;
      in original form 1998 January 00}

\pagerange{\pageref{firstpage}--\pageref{lastpage}}
\pubyear{1998}
\begin{document}
\maketitle
\begin{abstract} 

\noindent We analyse the performance of twelve different implementations
of Smoothed Particle Hydrodynamics (SPH) using seven 
tests designed to isolate key hydrodynamic elements of cosmological
simulations which are
known to cause the SPH algorithm problems.  In order, we consider a
shock tube, spherical adiabatic collapse, cooling
flow model, drag, a cosmological simulation, 
rotating cloud-collapse and disc
stability. In the implementations special attention is
given to the way in which force symmetry is enforced in the equations of
motion. We study in detail how the hydrodynamics are affected by
different implementations of the 
artificial viscosity including those with a shear-correction
modification. We 
present an improved first-order smoothing-length update algorithm that
is designed to remove instabilities that are present in the Hernquist
and Katz \shortcite{hk} algorithm. 
Gravity is calculated using the Adaptive Particle-Particle,
Particle-Mesh algorithm.

For all tests we find that the artificial viscosity
is the single most important factor distinguishing the results from
the various implementations. 
The shock tube and
adiabatic collapse problems show that the artificial viscosity
used in the current \Hydra\ code performs relatively poorly for simulations
involving strong shocks when compared to a more standard artificial
viscosity. The shear-correction term is shown to reduce the
shock-capturing ability of the algorithm and to lead to a spurious
increase  
in angular momentum in the rotating cloud-collapse problem. For the
disc stability test, the shear-corrected and current \Hydra\ artificial
viscosities are shown to reduce outward angular momentum
transport. The cosmological simulations produce comparatively similar
results, with the fraction of gas in the hot and cold phases varying
by less than 10\% amongst the versions. Similarly, the drag test shows
little systematic variation amongst 
versions. The cooling flow tests show that implementations using the
force symmetrization of Thomas and Couchman \shortcite{tc} are more
prone to accelerate the overcooling instability of SPH, although the
problem is generic to SPH.   
The
second most important factor in code performance is the way force
symmetry is achieved in the equation of motion. Most results favour
a kernel symmetrization approach. The exact method by which SPH pressure
forces are included in the equation
of motion appears to have comparatively little effect on the results.
Combining the equation of motion presented in Thomas
and Couchman \shortcite{tc} with a modification of the Monaghan and
Gingold \shortcite{jj} artificial viscosity leads to an
SPH scheme that is both fast and reliable.

\end{abstract}


\label{firstpage}

\section{Introduction}
Smoothed Particle Hydrodynamics (SPH) \cite{m1,l1} is a popular numerical
technique for solving gas-dynamical equations. SPH is unique among
numerical methods in that many algebraically equivalent -- but formally
different -- equations of motion may be derived. In this paper we report
the results of a comparison of several implementations of SPH in tests
which model physical scenarios that occur in hierarchical clustering
cosmology. 

SPH is fundamentally Lagrangian and fits well with gravity solvers that
use tree structures (Hernquist \& Katz 1989, hereafter HK89) and mesh
methods supplemented by short range forces (Evrard 1988; Couchman, Thomas
\& Pearce~1995, hereafter CTP95). In an adaptive form \cite{wood}, the
algorithm lends itself readily to the wide range of densities encountered
in cosmology, contrary to Eulerian methods which require the storage and
evaluation of numerous sub-grids to achieve a similar dynamic range. SPH
also exhibits less numerical diffusivity than comparable Eulerian
techniques, and is much easier to implement in three dimensions, typically
requiring 1000 or fewer lines of FORTRAN code. 

The main drawback of SPH is its limited ability to follow steep density
gradients and to correctly model shocks. Shock-capturing requires the
introduction of an artificial viscosity \cite{jj}.  A number of different
alternatives may be chosen and it is not clear whether one method is to be
preferred over another. The presence of shear in the flow further
complicates this question.

Much emphasis has been placed upon the performance of SPH with a small
number of particles (order 100 or fewer). Initial studies \cite{ge} of SPH
on spherical cloud collapse indicated acceptable performance when compared
to low resolution Eulerian simulations, with global properties, such as
total thermal energy, being reproduced well. A more recent study
(Steinmetz \& Muller 1993, hereafter SM93), which compares SPH to modern
Eulerian techniques (the Piecewise Parabolic Method \cite{caw}, and
Flux-Corrected-Transport methods) has shown that the performance of SPH is
not as good as initially believed, and that accurate reproduction of local
physical phenomena, such as the velocity field, requires as many as $10^4$
particles. In the context of cosmology with hierarchical structure
formation, the small-$N$ performance remains critical as the first objects
to form consist of tens -- hundreds at most -- of particles and form, by
definition, at the limit of resolution.  It is therefore of crucial
importance to ascertain the performance of different SPH implementations
in the small-$N$ regime.  Awareness of this has caused a number of authors
to perform detailed tests on the limits of SPH \cite{ov,gg}. To address
these concerns, some of the tests we present here are specifically
designed to highlight differences in performance for small $N$.

Our goal in this paper is to detail systematic trends in the results for
different SPH implementations. Since these are most likely to be visible
in the small-$N$ regime we concentrate on smaller simulations, using a
larger number of particles to test for convergence.  Because of the
importance of the adaptive smoothing length in determining the local
resolution we pay particular attention to the way in which it is
calculated and updated.  We are also concerned with the efficiency of the
algorithm.  Realistic hydrodynamic simulations of cosmological structure
formation typically require $10^4$ or more time-steps. Therefore when
choosing an implementation one must carefully weigh accuracy against
computational efficiency. This is a guiding principle in our
investigation.

The seven tests used in this study are:
\begin{itemize}
\item Sod shock (section \ref{shock})
\item Spherical collapse (section \ref{evrard})
\item Cooling near density jumps(section \ref{cool})
\item Drag on a cold clump(section \ref{drag})
\item Hierarchical structure formation (section \ref{cosmo})
\item Disc formation (section \ref{rotcloud})
\item Disc stability (section \ref{disk})
\end{itemize}

Each of these tests is described in the indicated
section. The tests investigate various aspects of the SPH algorithm
ranging from explicit tests of the hydrodynamics to investigations
specific to cosmological contexts. The Sod shock~(1978), although a
relatively simple shock configuration, represents the minimum flow
discontinuity that a hydrodynamic code should be able to
reproduce. The spherical collapse test \cite{ge}, although idealised,
permits an assessment of the resolution necessary to approximate
spherical collapse. It also allows a comparison with other
authors' results and with a high resolution spherically symmetric
solution. The remaining tests are 
more closely tied to the arena of cosmological simulations. The
cooling test looks at the problems associated with modelling different
gas phases with SPH. A cold dense knot of particles embedded in a hot halo
will tend to promote cooling of the hot gas because of the inability
of SPH to separate the phases. The drag test looks at the behaviour of
infalling satellites and the overmerging problem seen in SPH
simulations \cite{fews}. Finally, three tests consider the ability of
the SPH algorithms to successfully model cosmic structure. First we
look at the overall distribution of hot and cold gas in a hierarchical
cosmological simulation, followed by an investigation of disc formation
from the collapse of a rotating cloud (Navarro \& White~1993), and the
transport of angular momentum in discs. In each case we
compare the different algorithms and assess the reliability of the SPH
method in performing that aspect of cosmological structure formation.

The layout of the paper is as follows. Section \ref{sec2} reviews the
basic SPH framework that we use, including a description of the new
approach developed to update the smoothing length.  Next we examine
the equations of motion and internal energy, and discuss the 
procedure for symmetrization of particle forces.  In
section~\ref{tests} we present the 
test cases as listed above. Each of the subsections is self
contained and contains a description and motivation for the test
together with results and a summary comparing the relative merits of the
different implementations together with an assessment of the success
with which the SPH method can perform the test. Section~\ref{conc}
briefly summarises the main overall conclusions to be drawn from the
test suite, indicating where each implementation has strengths and
weaknesses and makes recommendations for the implementation which may
be most useful in cosmological investigations.

During final prepartion of this paper, a preprint detailing a similar
investigation was circulated by Lombardi \etal~\shortcite{lom}.

\section[]{Implementations of SPH}\label{sec2}

\subsection{Features common to all implementations} 
\label{kappa}

All of the implementations use an adaptive particle-particle-particle-mesh
(AP${}^3$M)  gravity solver \cite{h1}. AP${}^3$M is more efficient than
standard P${}^3$M, as high density regions, where the particle-particle
summation dominates calculation time, are evaluated using a further
P${}^3$M cycle calculated on a high resolution mesh. We denote the
process of placing a high resolution mesh `refinement placing' and the
sub-meshes are termed `refinements'. The algorithm is highly
efficient with the cycle time typically slowing by a factor of three under
clustering. 
The most significant drawback of AP${}^3$M
is that it does not yet allow multiple time-steps. However, the calculation 
speed of the global solution, compared with alternative methods 
such as the tree-code, more than outweighs this deficiency. Full
details of the adaptive scheme, in particular accuracy and timing 
information, may be found in Couchman \shortcite{h1} and CTP95.

Time-stepping is performed using a Predict-Evaluate-Correct (PEC) scheme.
This scheme is tested in detail against leapfrog and Runge-Kutta
methods in CTP95.  The value of the time-step, $dt$, is found
by searching the particle lists to establish the Courant conditions of the
acceleration, $dt_a$, and velocity arrays, $dt_v$. In this paper we introduce
a further time-step criterion, $dt_h$, which prevents particles travelling
too far within their smoothing radius, (see section \ref{tstep}). $dt$ is
then calculated from $dt<\kappa\, \min(0.4 dt_v, 0.25 dt_a, 0.2
dt_h)$ where  $\kappa$ is a normalisation constant that is taken equal to unity
in adiabatic simulations.  In simulations with cooling, large density
contrasts can develop, and a smaller value of $\kappa$ is sometimes
required. We do not have a Courant condition for cooling since we 
implement it assuming constant density (see below and CTP95 for
further details). 

SPH uses a `smoothing' kernel to interpolate local hydrodynamic quantities
from a sample of neighbouring points (particles). For a continuous system
an estimate of a hydrodynamic scalar $A(\bmath{r})$ is given by,
\begin{equation}
\<A(\bmath{r})\> = \int d^3 \bmath{r}' A(\bmath{r}')
W(\bmath{r}-\bmath{r}',h),
\end{equation}
where $h$ is the `smoothing length' which sets the maximum smoothing
radius and $W(\bmath{r},h)$, the smoothing kernel, is a function of
$|\bmath{r}|$. For a finite number of 
neighbour particles the approximation to this is,
\begin{equation}
\<A(\bmath{r})\> = \sum_j m_j 
{A(\bmath{r_j}) \over \rho(\bmath{r}_j)}\; W(\bmath{r}-\bmath{r}_j,h),
\end{equation}
where the radius of the
smoothing kernel is set by $2h$ (for a kernel with compact support). The
smoothing kernel used is the 
so-called $B_2$-spline \cite{ml}, 
\begin{equation}
W(\bmath{r},h)={W_s(r/h) \over h^3},
\end{equation}
where if $x=r/h$,
\begin{equation}
W_s(x)={1 \over 4\pi}\cases{
 4-6x^2+3x^3,  & $0\leq x\leq 1$; \cr
 (2-x)^3,  & $1<x\leq 2$;   \cr
0,  & $x>2$.
}
\end{equation}
The kernel gradient is modified to give a small repulsive
force for close particles \cite{tc},
\begin{equation}
{dW_s(x) \over dx}=-{1 \over 4\pi}\cases{
4, & $ 0 \leq x \leq 2/3 $; \cr
 3x(4-3x),  & $2/3 < x\leq 1$; \cr
 3(2-x)^2,  & $1<x\leq 2$;   \cr
0,  & $x>2$.
}
\end{equation}
The primary reason for having a non-zero gradient at the origin is to
avoid the artificial clustering noted by Sch\"{u}ssler \& Schmidt
\shortcite{ss1}.  (Some secondary benefits are discussed by Steinmetz~1996.)

In standard SPH the value of the smoothing length, $h$, is a constant for
all particles resulting in fixed spatial resolution. Fixed $h$ also
leads to
a slow-down in the calculation time when
particles become clustered -- successively more particles on average fall
within a particle's smoothing length. In the adaptive form of SPH
\cite{wood} the value of $h$ is varied so that all particles have a
constant (or approximately constant) number of neighbours. This
leads to a resolution scale dependent upon the local number density of
particles. It also removes the slow-down in calculation time since the
number of neighbours is held constant provided that the near
neighbours can be found efficiently. In this work we attempt to smooth
over 52 neighbour particles, and tests on the update algorithm we use (see
section \ref{tstep}) show that this leads to a particle having between 30
and 80 neighbours, whilst the average remains close to 50. 

A minimum value of $h$ is set by requiring that the SPH resolution
not fall below that of the gravity solver. We define the
gravitational resolution to be twice the S2 softening
length~\cite{he}, $\epsilon$, as at this radius the force is closely
equivalent to the $1/r^2$ law. We quote the equivalent Plummer softening
throughout the paper, as this is the most common force softening
shape. Since the minimum resolution of the SPH kernel is
the diameter of the smallest smoothing sphere, $4h_{min}$, equating this
to the
minimum gravitational resolution yields, 
\begin{equation}
2 \epsilon = 4  h_{min}.
\end{equation}
Unlike other authors, once this minimum $h$
is reached by a particle, smoothing occurs over {\em all\/}
neighbouring particles within a radius of $2h_{min}$. 
As a result setting a lower $h_{min}$ increases
efficiency
as fewer particles contribute to the sampling, but this leads to a mismatch
between hydrodynamic and gravitational resolution scales which we
argue is undesirable.

When required, radiative cooling is implemented in an integral form
that assumes constant density over a time-step
\cite{tc}. The change in
the specific
energy, $e$, is evaluated from. 
\begin{equation}
\int^{e-\Delta e}_e {de \over \Lambda} = -{n_i^2
\over \rho_i} \Delta t,
\end{equation}
where $\Delta t$ is the time-step, $n_i$ the number density and $n_i^2 \Lambda$
is the power radiated per unit volume. 
In doing this we circumvent having
a Courant condition for cooling, and hence it never limits the time-step.
  
\subsection{An improved first-order smoothing-length update
algorithm}\label{tstep}

 \begin{figure*}
 \vspace{135mm}
\begin{minipage}{170mm}
\includegraphics{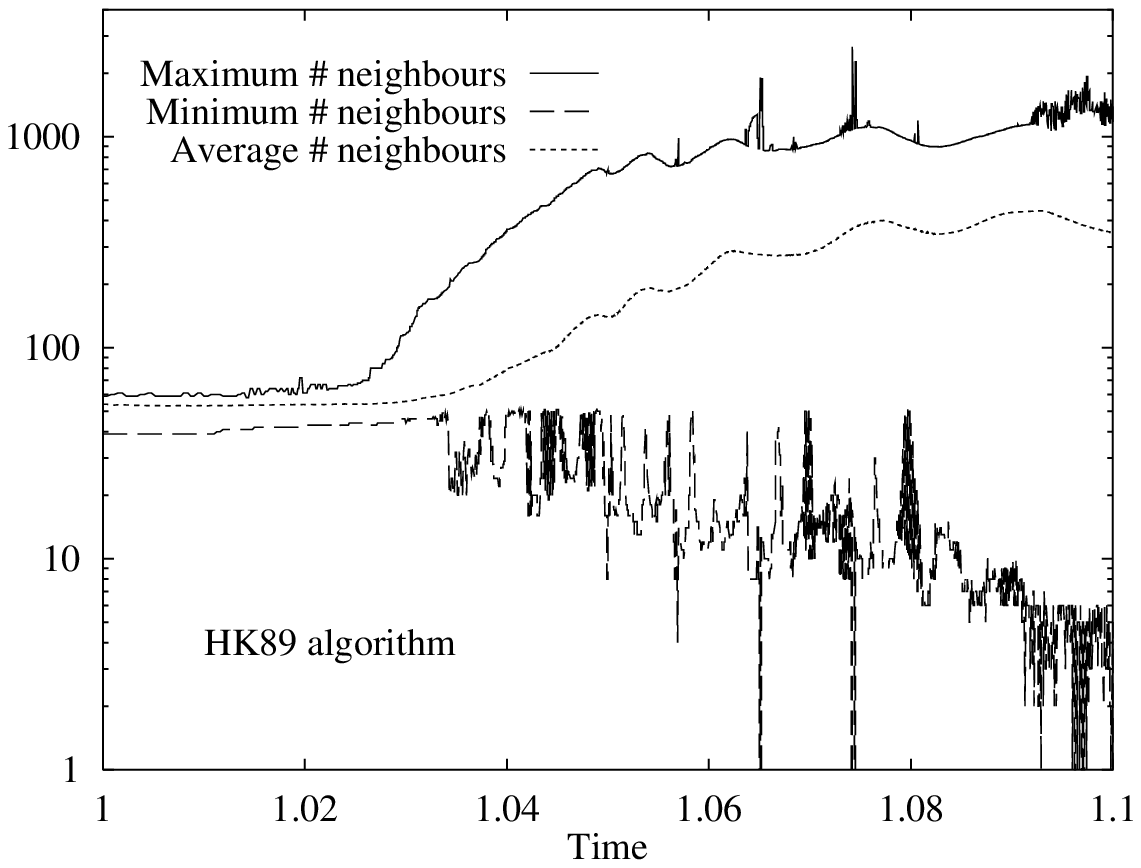}
\includegraphics{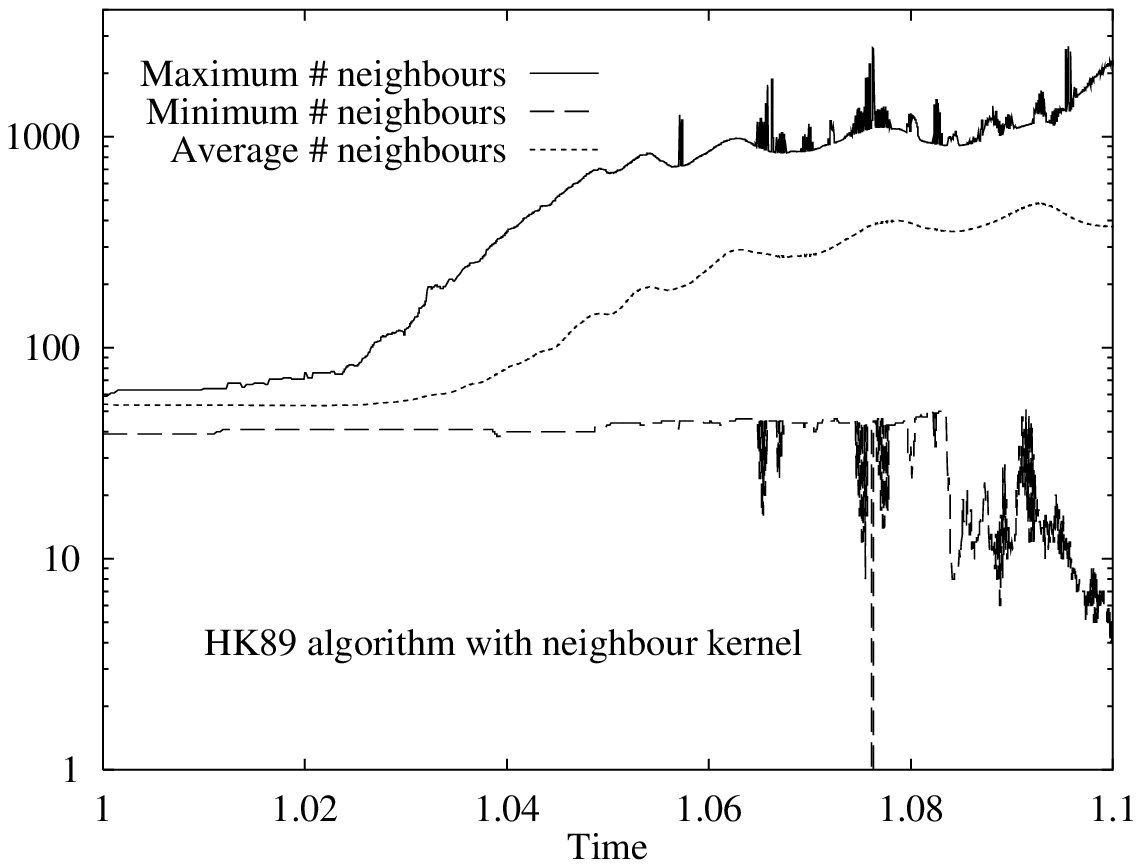}
\includegraphics{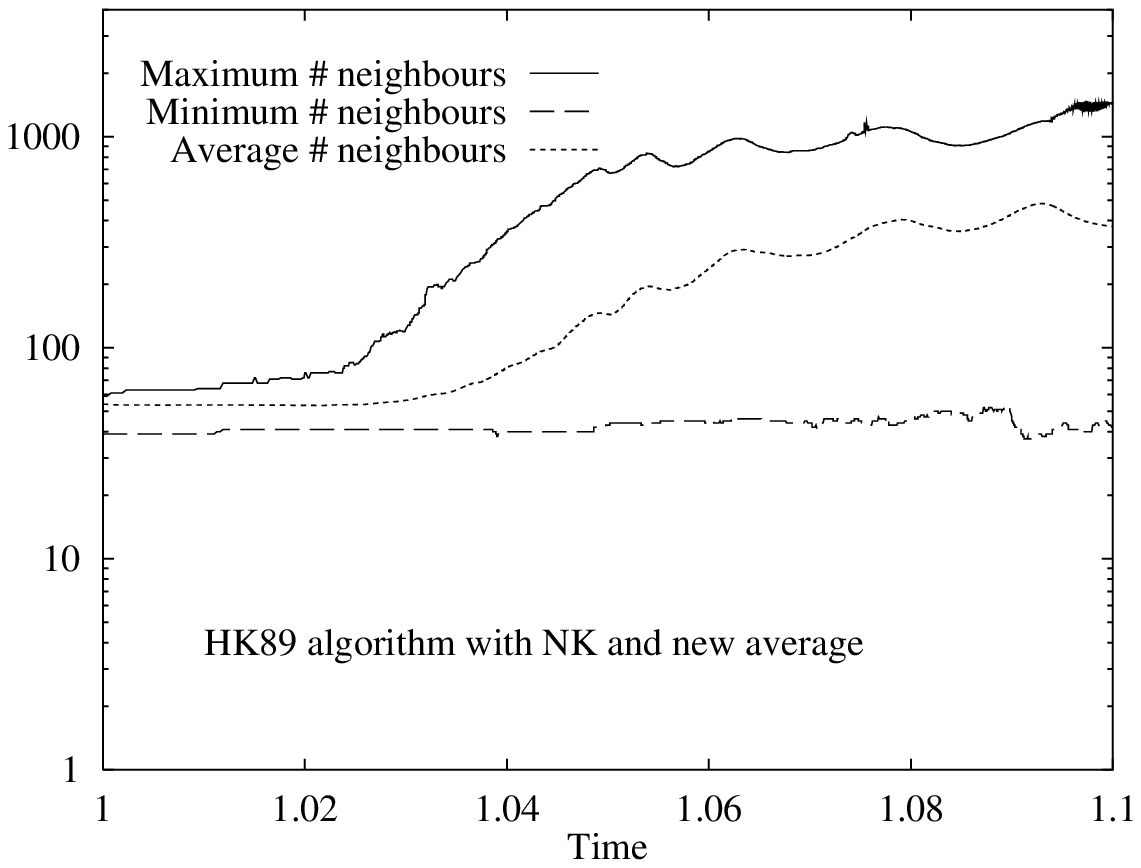}
\includegraphics{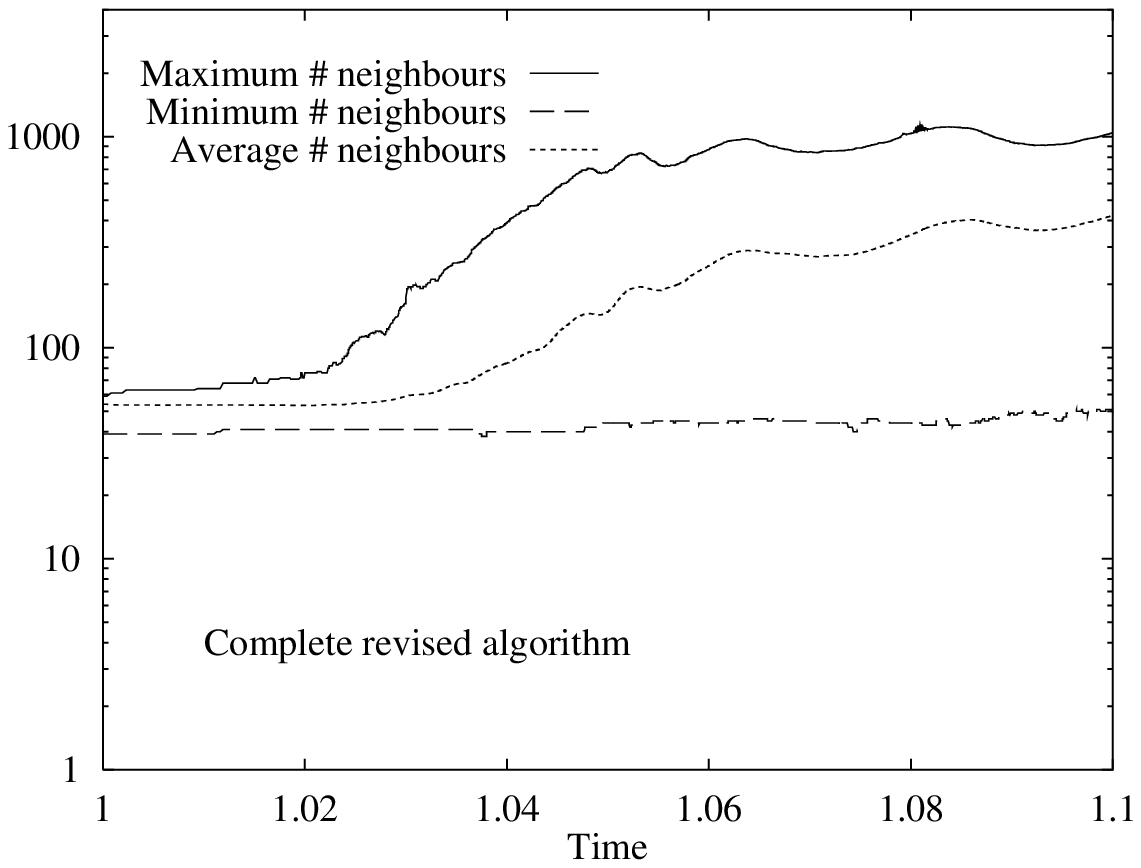}

\caption{Improvement in the neighbour counts as each component of the new
algorithm is added. A rotating cloud collapse problem
(see section \ref{rotcloud}) was repeated with each of our modifications
to the HK89 algorithm. The improvement from the first panel (upper
left) to the third panel (bottom   
left) is clear, the final panel shows a slight reduction in the 
range of neighbours and in the oscillation of the counts. The
large increase in the maximum number of neighbours is due to the
$h_{min}$ limit being reached. Time units are given in Gyr.}
\label{f2}\end{minipage} \end{figure*}

As stated earlier, in the adaptive implementation of SPH the smoothing
length, $h$, is
updated each time-step so that the number of neighbours is held close to,
or exactly at, a constant. Ensuring an exactly constant number of
neighbours
is computationally expensive (requiring additional neighbour-list searching)  and
hence many researchers prefer to update $h$ using an
algorithm that is closely linked to the local density of a particle. 

We have used two guiding principles in the design of our new algorithm.
First, since gravitational forces are attractive the algorithm will
spend most of its time decreasing $h$ (void evolution is `undynamic' and
poses little problem to all algorithms). Second, smoothing over too many
particles is generally preferable to smoothing over too few.
Despite some loss of spatial resolution and computational efficiency,
it does not `break' the SPH algorithm. Smoothing over too few particles
can lead to unphysical shocking.

A popular method for updating the smoothing length is to
predict
$h$ at the next time-step from an average of the current $h$
and the $h$ implied by the number of
neighbours found at the current time-step (HK89). This is expressed 
as,
\begin{equation}
\label{1}
h_i^n={h_i^{n-1}\over 2}\left[1 + \left( {N_s \over N_i^{n-1}} \right)^{1/3} \right],
\end{equation}
where $h_i^n$ is the smoothing length at time-step $n$ for particle $i$,
$N_s$ is the desired number of neighbours and $N_i^{n-1}$ is the number of
neighbours found at step $n-1$. The performance of this algorithm for a
rotating cloud collapse problem (see section \ref{rotcloud}) is shown in
the top left panel of \fig~\ref{f2}. The rotating cloud collapse is a
difficult problem for the $h$ update algorithm to follow since the
geometry of the cloud rapidly changes from three to two dimensions. The
plot shows that both the maximum and minimum number of neighbours
(selected from all the SPH particles in the simulation)  exhibit a
significant amount of scatter. Clearly, the maximum number of neighbours
increases rapidly once the $h=h_{min}$ limit is reached. 

The rotating cloud problem demonstrates how this algorithm may
become unstable when a particle approaches a high density region 
(the algorithm is quite stable where the density gradients
are small). If the current $h$ is too large, the neighbour count will
be too high leading to an underestimate of $h$. At the next step too
few neighbours may be found. The result is an oscillation in the estimates of
$h$ and in the number of neighbours as a particle accretes on to a
high density region from a 
low density one. This oscillatory behaviour is visible in the fluctuations
of the minimum number of neighbours in top left panel of \fig~\ref{f2}.

The instability can be partially cured by using an extremely small
time-step.  However, this is not practical as simulations currently take
thousands of time-steps.  A solution suggested by Wadsley~\shortcite{jaw}
that helps
alleviate the discontinuity in the number of neighbours, is to `count'
neighbours with a smoothed kernel rather than the usual tophat. This
(normalised) weighted neighbour count is then used in the $h$ update
equation rather than $N_i^{n-1}$. The instability in the standard HK89
algorithm is caused by the sharp discontinuity in the tophat at $r=2h$.
Hence we require that the new kernel smoothly approach 0 at $2h$.
Secondly, we wish to count the most local particles
within the smoothing radius at full
weight. Hence we consider a kernel that is unity to a certain radius
followed by a smooth monotonic decrease to zero. We choose,
\begin{equation}
W_{nn}(r/h)=\cases{1, &$0 \leq r/h < 3/2$; \cr 
 \pi W_s(4(r/h-3/2)),  & $3/2 \leq r/h \leq 2$, 
}
\end{equation} 
where $W_s(x)$ is the normalised $B_2$-spline kernel. We have experimented
with changing the radius at which the counting kernel switches over to the
spline, and a value of $r=3h/2$ has proven to be optimal, providing a good
balance between the smoothness of the variation of the estimate and the
closeness of 
actual number of neighbours to the desired number. At this value 
approximately half
of the smoothing volume is counted at full weight.  For smaller values the
smoothed estimate becomes progressively more unreliable. Conversely as the  
limiting radius
is increased the gradient of the kernel at $r\simeq 2h$ becomes too steep.
The improvement in fluctuation of the maximum and minimum number of
neighbours can be seen in the top right panel of \fig~\ref{f2}.

The next step in the construction of the new algorithm is to adjust the
the average 
used to update $h$. The primary reason for doing this is to avoid sudden
changes in the smoothing length. Whilst the neighbour counting kernel
helps to alleviate 
this problem, it does not remove it entirely. Secondly, since we
shall limit the time-step by only allowing the particles to move a
certain fraction of $h$ it is also useful to limit
the change in $h$. The motivation here is that a particle which is
approaching a high density region, for example, and is restricted to move
$0.2h$ per time-step should only be allowed to have $h\rightarrow
0.9h$ (since the smoothing radius is $2h$). However, a particle at the
centre of homologous flow must be able to update faster since collapse
occurs from all directions. To account for this it is helpful to permit a
slightly larger change in $h$, $h\rightarrow0.8h$ for example.

\begin{figure}
 \vspace{140pt}
\includegraphics{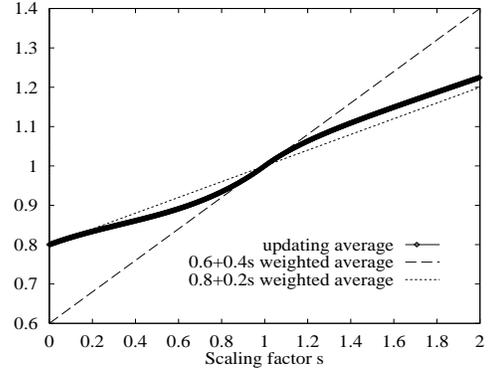}
 \caption{Weighting function compared to simple weighted averages.}
\label{f1}
\end{figure}

Setting $s=(N_s/N_i^{n-1})^{1/3}$, we may express equation~\ref{1} as
\begin{equation}
h_i^n=h_i^{n-1}(1-a+a s),
\end{equation}
where $a$ is a weighting coefficient, and for equation~\ref{1},
$a=0.5$. 
We have tested the performance of this average for $a\in
[0.2,0.5]$. A value of $a=0.4$ proved optimal, reducing scatter
significantly yet
allowing a sufficiently large change in $h$. A problem remains that
if $s\simeq0$ then $h_i^n=0.6h_i^{n-1}$, which represents a large change
if one limits the time-step according to an $h/v$ criterion. Hence we have  
implemented
a scheme that has the asymptotic property $h_i^n=0.8h_i^{n-1}$, but
for small changes in $h$ it yields $h_i^n=h_i^{n-1}(0.6+0.4s)$. The
function we
use for determining the weighting variable $a$ is,
\begin{equation}
a=\cases{0.2(1+s^2), &$ s<1$; \cr
         0.2(1+1/s^3),  & $s\geq1$.
}
 \end{equation}
A plot of this function compared to the 0.6,0.4 weighted average can
be
seen in \fig~\ref{f1}. The lower left panel of \fig~\ref{f2} shows
that introduction of this average reduces the scatter in both the maximum
and minimum number of neighbours.

Even with the improvements made so far it remains possible that the
particle may move very quickly on to a high density region causing a
sudden change that cannot be captured by the neighbour counting kernel.
Thus it is sensible to limit the time-step according to a `Courant
condition' which is the smoothing radius divided by the highest velocity
within the smoothing radius. Further, the velocity $\bmath{v}_r$, must be
measured within
the frame of the particle under consideration,
$\bmath{v}_r=\bmath{v}_i-\bmath{v}_j$. The new time-step criterion
may
be summarised,
\begin{equation}
dt=C_h \min_i(h_i/\max_j(|\bmath{v}_i-\bmath{v}_j|),
\end{equation}
where $C_{h}$ is the Courant number and the $i,j$ subscripts denote a
reduction over the variable indicated. We have chosen 
$C_h=0.2$, which is usually the limiting factor in time-step
selection. The gains in introducing this condition are marginal
for the neighbour count in this test (maximum and minimum values are
within a tighter range and show slightly less oscillation).

When taken together all of these adjustments combine to make a scheme
that is both fast and very stable. The final result of including these
improvements is seen in the lower right hand panel of \fig~\ref{f2}.
 
 \subsection{Equations of motion}
\label{eqnmot}
The SPH equation of motion is derived from,
\begin{equation}\label{eqnmotion}
{dv \over dt}=-{1\over \rho} \nabla P,
\end{equation}
using identities involving the
pressure and density. An excellent review of this derivation, and why so
many different schemes are possible, may be found
in Monaghan \shortcite{jm}. In short, different identities produce
different
equations of motion and a clear demonstration of this can be seen by
comparing the equations of motion of HK89 to those of SM93.

Once an adaptive scheme is implemented in SPH, neighbour smoothing
develops a dualism -- the smoothing may be
interpreted as either a ``gather'' or a  
``scatter'' process.  If we attempt to smooth a field at the position
of particle $i$, then the contribution of particle $j$ 
to
the smoothed estimate may be evaluated using the value of $h$ from either
particle $i$ (gather) or particle $j$ (scatter). For constant $h$
SPH the schemes are equivalent.

The uncertainty in the neighbour smoothing may be circumvented by
using a
formalism based upon the average of the two smoothing lengths
\cite{ge,wb}. Under this prescription, the gather and scatter
interpretations are equivalent since the smoothing length for particles
$i$
and $j$ are the same. We call this prescription $h$-averaging. The density
estimate under this prescription is given by,

\begin{equation}
\<\rho(\bmath{\bmath{r}_i})\>=\sum_{j=1}^N m_j
W(\bmath{r}_i-\bmath{r}_j,h_{ij}),
\end{equation}
where,
\begin{equation}
h_{ij}=(h_i+h_j)/2,
\end{equation}
and the neighbour search is conducted over particles for which
$r_{ij}<2h_{ij}$.

Most $h$-averaging schemes use the arithmetic mean of the smoothing
lengths
but it is possible to consider other averages (any `average' that is
symmetric in $i-j$ is potentially acceptable). Two other averages of
interest
are the geometric mean and the harmonic mean. Note
that the 
harmonic and geometric means both pass through the origin where as the
arithmetic mean does not. This has potentially important consequences when
particles with large $h$ interact with particles that have a
small $h$. This situation occurs at the boundary of high density regions
in simulations with radiative cooling (see section \ref{cool} for a
detailed
discussion of this). 
 
An alternative way of circumventing the smoothing uncertainty
(HK89) 
is to combine the ``gather'' and ``scatter'' interpretations
into one hybrid framework by averaging the kernels. In this scheme the
density estimate is given by,
\begin{equation}
\<\rho(\bmath{r}_i)\>= \sum_{j=1}^N m_j
[W(\bmath{r}_i-\bmath{r}_j,h_i)+W(\bmath{r}_i-\bmath{r}_j,h_j)]/2,
\end{equation}
and we denote this scheme kernel averaging. The averaged kernel
then replaces
the normal kernel in all equations. Rewriting equation \ref{eqnmotion}  
using the identity, 
\begin{equation}\label{ident}
{ \nabla P \over \rho}=\nabla {P \over \rho} + {P \over \rho^2} \nabla
\rho,
\end{equation}
the SPH equation of motion with kernel averaging becomes,
\[
{d \bmath{v}_i \over dt}=- \sum_{j=1}^N m_j \left({P_i \over
\rho_i^2}+{P_j \over \rho_j^2}\right) \times 
\]
\begin{equation}\label{3}
\;\;\;\;\;\;\;\;\;\; \nabla_i 
[W(\bmath{r}_i-\bmath{r}_j,h_i)+W(\bmath{r}_i-\bmath{r}_j,h_j)]/2.
\end{equation}
This equation of motion is used in SM93.

Thomas \& Couchman (1992, hereafter TC92), 
present an alternative prescription where
the force term is symmetrized and the density is calculated under the
gather interpretation. Using the standard pressure and density identity
in equation \ref{ident},
the acceleration for particle $i$ is written,
\[
{d \bmath{v}_i \over dt}=\sum_{j=1}^N{\bmath{f}_{ij} \over m_i}=
- \sum_{j=1}^N m_j {P_i \over \rho_i^2} \nabla_i 
W(\bmath{r}_i-\bmath{r}_j,h_i)
\]
\begin{equation} \label{2}
\;\;\;\;\;\;\;\;\;\;\;\;\;\;\;\;\;\;\;\;\;\;\;\; + \sum_{j=1}^N m_j {P_j \over \rho_j^2} \nabla_j 
W(\bmath{r}_i-\bmath{r}_j,h_j).
\end{equation}
This symmetrization is used in the current implementation of the
publically available code, \Hydra. 
When combined with an artificial viscosity that does not require the
pre-computation of all density values this scheme is extremely efficient. 

In deriving equation~\ref{2}
the approximation,
\begin{equation}
\nabla_i W(\bmath{r}_i-\bmath{r}_j,h_i) \simeq -\nabla_j
W(\bmath{r}_i-\bmath{r}_j,h_j)
\end{equation}
was used. This approach to symmetrization of the equation of motion is
fundamentally different to the other two schemes which both result in
a kernel symmetric in $i$ and $j$. The approximation is
not
correct to first order in $h$ and may introduce small errors. If this
symmetrization is supplemented by either $h$-averaging or kernel averaging
(there is no argument against this) then the
substitution is correct to first order in
$h$ (but not in $\nabla h$). 

When SPH is implemented in its adaptive form equation~\ref{3} will not
guarantee conservation of momentum. To do so one must alter the neighbour
list so that for any particle $i$ which smooths over particle $j$ the
converse is also true. This will not necessarily be true if one smooths
over the nearest 52 neighbours for each particle. Note that
equation~\ref{2} does conserve momentum regardless of the neighbour list
used. 

In the SPH--AP$^3$M solver an
SPH particle only has gas forces evaluated within a refinement if it
has a sufficiently small $h$ (see CTP95 for a detailed discussion of
this topic). The introduction of $h$-averaging complicates refinement
loading since 
a particle no longer has a well defined $h$ (one must 
consider all $h_{ij}$ values). To
establish which particles are to be evaluated, one must calculate all the
$h_{ij}$ values for a particle, which is computationally costly. 
As a
compromise, we place particles in a refinement only
if $1.4\times2h$ is smaller than the sub-mesh search length (note that 
for arithmetic $h$-averaging a search to $2\times2h$ will ensure that all
neighbours are found and we use this search radius in sections
\ref{cosmo}, \ref{rotcloud} \& \ref{disk}). This does not
absolutely guarantee that all the neighbour data for a particle will be
placed
in the refinement data arrays but we have found no detectable difference
between this method and the full $h_{ij}$ method. 

\subsection{Internal energy equation}

The SPH internal energy equation is derived from,
\begin{equation}
{du \over dt}=-{P \over \rho} \nabla . \bmath{v}.
\end{equation}
The SPH estimate for $\nabla . \bmath{v}$ may be used
to
calculate $du/dt$ directly, yielding the internal energy equation
for particle $i$,
\begin{equation} \label{etc92}
{d \epsilon_i \over dt}=- {P_i \over \rho_i} \nabla.\bmath{v}_i.
\end{equation}
Explicitly writing the summation for the SPH estimate gives,
\begin{equation}\label{e1}
{d \epsilon_i \over dt}=\sum_j m_j {P_i \over \rho^2_i}
(\bmath{v}_i-\bmath{v}_j).\nabla_i W(\bmath{r}_i-\bmath{r}_j,h).
\end{equation}
Substituting $h=h_{ij}$ and inserting the artificial viscosity 
$P_i\rightarrow P_i + \rho_i^2 \Pi_{ij}/2$ (see equation~\ref{monvis}), yields the internal energy
equation used in SM93. Whilst
not strictly compatible with equation~\ref{3} (equation~\ref{e1}
has no
dependence upon $P_j$) it can nevertheless be shown that energy will be
conserved \cite{wb}.

An internal energy equation may be constructed
using the same symmetrization as equation \ref{2}, yielding,
\[
{d\epsilon_i \over dt} ={1\over2}\sum_{j=1}^N m_j
(\bmath{v}_i-\bmath{v}_j).{P_i \over \rho_i^2} 
\nabla_i W(\bmath{r}_i-\bmath{r}_j,h_i)
\]
\begin{equation} \label{5}
\;\;\;\;\;\;\;\;\;\;\; - {1\over2}\sum_{j=1}^N m_j
(\bmath{v}_i-\bmath{v}_j).{P_j \over \rho_j^2}
\nabla_j   
W(\bmath{r}_i-\bmath{r}_j,h_j).
\end{equation}
A similar equation was considered in
TC92, and was shown to exhibit excellent
energy conservation. In CTP95 the entropy scatter produced by this
equation was compared to that of equation \ref{etc92} and was shown to be
significantly larger. 
Therefore to avoid
this problem we only consider the energy equation \ref{etc92}.   

It has been shown by Nelson \& Papaloizou (1993,1994) and Serna,
Alimi \& Chieze \shortcite{se}, that inclusion of the $\nabla h$ terms, in
both the equation of motion and internal energy, is necessary to ensure
conservation of both energy and entropy. We chose to neglect these terms
due to the overhead involved in computing them. 

\subsection{Artificial viscosity algorithms}
\label{artvisc}
An artificial viscosity is necessary in SPH to dissipate convergent 
motion and hence prevent interpenetration of gas clouds \cite{jj}. 

We investigate four different types of artificial viscosity. The
first considered is the implementation of TC92, where an additional
component is added to a particle's pressure,
\begin{equation}\label{petersa}
P_i \rightarrow \cases{
P_i + \rho_i[-\alpha c_i h_i \nabla . \bmath{v}_i + \beta
(h_i\nabla.\bmath{v}_i)^2 ], &
$\nabla .\bmath{v}_i<0$; \cr
P_i, & $\nabla . \bmath{v}_i \geq 0$,
}
\end{equation}
where $c_i$ is the sound speed for the particle. 
Typically, the viscosity coefficients are $\alpha=1$ and $\beta=2$. This
algorithm is a combination of the Von Neumann-Richtmeyer and bulk
viscosity prescriptions (see Gingold \& Monaghan~1983). A
notable feature is that it uses
a $\nabla.\bmath{v_i}$ `trigger' so that it only applies to particles for
which the local velocity field is convergent. This is different from most
other implementations, which use an $\bmath{r}_{ij}.\bmath{v}_{ij}<0$ trigger. 
With this artificial viscosity the first term in 
equation \ref{2} does not depend upon the
density of particle $j$. This is advantageous numerically since it is 
not necessary to construct
the neighbour lists twice.  In this circumstance one can reduce the SPH search
over particles to a primary loop, during which the density is
calculated and the neighbour list is
formed and stored, followed by a
secondary loop through the stored neighbour list to calculate the
forces and internal energy.

A modification of this artificial viscosity prescription is to
estimate the velocity divergence over a smaller number of neighbours
found by searching to a reduced radius. This was motivated by the
observation that in some circumstances the gas does not shock as
effectively as when a pairwise viscosity is employed. The reduced
search radius leads to a higher resolution (but likely noisier)
estimate of $\nabla .\bmath{v}$. 

The third artificial viscosity we consider is the standard
`Monaghan' artificial viscosity \cite{jj}. This artificial viscosity 
has an explicit $i-j$ particle label symmetry which is
motivated to fit with equation \ref{3}. In this algorithm 
the summation of $P/\rho^2$ terms in equation \ref{3} is extended to
include
a term
$\Pi_{ij}$, which is given by,
\begin{equation}
\Pi_{ij}={ -\alpha \mu_{ij} \bar{c}_{ij} + \beta \mu_{ij}^2 \over
\bar{\rho}_{ij}},
\label{monvis}
\end{equation}
where,
\begin{equation}\label{muij}
\mu_{ij}=\cases{
 \bar{h}_{ij} \bmath{v}_{ij}.\bmath{r}_{ij} / (r_{ij}^2+\nu^2), &
$\bmath{v}_{ij}.\bmath{r}_{ij}<0$;
\cr
0, & $\bmath{v}_{ij}.\bmath{r}_{ij} \geq 0$,
}
\end{equation}
where the bar denotes arithmetic averaging of the quantities for particles
$i$
and $j$ and $\nu^2=0.01\bar{h}_{ij}^2$ is a term included to prevent
numerical divergences. Again, typical values for the coefficients
are $\alpha=1$
and $\beta=2$ although for problems involving weak shocks values of
0.5 and 1, respectively, may be preferable. This artificial viscosity has
the
same functional dependence upon $h$ as the $\nabla . \bmath{v}$ version. 
Since this
algorithm utilises a pairwise
evaluation of the relative convergence of particles it is capable of
preventing interpenetration very effectively. A drawback is that it requires
the neighbour list for particles to be calculated twice  since the force on
particle $i$ depends explicitly on the density of particle $j$ (storing
all the neighbour lists
is possible, but in practice too memory consuming). 

One major concern about this algorithm relates to its damping effect on
angular momentum. In the presence of shear flows, $\nabla .\bmath{v}=0$,
$\nabla
\times \bmath{v}>0$ the pairwise $\bmath{r}_{ij}.\bmath{v}_{ij}$ term can
be non zero
and hence the viscosity does not vanish. This leads to a large shear
viscosity which is highly undesirable in simulations of disc formation,
for example. A way around this problem is to add a shear-correcting term
to the artificial viscosity~\cite{ba}. The
modification we consider is given by Steinmetz (1996);
\begin{equation}
\Pi_{ij} \rightarrow \tilde{\Pi}_{ij}=\Pi_{ij}\bar{f}_{ij},
\end{equation}
where,
\begin{equation}
\bar{f}_{ij}={f_i+f_j \over 2},
\end{equation}
and,
\begin{equation}
f_i={|\<\nabla . \bmath{v}\>_i| \over |\<\nabla .\bmath{v}\>_i| + |\<\nabla
\times \bmath{v}\>_i| +
0.0001c_i/h_i }.
\end{equation}
For pure compressional flows $f=1$ and the contribution of
$\Pi_{ij}$ is unaffected whilst in shear flows $f=0$ and the viscosity
is zero. This modification has been studied by Navarro \& Steinmetz
\shortcite{sn} who found that the dissipation of angular momentum is
drastically reduced in small-$N$ problems using this method. However, of
concern is whether this modification leads to poorer shock-capturing.
This may arise due to sampling error in the SPH estimates of the velocity
divergence and curl. We compare schemes using the shear-corrected viscosity
against the standard Monaghan viscosity.

The efficiency gained from having an artificial viscosity that does not
depend upon the density of particle $j$ has motivated us to consider the
following modification of the Monaghan viscosity: we use the same
quantities as equation \ref{monvis}, except that,
\begin{equation}
\bar{\rho}_{ij} \rightarrow \tilde{\rho}_{ij} = \rho_i(1+(h_i/h_j)^3)/2,
\end{equation}
which provides an estimate of $\bar{\rho}_{ij}$. The estimate is based
upon the approximation $\rho_j \simeq \rho_i h_i^3/h_j^3$. We have
plotted this estimate against the real $\rho_j$ in the spherical
collapse test (see section \ref{evrard}) and find the maximum error
to be a
factor of ten. In practice the maximum error is dependent upon the problem
being studied, but in general we have found the error to usually fall in
the range of a factor of two. Note that the shear-free single-sided 
Monaghan variant would have to implemented as,
\begin{equation}
\tilde{\Pi}_{ij}= { -\alpha \mu_{ij} \bar{c}_{ij} + \beta \mu_{ij}^2 \over
\tilde{\rho}_{ij}}f_i,
\end{equation}
thereby removing the $j$ index. Despite the estimated quantity
$\tilde{\rho}_{ij}$, this artificial viscosity, when used in
conjunction with the equation of motion in
equation~\ref{2}, still results in a momentum-conserving scheme. 

In principle the $\nabla . \bmath{v}$-based artificial viscosity
should not suffer the problem of damping during pure shear flows,
since the artificial viscosity only acts in compressive flows. A
useful test is to supplement this artificial viscosity with the
shear-correction term. This enables an estimate to be made of the
extent to which the correction term under-damps due to SPH sampling
errors.

\section{Test scenarios}\label{tests}
\begin{table*}
 \begin{tabular}{@{}cllc}
  Version & Artificial Viscosity & symmetrization &
equation of motion \\
\hline
 1 & TC92 & TC92 & TC92 \\
 2 & TC92+shear correction & TC92 & TC92 \\
 3 & local TC92 & TC92 & TC92 \\
 4 & Monaghan & arithmetic $h_{ij}$ & SM93 \\
 5 & Monaghan & harmonic $h_{ij}$ & SM93 \\
 6 & Monaghan & kernel averaging & SM93 \\
 7 & Monaghan+shear correction & arithmetic $h_{ij}$ & SM93 \\
 8 & Monaghan+shear correction & harmonic $h_{ij}$ & SM93 \\
 9 & Monaghan+shear correction & kernel averaging & SM93 \\
 10 & Monaghan       & TC92 & TC92 \\
 11 & Monaghan $\tilde{\rho}_{ij}$ & TC92 & TC92  \\
 12 & Monaghan $\tilde{\rho}_{ij}$ & TC92 + kernel av & TC92 \\
\hline
 \end{tabular}
 \caption{Summary of the implementations examined. `Monaghan' is the
artificial viscosity of equation~\ref{monvis}.
Steinmetz-type shear correction~Steinmetz~(1996) is applied where noted.
The remaining terms are discussed in the text.
}
\medskip
 \label{t1}
\end{table*}
The SPH implementations that we examine are listed in Table \ref{t1}.
The list, whilst not exhaustive, represents a range of
common variants for the SPH algorithm. Our motivation is to determine the
extent to which the different implementations affect behaviour in
cosmological settings. In all cases we employ the full $h$-updating
scheme described in
section~\ref{tstep}. 

We have chosen twelve different combinations of artificial viscosity,
symmetrization and equation of motion. The first is that
employed by the \Hydra\ code, (Couchman, Pearce \& Thomas~1996).
We then
add a shear correction term to this and also try a localised
artificial viscosity (as discussed in section~\ref{artvisc}),
resulting in three versions close to the TC92
formalism. We then have three versions based on the popular Monaghan
viscosity (again see section~\ref{artvisc}) each with and without
shear correction. These employ different $h$-averaging schemes
or kernel averaging as discussed in section~\ref{eqnmot} and 
are the most common schemes found in the literature. Finally
we have three new schemes which attempt to combine the best features
of the other implementations. 

\subsection{Shock tube}\label{shock}
The shock tube is an environment which provides an abundant source of
simple tests fundamental to hydrodynamical simulations. 
One well-studied test \cite{jj,hk,rs}
is the Sod shock \cite{sod} for which analytical solutions are
given by Hawley, Smarr \& Wilson \shortcite{ha} and Rasio \& Shapiro
\shortcite{rs}. 
In this test two regions 
of uniform but different density gas are instantaneously 
brought into contact. If initially high
density and pressure gas is to the left and the low density 
and pressure gas is to the right then a rarefaction wave will
propagate left into the high density gas whilst a shock
wave will propagate rightwards into the low density gas. Between
these there will be a contact discontinuity where the pressure
is continuous but the density jumps. This behaviour is shown in
\fig~\ref{shock.sod} where the analytic solution 
is superimposed on the simulation results.

Many authors have carried out this test in 1-dimension, but
this is of limited use as most interesting astrophysical
phenomena are 3-dimensional. The 1-dimensional results do
not automatically carry over to 3-dimensions. The
core of the SPH approach -- the smoothing kernel -- must be
altered to take account of the different volume elements.
Otherwise far too much emphasis will be placed on the central
region in the 1-dimensional tests. Additionally, in 1-dimension,
interpenetration is reduced.

Rasio \& Shapiro (1991) perform this test in 3-dimensions and note the
large increase in numerical scatter and interpenetration
this produces. This is partly because of their choice of
initially randomly distributed particles which introduces
large fluctuations in the supposedly uniform initial density and 
pressure. Contrary to their assertion, this is not a particle
distribution encountered in the course of a typical SPH
simulation. SPH is adept at equalising density and
pressure differences and rapidly relaxes from such an
initial state. 

\subsubsection{Initial conditions}
We have altered our standard cubical simulation volume to
match the geometry of a shock tube. The tube has an
aspect ratio of 16 and all the boundaries are periodic. 
We take the high density gas to have $\rho_l=4$ and $p_r=1$. The
low density gas has $\rho_r=1$ and $p_r=0.1795$. In accordance with the
rest of our tests we set $\gamma={5 \over 3}$. 

In this test there are 4096 low density 
particles and 16384 high density particles.
Before starting the test itself 
each of the regions is independently evolved at constant
temperature until induced random fluctuations have damped
away. This reduces the spurious initial fluctuations mentioned
above and closely mimics the conditions typical of a cold
flow in an SPH simulation. 

\begin{figure*}
 \centering
\psfig{file=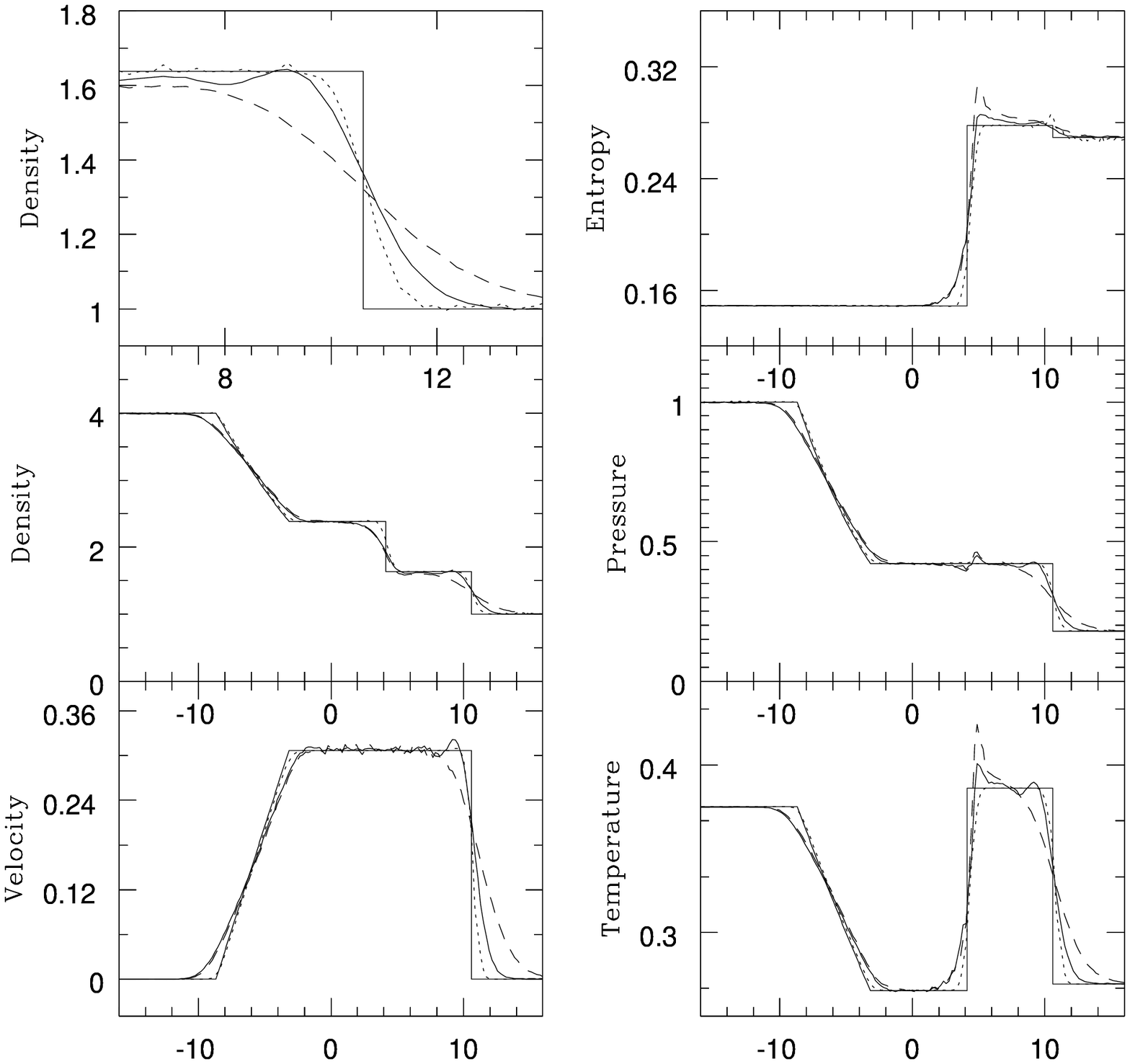,height=18cm}
\caption{A comparison between the analytic result (solid abrupt line)
and those from version 1 (dashed line) and version 12 (solid
line). Also shown is the ideal result (dotted line)
obtained by smoothing the theoretical result with the SPH kernel. Codes
2 and 3 match the profile of 1 closely, the remainder match that of 12.
The x-axis is in units of the SPH smoothing length, $h$,
in the low density gas and the zero point is the original interface
between the high- and low-density gas. The top-left panel shows a zoom
into the 
right-moving shock, the area where the largest difference between the  
codes is seen. 
}
\label{shock.sod}
\end{figure*}

Note that even if the particles resided in exactly the correct
positions to produce the theoretical curve, it is impossible to
reproduce the discontinuities due to the inherent smoothing in SPH
estimates. The smoothed theoretical
curves are shown in \fig~\ref{shock.sod}.

\subsubsection{Shock evolution and results}
For all versions
only around 35 time-steps are required to reach the state shown in
\fig~\ref{shock.sod}, by which time the shock front and the front of
the rarefaction wave have moved around $10h_{low}$. The asymptotic
solutions are recovered around each of the density jumps and even the
small drop in entropy at the shock front is well modelled. Although
the correct jump conditions are captured for the rarefaction wave and
none of the versions introduces spurious entropy, the  
density gradient is too shallow. This effect has also been seen previously
\cite{rs,cp}. At the contact discontinuity the entropy jump
is overshot by all the codes, with the problem worse for versions
1--3. The blip in the ideally uniform pressure at this point is due to
the SPH smoothing of the discontinuous density \cite{jj}.  The
top-left panel of \fig~\ref{shock.sod} shows that the right-moving shock
front in versions 1--3 is broadened 
over a region about twice that seen for versions 4--12 which
themselves have about twice the width of the smoothed
theoretical solution.

\begin{figure}
 \centering
\psfig{file=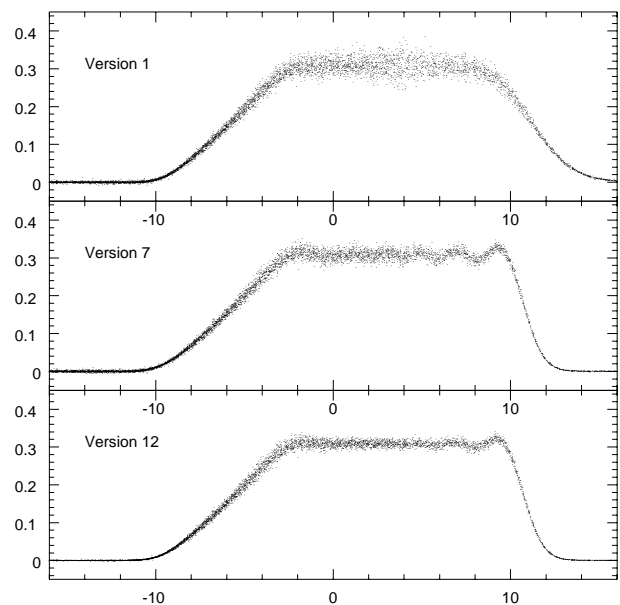,height=9cm}
\caption{The scatter in velocity perpendicular to the shock tube
for versions 1, 7 \& 12. The x-axis is
in units of the SPH search length in the low
density gas. All the particles in the region of interest are shown
as single points. The scatter is much higher and the profile is
clearly different for version 1 than
for either of the other versions. Version 12 exhibits notably less
post-shock ringing than 7. Of the other codes,
Versions 2 and 3 look like version 1, versions 8 and 9 look
like version 7 and versions 4, 5, 6, 10 and 11 look like
version 12.
}
\label{shock.xv}
\end{figure}

\subsubsection{Particle interpenetration}
We have also used this test to study the level of flow
interpenetration. 
Ideally there should be no
interpenetration 
and the flow should remain smooth with the particles in well
defined slabs at the end. As \fig~\ref{shock.xv} shows for versions
1, 2 and 3 the flow is far from smooth, with large differences in
velocity at any one point along the shock tube. Versions 7, 8 and 9
show much less scatter but produce noticeable post-shock oscillation
whilst the remaining versions all produce a very smooth flow with
little velocity scatter.  Due to the large intrinsic scatter,
versions 1--3 of the code suffer from much greater interpenetration
than the other versions.  These effects are exacerbated in the
presence of some initial turbulence. The locally averaged nature of the
viscosity employed by versions 1--3 allows regions to
interpenetrate. The shear-corrected versions (7--9), despite producing
just as sharp a 
density jump as the remaining versions, also perform poorly in the
interpenetration test because 
interpenetrating streams of particles can mimic shear.

\subsubsection{Summary}
For this test the versions split into 2 main groups with the
more accurate codes (4--12) further sub-dividing into two
groups depending upon whether or not the viscosity is shear
corrected. Versions 1--3 produce broader shocks
and suffer from greater interpenetration because these codes
employ a locally averaged method for calculating the viscosity. All
the other versions employ a particle-particle approach to
the viscosity calculation and hence resolve shorter
scales. The addition of a shear-correction term to prevent 
transport of angular momentum degrades the performance of these codes
for this particular test. Although the shock profiles are very
similar (see \fig~\ref{shock.sod}) they produce a flow which
is less smooth (see \fig~\ref{shock.xv}) and suffers
from greater interpenetration than versions 4, 5, 6, 10, 11 and 12.

\subsection{Collapse of a spherical cloud}\label{evrard}
\begin{figure}
\vspace{60mm}
\includegraphics{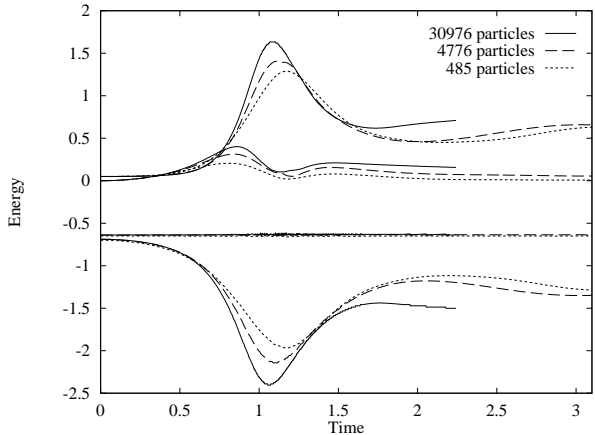}
\caption{Convergence of energy values with particle number for
the Evrard collapse test using version 12. Energy is plotted on the y
axis and time along the x axis 
in normalised units. The sets of curves are, from top to bottom, the
thermal, kinetic, total and potential energies. 
There is comparatively
little difference between the 4776 and 485 particle collapse because they were run with
the same softening length. The 30796 particle run matches the solution calculated in
SM93 very accurately.}  
\label{ecomp}
\end{figure}

This test examines the adiabatic collapse of an initially isothermal
spherical gas cloud. We denote this test the `Evrard collapse' since it
first appears in Evrard (1988), and since then has proved itself as a useful
test case for combined SPH-gravity codes (e.g. HK89, SM93, Serna \etal 1996,
Hultman \& Kallander 1997). 

\subsubsection{Units and initial conditions}
Test results are presented in normalised units.
The density, internal
energy, velocity, pressure and
time are normalised by $3M/4\pi
R^3$, $GM/R$,
$(GM/R)^{1 \over 2}$, $\rho u$ and
$(R^3/GM)^{1
\over 2}$,
respectively, where $R$
denotes the initial radius and $M$ the total mass. The initial
physical density
distribution is given by,
\begin{equation}
\rho(r)={M(R)\over 2\pi R^2}{1 \over r},
\end{equation}
which is achieved by applying a radial stretch to an initially uniform
grid \cite{ge}. We prefer this configuration over a random one since it
has significantly less sampling error than a random distribution
\cite{sw}. The internal energy of the system is chosen to be 0.05$GM/R$,
and the adiabatic index $\gamma=5/3$. The softening length used for each
simulation is given in the caption to Table~\ref{t3}.

\subsubsection{System evolution}
The evolution of the system is 
shown in \fig~\ref{ecomp}. As the collapse occurs the gas is
heated until the temperature of the core rises sufficiently to cause a
`bounce', after which a shock propagates outward through the gas. After
the shock has passed through the majority of the gas, the final state of
the system is one of virial equilibrium.
Along with placing emphasis on the performance of the implementations for
small-$N$ systems we also study convergence at larger values of $N$. A
summary of the runs performed is presented in Table~\ref{t3}.

\begin{figure*}
\vspace{205mm}
\begin{minipage}{170mm}
\includegraphics{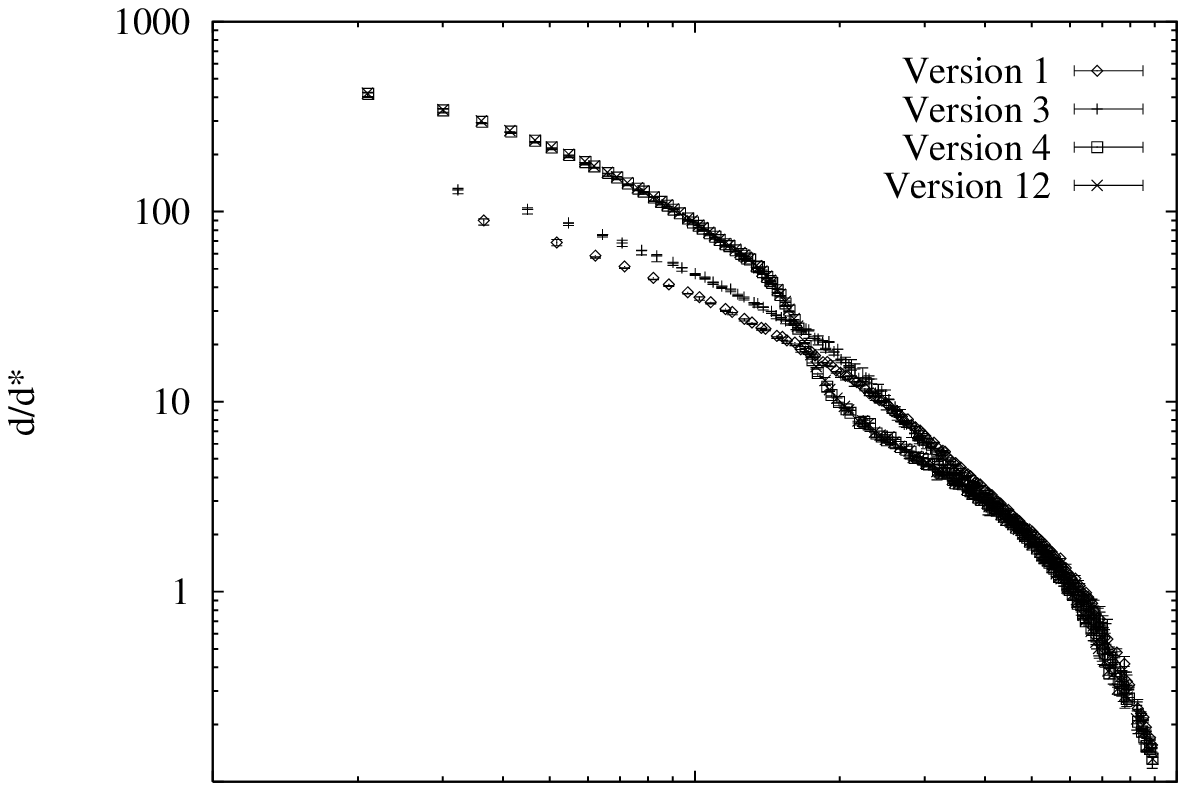}
\includegraphics{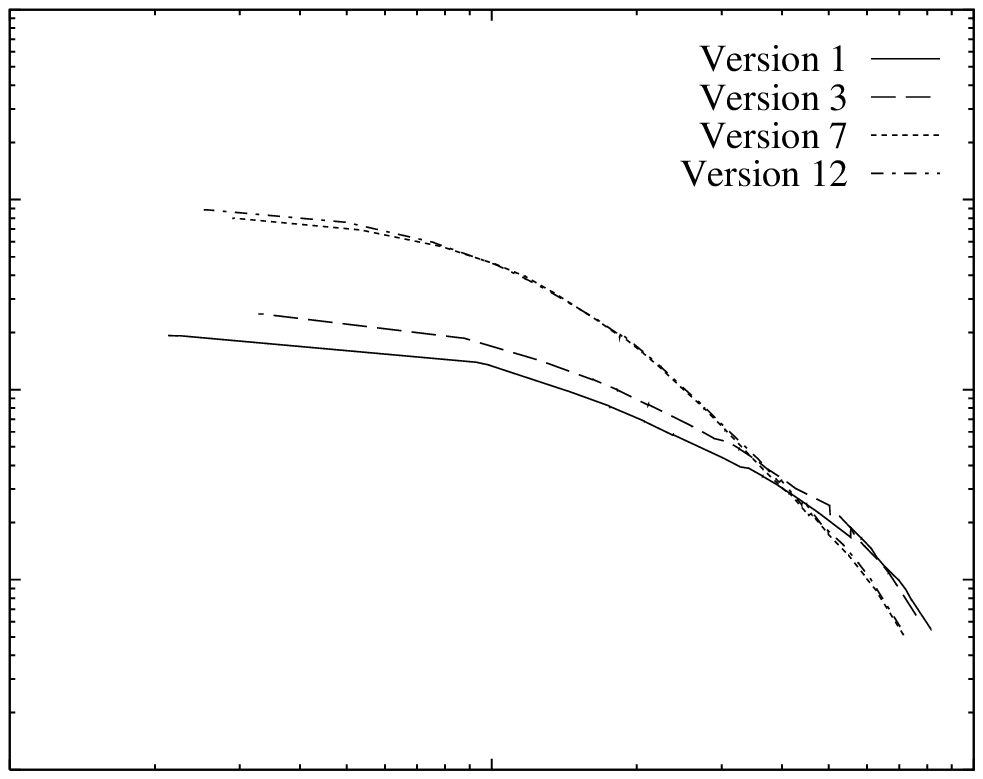}
\includegraphics{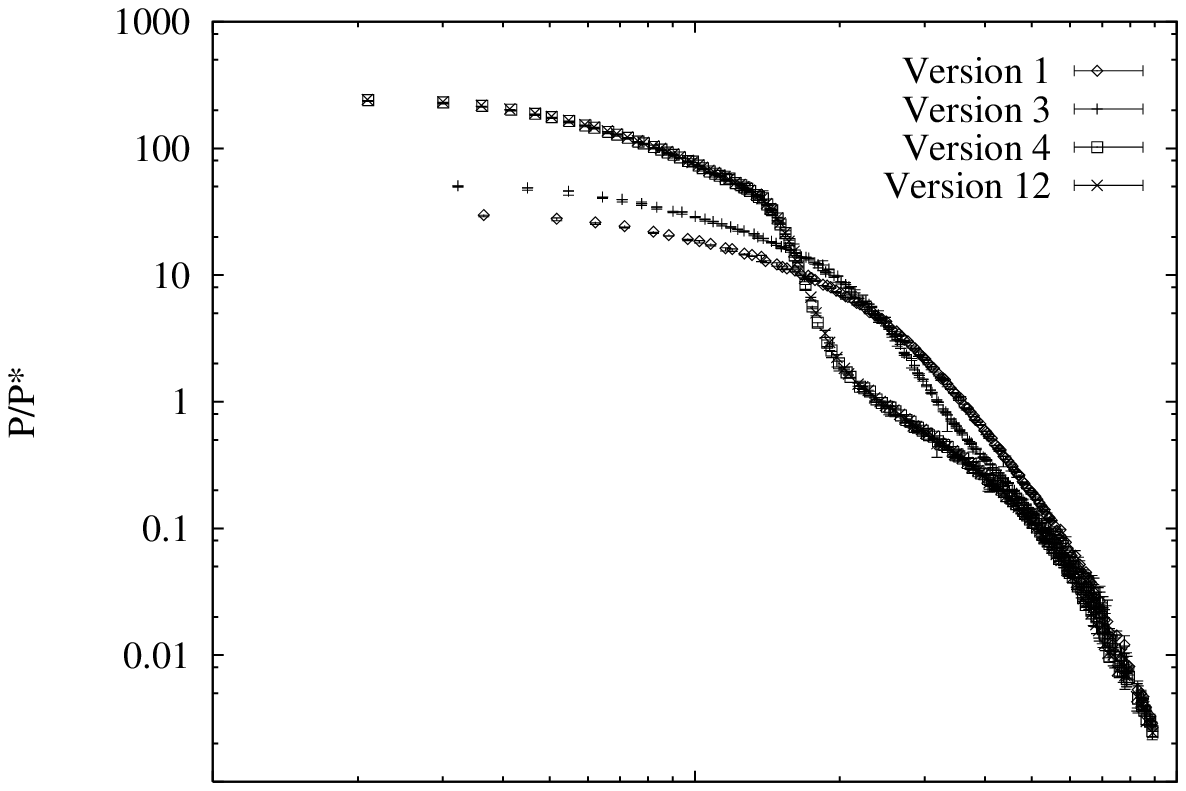}
\includegraphics{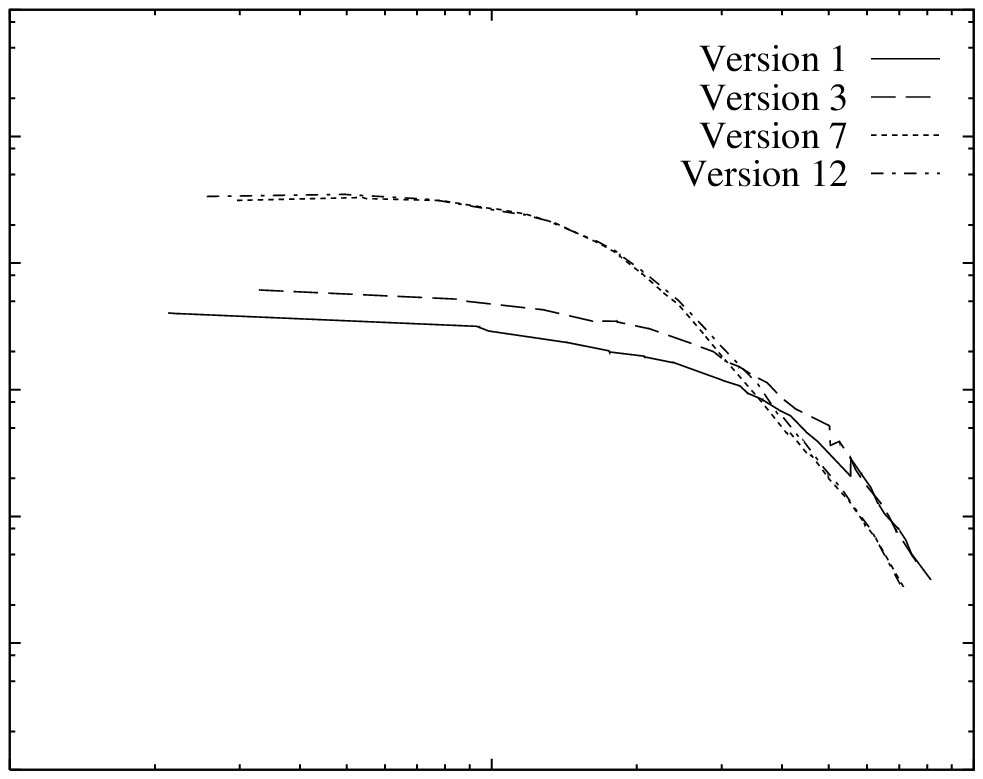}
\includegraphics{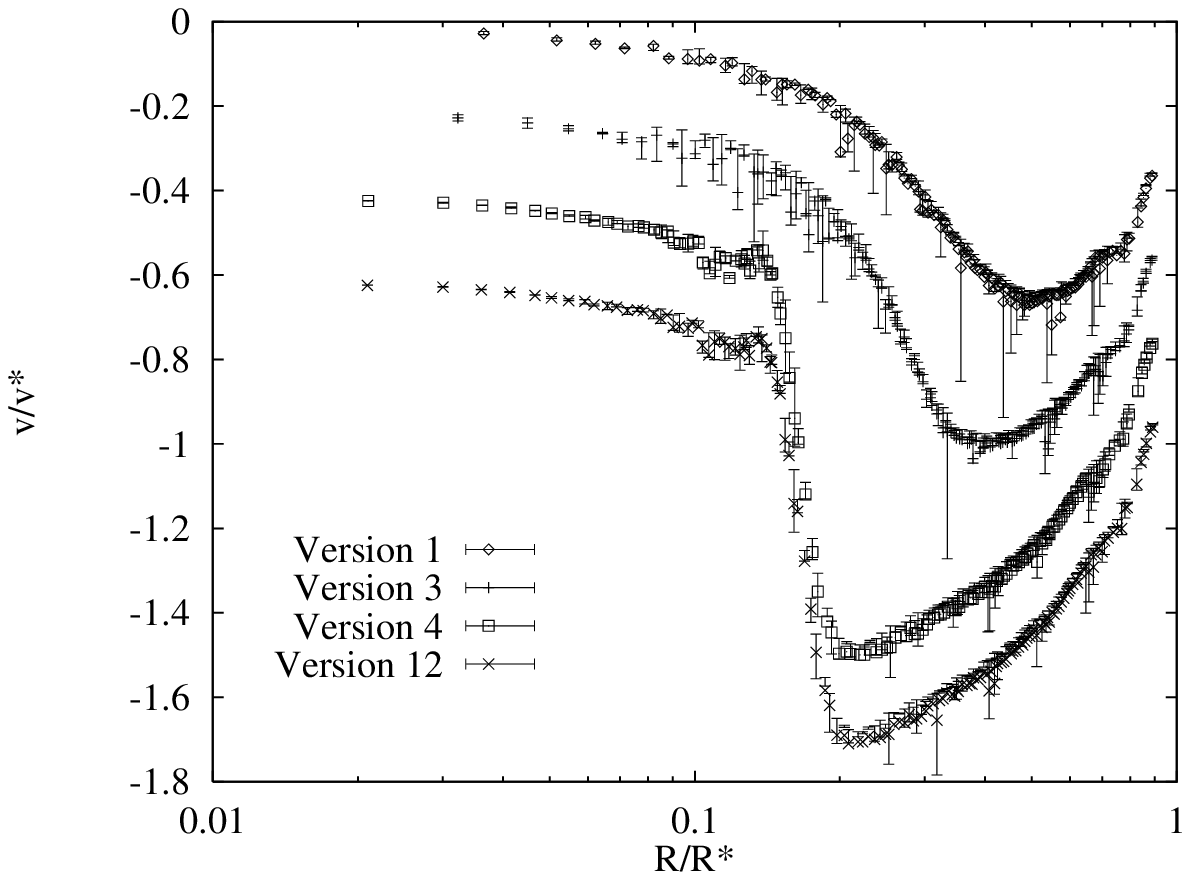}
\includegraphics{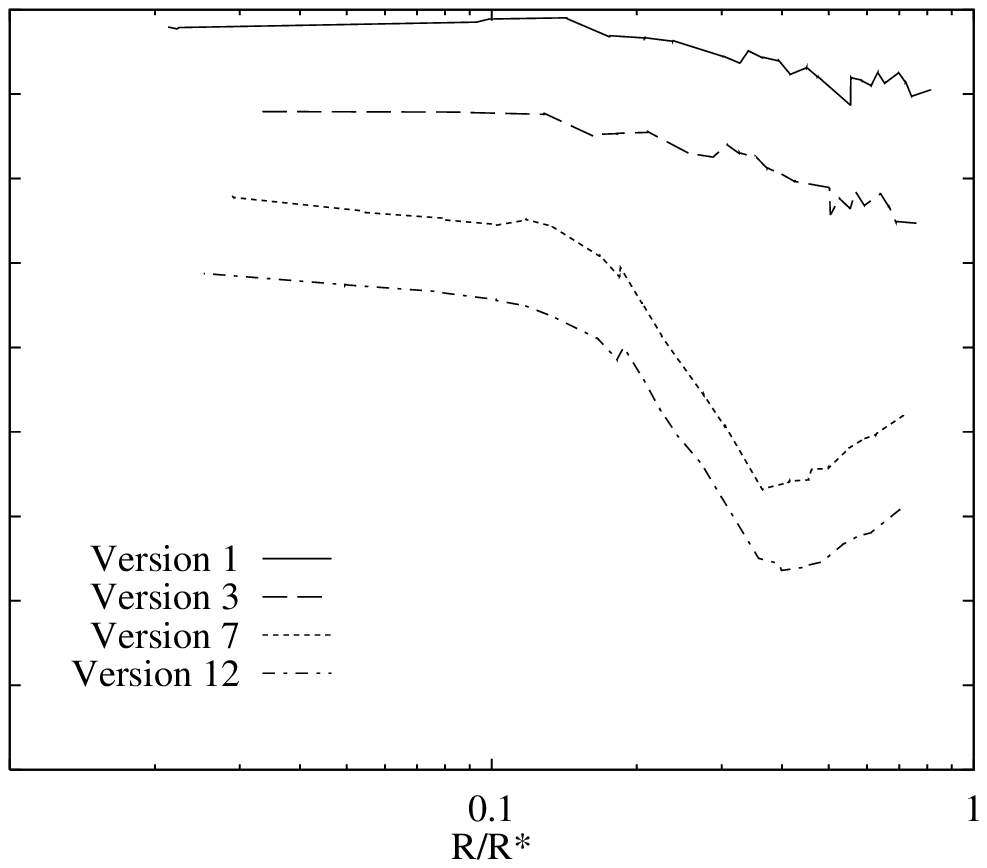}
\caption{Radial profiles for the 30976 and 485 particle collapses at t=0.8 are
displayed in the left and right hand columns respectively. For the
30976 particle tests the density,
pressure and velocity are plotted using Lagrangian bins of size
$4\times52=208$ particles, corresponding to $4\,N_{smooth}$ 
and for the 485 particle collapses a single line connects
all particles.  Error bars on the 30976 particle plots show the {\em
maximum} variation within bins. The velocity plots are displaced by
intervals of -0.2 vertically for easier interpretation.  Notably versions
1 and 3 show significant differences to versions 4, 7 and 12, in
particular
both the density and pressure are underestimated (as compared to the
solution calculated in SM93). Version 3 shows a better capture of the
collapsing shock front than version 1, this effect being most evident in
the pressure plot. Versions 4, 7 and 12 show excellent agreement, being
difficult to distinguish at all radii.  Version 7 shows that the inclusion
of the shear-free term has little effect on the radial profile. }\label{pro30976} 
\end{minipage}
\end{figure*}

\begin{table}
 \caption{Results of Evrard collapse test.}
 \begin{tabular}{@{}ccccc}
  Version & $N_{steps}$ & $\Delta E/E (\times 10^{-3})$& $\Delta L (\times 10^{-7})$ & $N_{par}$  \\
\hline
 1 & 62 & $6.2$  & $49$ & 485\\
 2 & 62 & $3.3$ & $31$ & 485\\
 3 & 66 & $1.1$ & $38$ & 485\\ 
 4 & 94 & $2.5$ & $11$ & 485\\
 5 & 88 & $1.0$ & $39$ & 485\\
 6 & 129 & $1.7$ & $23$ & 485\\
 7 & 104 & $0.8$ & $15$ & 485\\
 8 & 95 & $0.8$ & $65$ & 485\\
 9 & 135 & $1.2$ & $28$ & 485\\
 10 & 90 & $1.5$ & $62$ & 485\\
 11 & 84 & $1.5$ & $40$  & 485\\
 12 & 103 & $0.6$ & $29$ & 485\\
      &     &                     &                      &    \\
 1 & 82 & $6.1$  & $2.8$ & 4776\\
 2 & 101 & $6.1$ & $6.7$ & 4776\\
 3 & 98 & $1.9$ & $14$ & 4776\\ 
 4 & 240 & $3.1$ & $45$ & 4776\\
 5 & 241 & $3.2$ & $24$ & 4776\\
 6 & 231 & $3.5$ & $63$ & 4776\\
 7 & 243 & $3.2$ & $68$ & 4776\\
 8 & 252 & $3.1$ & $38$ & 4776\\
 9 & 242 & $3.7$ & $64$ & 4776\\
 10 & 234 & $2.9$ & $27$ & 4776\\
 11 & 227 & $2.8$ & $29$  & 4776\\
 12 & 241 & $2.8$ & $7.0$ & 4776\\
      &     &                     &                      &     \\
 1 & 289 & $0.3$ & $1.3$ & 30976\\ 
 3 & 218 & $1.9$ & $0.91$ & 30976\\
 4 & 418 & $4.4$ & $1.4$ & 30976\\ 
 6 & 411 & $4.5$ & $5.5$ & 30976\\
 11 & 405 & $3.6$ & $7.0$ & 30976\\ 
 12 & 489 & $4.7$ & $2.9$ & 30976\\ 
\hline
\end{tabular}
\medskip

Values for the 485 particle test are given at t=4.3, for the 4776 test at
t=3.4 and for the 30976 test at t=2.0. $\Delta L$ is measured in internal
code units, normalising by $L_{start}$ is not possible since
$L_{start}=0$. The 485 and 4776 particle tests used a softening length of
0.05R, the 30976 particle tests used a softening length of 0.02R. 
\label{t3}
\end{table}

\subsubsection{Results from the 485 particle collapse} 
With the $N$ considered here, the standard \Hydra\ code, and variants of it,
do relatively poorly in this test. A lack of thermalization is clearly
visible in \fig~\ref{land485}, where the difference in energy
between versions 1--3 of the code and version 12 are shown
along the top row. These implementations all
have thermal energy peaks that are half
that of the other codes.  The other versions perform reasonably
similarly with minor differences being seen in the peak thermal energy and
in the strength of post bounce oscillation (note the change in
$y$-axis scale on the bottom two rows of \fig~\ref{land485}).

\begin{figure}
\vspace{60mm}
\includegraphics{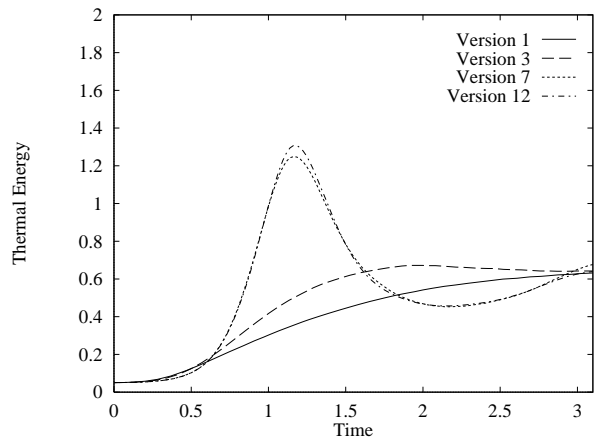}
 \caption{Thermal energy plot for the 485 particle collapse, comparing
versions 1, 3, 7 and 12. A lower peak thermalization is clearly visible
in
versions 1 and 3, whilst 7 and 12 show similar profiles. A
comparative plot of the energy difference between version
12 and the other codes is shown in \fig~\ref{land485}.}
\label{485therm}
\end{figure}  

\begin{figure*}
\vspace{220mm}
\begin{minipage}{170mm}
\includegraphics{land485.ps}
\end{minipage}
\caption{}
\label{land485}
\end{figure*}

\begin{figure*}
\vspace{220mm}
\begin{minipage}{170mm}
\includegraphics{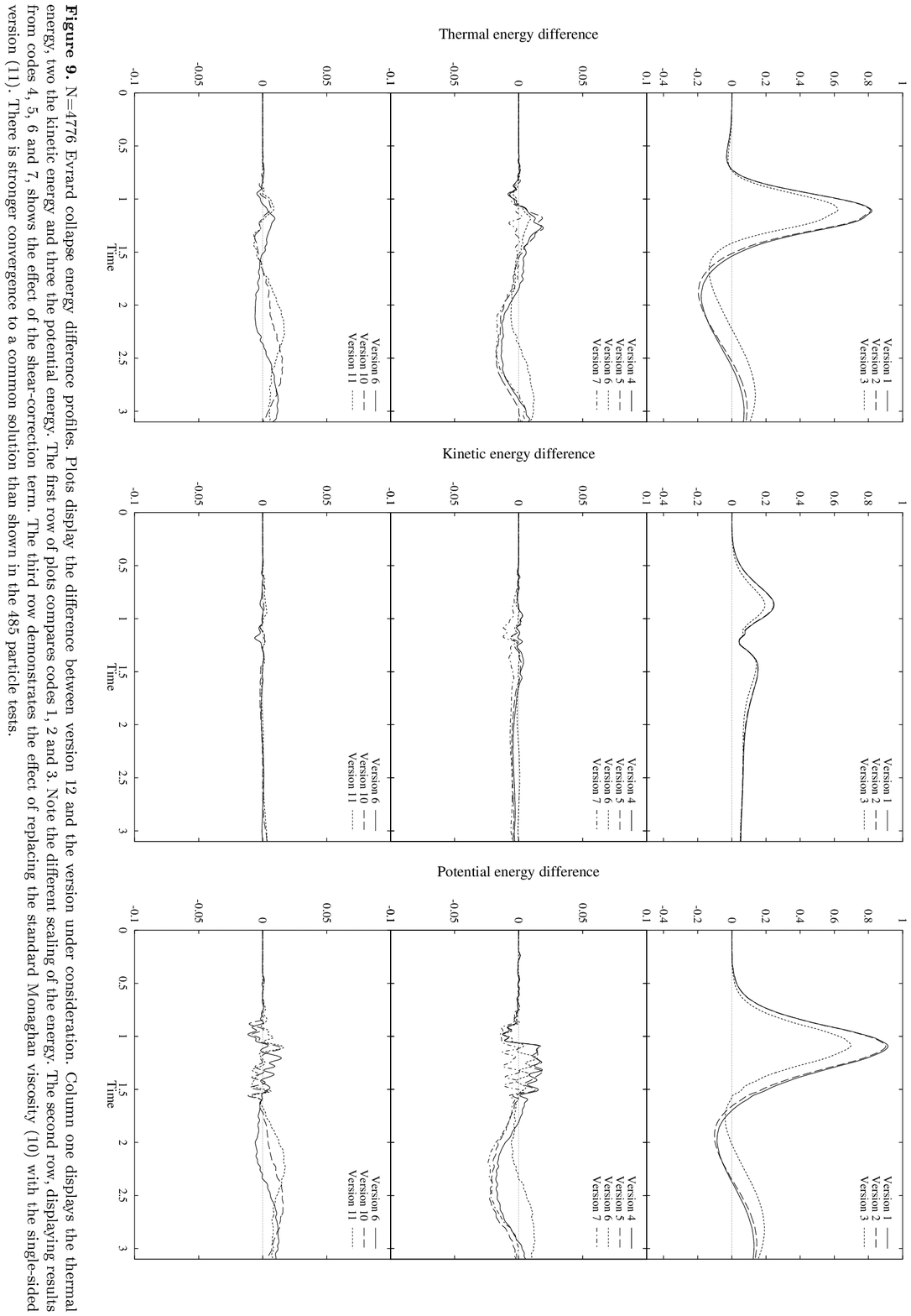}
\end{minipage}
\caption{}
\label{land4776} 
\end{figure*}

The modified viscosity variant, 3, is slightly better at thermalization
than versions 1 \& 2 but still much worse than any of the other
versions. This indicates that the
artificial viscosity is the primary factor in deciding the amount of
kinetic energy that is thermalised (as expected).  The more local
estimate of $\nabla . \bmath{v}$ used in version 3 captures the strong flow
convergence near the bounce better than the standard estimate, leading to 
greater dissipation. At the other extreme the pairwise Monaghan viscosity,
which uses the $\bmath{r}.\bmath{v}$ trigger, leads to far more dissipation at
bounce. It is important to note that the final
energy values for the virialised state are very similar for all codes,
even though their evolution is very different in some cases.

Similar characteristics can be seen in the kinetic energy
graphs. The gas in versions 1--3 develops very little kinetic energy. 
The primary cause of this is the $\nabla . \bmath{v}$ artificial viscosity
which trips during the early stages of collapse (prior to t=0.6) and
causes an increase in the thermal energy for all particles. This acts to
decrease the compressibility of the gas.  For all the other codes
which use the Monaghan viscosity (or variant), the
$\bmath{r}.\bmath{v}$ trigger produces far less dissipation during the
early stages of evolution, and the gas develops more kinetic energy.

For the $N=485$ test, \figs~\ref{land485}, \ref{485therm} and \ref{pro30976}
demonstrate that
there is no clear optimal implementation, but the general comparison of
versions 1, 3, 7 and 12 in \fig~\ref{485therm}, indicates that some
perform marginally better than others. Notable features of the high
resolution radial PPM solution
in SM93 are a strong initial peak in the thermal
energy and little post bounce oscillation. If we choose a model on the
basis of these criteria then version 12 performs best, although 
it is difficult to differentiate versions 4--12 in \fig~\ref{land485}.
Version 12 has both a high initial peak and very little 
post bounce oscillation. It also conserves energy well.
The shear-correction term does have some effect
(middle row of \fig~\ref{land485}), in agreement with
the observations in section \ref{shock}. The general influence of the
shear-correction is to increase
the peak thermal energy at bounce and introduce slightly more post-bounce
oscillation, although, again, 
this is not a significant effect. The term has little
effect on the radial profile. 

The effect of $\tilde{\rho}_{ij}$ replacement
can be seen in the bottom row of \fig~\ref{land485}.
Versions 10 and 11 differ only by this substitution and there is very little
to choose between them, the scatter between the kernel-averaged version 6
and version 12 being as large.
None of the implementations considered show poor energy conservation, and
all show excellent conservation of angular momentum.

\subsubsection{Effect of numerical resolution}
\begin{figure}
\vspace{60mm} 
\includegraphics{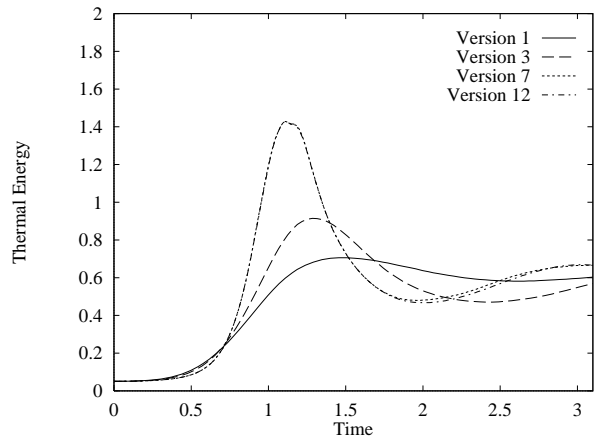}
 \caption{Thermal energy plot for the 4776 particle collapse, comparing
versions 1, 3, 7 and 12. As in \fig~\ref{485therm}, versions 1 and 3   
exhibit a lower peak
thermalization although the peak value is marginally higher. Versions 7
and 12 again have similar profiles and agree very accurately on the peak
thermal energy. A
comparative plot of the energy
difference
between version 12 and the other codes is shown in \fig~\ref{land4776}.}
\label{4776therm}
\end{figure}

Increasing $N$ to 4776 produces the expected results. The implementations
with
pairwise artificial viscosity converge to very similar energy
profiles, see \figs~\ref{land4776} and~\ref{4776therm}. The 10\% spread
seen in the $N=485$ test is reduced to close 
to  
1\% and the limited scatter visible in the radial profiles is 
further reduced. The shear-correction term also has much less effect on
the radial energy
profile.  For comparison, in \fig~\ref{ecomp} we show the convergence of runs
performed with different particle number using code version 12. This plot should be
compared with \fig~6 of SM93. Clearly for a Monaghan-type
viscosity the differences caused by particle
number and softening parameter are much larger than those caused
by the choice of SPH implementation for this range of particle number.

Versions 1, 3, 4, 6, 11 and 12 were run with 30976 particles to check for
convergence of the implementations at high resolution. Radial
profiles at t=1.4 are plotted in \fig~\ref{pro30976}. It is evident from
these profiles that the solutions are much closer than the radial profiles
for the 485 particle collapse. However, the difference between the
versions with the standard hydra viscosity and the Monaghan variants
remains comparatively large -- a factor of two at the centre in the
pressure
and density at t=1.4. (The relatively large energy error is a product
of our
choosing a longer timestep normalization, $\kappa=1.5$ versus 1, to run
the simulation in a shorter wall-clock time.) The profiles of the
Monaghan variants all compare
well to the radial PPM solution presented in SM93.

\subsubsection{Summary}
We conclude that the relatively poor shock capturing
ability of the $\nabla .\bmath{v}$ viscosity of TC92
is a severe impediment to correctly
calculating the evolution of this system. In contrast, the Monaghan
viscosities (including the shear-corrected variants) correctly follow the
evolution. 
All the Monaghan variants perform well enough in this test to
be acceptable algorithms, and when 30976 particles are used in the
test it is almost impossible to differentiate between methods (at least
in the well resolved core regions). At low resolution (485 particles) the
kernel-averaging variants produce a slightly higher thermal peak,
although some additional ringing is visible for version
6, but not for 12. At medium resolution (4776 particles) convergence is
stronger than the low resolution runs and the difference in post shock
ringing is removed.
On this basis version 12, when combined with its extremely fast solution
time, is the preferable implementation.

\subsection{Cooling near steep density gradients}
\label{cool}

Large density gradients occur in the gas in cosmological simulations as a
result of radiative cooling. In simulations these occur when cold dense knots
of gas form within hot haloes (see section \ref{cosmo}). If the smoothing
radius of a hot halo particle encompasses the cold clump then it is likely
that it will smooth over an excess number of cold gas particles leading to an
over-estimate of the particle's density. Consequently the radiative cooling
for the hot particle is over-estimated, which allows the hot particle to cool
and accrete on to the cold clump even though, physically, the two phases would
be essentially decoupled.  This situation should be alleviated by
implementations that smooth over a fixed number of particles. We term this
effect {\em overcooling}. It should not be confused with the overcooling
problem (or `cooling catastrophe') in simulations of galaxy formation. 

\subsubsection{Description of the halo-clump systems\label{cool.systems}}

To examine this phenomenon we created core-halo systems each
consisting of a cold clump of gas surrounded by a hot
halo both being embedded in a dark-matter halo. 
The dark-matter and hot gas system was extracted directly from a
cosmological simulation. The cold clump was created by randomly placing
particles inside a sphere of size equal to the gravitational softening length and 
allowing this system to evolve to a relaxed state. The cold clump was then
placed in the hot gas and dark matter system. Two core-halo
systems -- designed to resemble galaxy clusters -- were 
created to test the effect of mass and linear scale dependence, with 
total masses $5\expd{14}\Msun$ and $5\expd{15}\Msun$. The parameters
of the systems are listed in Tables~\ref{Eric.Tab.cool.Clusters}
and~\ref{Eric.Tab.cool.Media}.

\begin{table}
\caption{ Cluster parameters.} 
\begin{tabular}{ccc}
Cluster                  & $5\expd{14}\Msun$  & $5\expd{13}\Msun$  \\
\hline
$R_{clump}$ ($\kpc$)         & 20                   & 9                      \\
$m_{gas}$   ($10^{10}\Msun$) & 1                    & 0.1                    \\
$m_{dark}$  ($10^{10}\Msun$) & 9                    & 0.9                    \\
$\epsilon$  ($\kpc$) & 20 & 10 \\
\hline
\end{tabular}
\medskip

$R_{clump}$ is the radius of the cold clump, $\epsilon$ is the  
gravitational softening length and
$m_{gas}$ and $m_{dark}$ are the mass of a gas and dark-matter
particle respectively.
\label{Eric.Tab.cool.Clusters}
\end{table}

\begin{table}
\caption{Cluster parameters in common.}
\begin{tabular}{cccc}
               & Cold clump    & Halo gas             & Halo dark-matter \\
\hline
$N$            & 500           & 4737                 & 4994             \\
$T$ ($\K$)     & $10^4$        & $2\times10^5 - 10^8$ & N/A              \\
$\rho/\rho_c$  & $2\times10^6$ & $10^2 - 10^5$        & $10^3 - 10^6$    \\
\hline
\end{tabular}
\medskip

$N$ is the particle number, $T$
the temperature range and
$\rho/\rho_c$ the ratio of the density to the critical density.
\label{Eric.Tab.cool.Media}
\end{table}

Because the time-step criterion makes no reference to the Courant condition
for cooling, it is possible that hot halo particles may not cool correctly
as they accrete on to the cold clump. Tests showed that choosing the time-step
normalization $\kappa \le 0.5$ was sufficient to avoid this problem.

\subsubsection{Testing the overcooling phenonomenon}
\label{cool.description}

\begin{figure}
\epsfxsize=8.5cm
\centerline{\epsffile{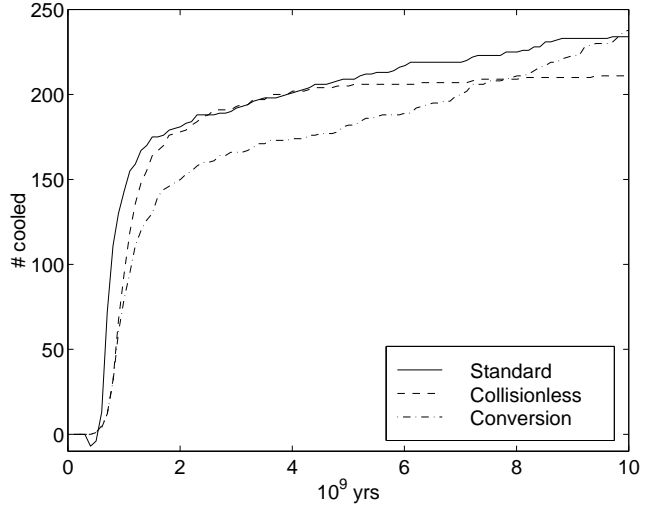}}
\caption{The number of cooled particles in the three experiments as a
function of time. If the hot gas did not interact hydrodynamically with
the cold dense clump in the simulations these lines would overlap.}
\label{Eric.Fig.cool.3schemes}
\end{figure}

For each cluster, we prepared three experiments
that examine how the nature of the 
central clump changes the overall cooling rate. For the first experiment
the central clump was left as a cold knot of gas, for the second it was
turned into collisionless matter and for the third experiment the hot gas
was allowed to become collisionless once it cooled below $2\expd{4}\K$. 
We denote these tests as `standard', `collisionless' and `conversion',
respectively.  If the hot gas did not interact with cold gas then these
tests would all give the same result. The number of particles cooled
over time for these three tests for the $5\expd{14}\Msun$ cluster is shown in
\fig~\ref{Eric.Fig.cool.3schemes}.  

The behaviour at early times ($<2\expd{9}\yrs$) is dominated by a sudden
rise in the number of cold particles. This is due to the hot gas within
$2h$ of the dense clump responding to its sudden introduction. In the
standard test, gas particles comprise the dense clump and hence the
densities calculated for the hot gas rise suddenly, causing rapid cooling.
For both the collisionless and conversion tests, the dense clump is
collisionless;  the hot-gas densities rise only in response to contraction
of the halo about the clump.  Overcooling does not become significant in
the collisionless test until a sufficient number of gas particles
(approximately 40) have cooled and contracted in the central region to
provide a 'seed' clump. Once formed, this seed clump permits the rapid
overcooling of the rest of the hot gas particles in its immediate
vicinity.  Consequently, a comparable number of particles are cooled
during this period as in the standard test. By construction the conversion
test never forms this seed clump and hence overcooling is never initiated.
Cooling occurs only by the contraction of the hot gas halo. 
 
At later times ($>2\expd{9}\yrs$), the greater number of gas particles in
the central region for the standard test (600 versus 200 for the
collisionless test) creates a larger gas-density gradient and hence more
efficient overcooling. The accretion is fed by a contracting halo.  It is
this quasi-steady state which is examined in Section~\ref{cool.results}.
For the conversion test, the increasing central mass density, coupled with
the absence of a central gas clump to provide pressure support, leads to a
progressively increasing cooling rate which is not directly related to the
overcooling phenomenon. 

\begin{figure*}
\epsfxsize=15cm
\centerline{\epsffile{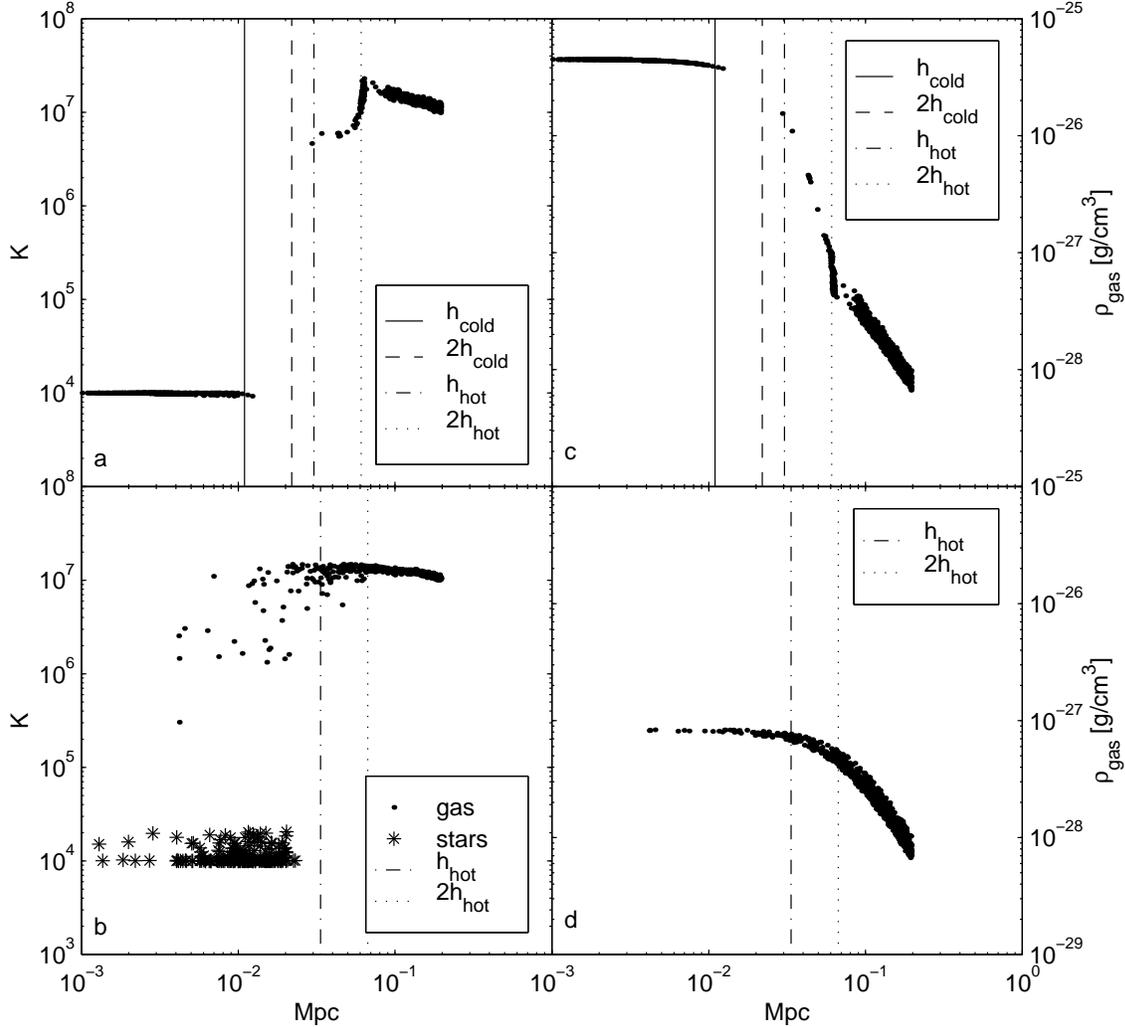}}
\caption{Temperature and density profiles for the standard test (a, c)
and conversion test (b, d).
The mean smoothing radii for both the hot and cold
particles are represented by the vertical lines.  The line for $h$ shows
the point internal to which the bulk of the kernel weighting is
accumulated.  The $2h$ line gives the outer limit of the smoothing.  The
inner pair of lines (left) are for the cold gas, while the
outer pair are for the hot gas. In the lower panel, the cold particles
are collisionless, and hence have no smoothing length. For these data,
the initial conditions of the $5\expd{14}\Msun$ mass cluster were
used. In the conversion test the stars are plotted at the temperature
to which they had cooled just before conversion to star particles.} 
\label{Eric.Fig.cool.Profiles_4Panel}
\end{figure*}

The a-physical
drop in temperature of the `hot' gas particles when the cold clump falls
within twice their smoothing length is clearly illustrated by
comparing the temperature profile of the gas produced in the standard test
with that produced by the conversion test 
(\figs~\ref{Eric.Fig.cool.Profiles_4Panel} (a) and (b)).  
The conversion test produces
a profile that approximately resembles that to be expected in the absence
of hot and cold gas phase interaction. For this test the gas is
approximately isothermal except in the core, where the increased density
has caused the gas to cool -- this is clearly different to the standard
test. Note that in the conversion test the central core is more extended
than in the standard test, but remains within a sphere smaller than the hot
gas smoothing radius. 

The behaviour of the gas at the interface between the gas phases is
illustrated in
\fig~\ref{Eric.Fig.cool.Profiles_4Panel}.  For the
standard test the smoothing process forces the density to rise very
abruptly from the halo to the core, whilst for the conversion test the
lack of a cold gas core removes this imperative.  Consequently in the
standard test, particles outside the dense core, but within $2h_{hot}$,
have a high density and cooling rate. Thus once particles fall within
$2h_{hot}$ an abrupt temperature decrease results as the cooling rate
increases -- as seen in panel (a).  In the conversion test the flatter
density profile does not lead to excessively high cooling rates, and there
is no abrupt temperature drop.

\begin{figure}
\epsfxsize=8.5cm
\centerline{\epsffile{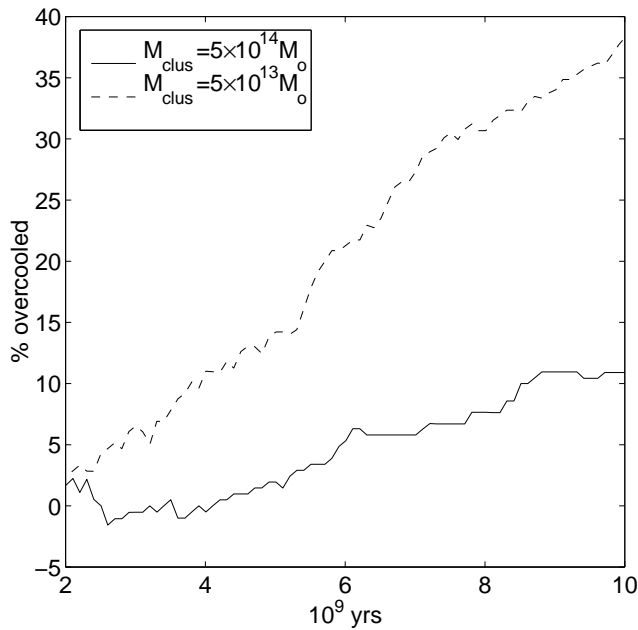}}
\caption{The amount of overcooled gas for two clusters of different
mass. The y-axis is the percentage excess of cooled gas for a
`standard' run  relative to
the amount cooled in the corresponding `collisionless' run.} 
\label{Eric.Fig.cool.MassRes}
\end{figure}

The cooling rate of the gas is also affected by the virial temperature
of the halo gas. The overcooling for the two different clusters are
compared in \fig~\ref{Eric.Fig.cool.MassRes}.

\subsubsection{Results of the different SPH implementations}
\label{cool.results}

Experiments were run with all 12 different SPH implementations. The
collisionless tests produced cooling rates which were essentially
identical. We report on implementation sensitivities for the standard
test. In order to assess the significance of the observed variations
we note that four different realisations of the same test produced a
significant variation of up to 40 cooled particles after $2\Ga$.
We expect, however, that for a given realisation the general trend
amongst the SPH variants would be the same.

\begin{figure*}
\epsfxsize=15cm
\centerline{\epsffile{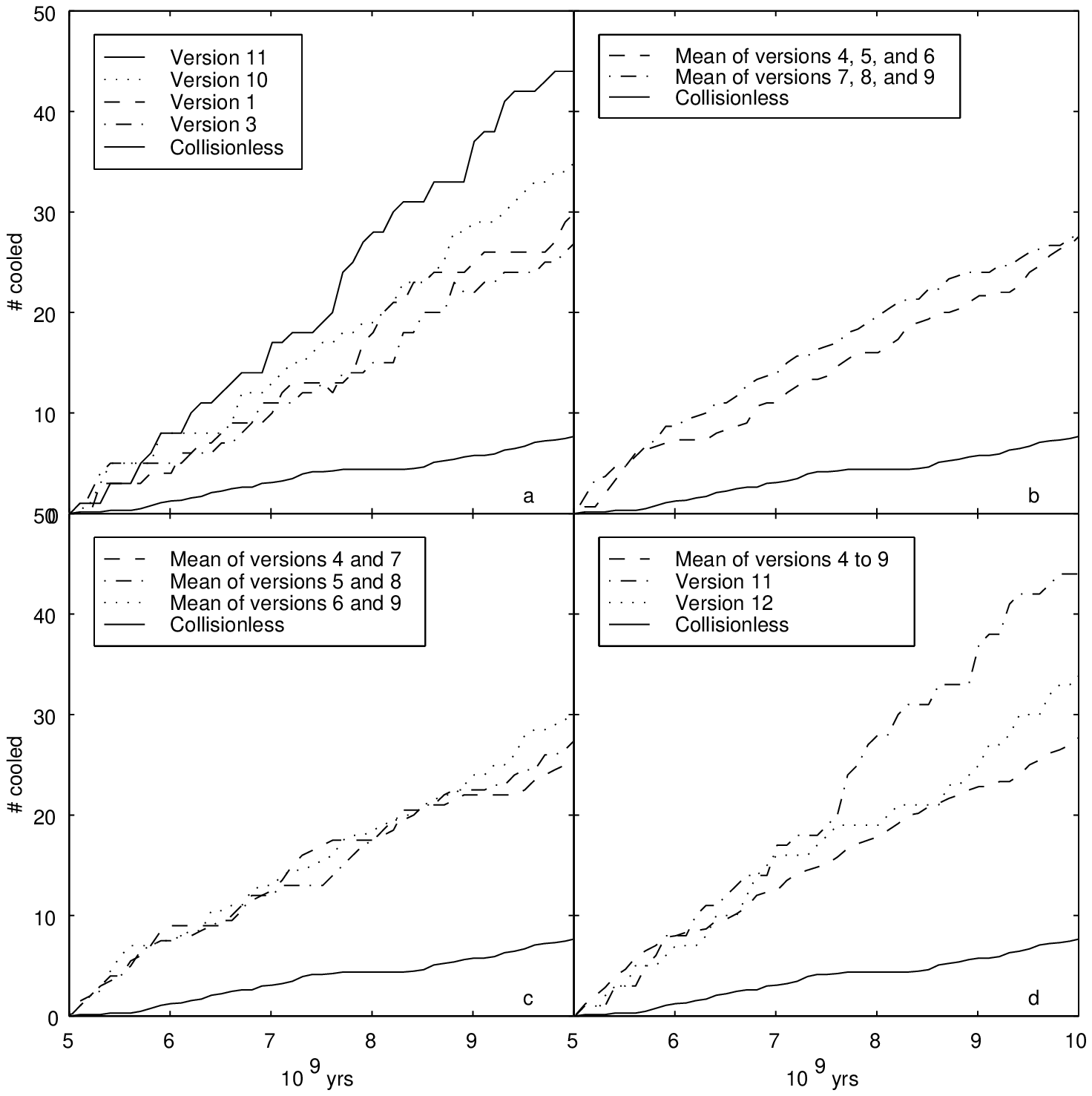}}
\caption{ Variation with time in the number of cold particles ($ T < 10^5
\K$) due to:  a) different implementations of the artificial viscosity. 
The viscosity given by TC92, both in original forms (dashed line) and in
the more localized implementation (dot-dashed line) lead to a similar
amount of cooling as the artificial viscosity given by Monaghan (dotted
line).  However, the one-sided version of the Monaghan viscosity (upper
solid line) does produce an extra excess of cold particles.  For
comparison, the amount of cooling produced in the run with an initially
collisionless dense core is
 given by the solid line.  b) the presence of a Balsara term in the
artificial viscosity.  c) the symmetrization scheme. Those implementations
using arithmetically averaged values of the smoothing lengths, $h$,
produce the mean cooling curve given by the dashed line.  The mean curved
produced by those that use a harmonically averaged value of $h$ give the
dot-dashed line.  If the kernels themselves are averaged, the dotted line
is the result.  d) the symmetrization of TC92 and its variant.  The mean
of the rates given in panel c) (dashed line) is just slightly lower than
the rates produced by the code version which uses the symmetrization
procedure of TC92 as well as that of its more localized
variant, both of which are essentially the same.
These data are for the $5\expd{14}\Msun$ cluster.
}
\label{Eric.Fig.cool.4Panel}
\end{figure*}

The primary source of variation in the overcooling rate among the 
implementations of SPH is the viscosity (\fig~\ref{Eric.Fig.cool.4Panel}
a).  The $\nabla.\bmath{v}$-based artificial viscosities produce cooling
rates that are essentially indistinguishable, while the
Monaghan variants lead to a significantly greater cooling (about  
$50\%$ more).  This difference is probably related to the $\nabla .
\bmath{v}$ variants providing somewhat greater pressure
support in the core (see section \ref{evrard}). The inclusion of a
shear-correction term
in the artificial viscosity (\fig~\ref{Eric.Fig.cool.4Panel} b) has
little effect on the cooling rate -- as expected.

Since the overcooling effect is caused by the large difference in kernel
sizes associated with the hot halo particles and the cold clump
particles, it might be expected that the symmetrization has a role to play
in determining the cooling rate. Consider the $h$-averaging schemes: the
arithmetic mean is limited to having a minimum value of $h_{large}/2$,
while the harmonic mean is zero if any particle interacts with another
particle having $h=0$. In practice, as is shown in
\figs~\ref{Eric.Fig.cool.4Panel} c) and d), there is little difference
between all symmetrization schemes, with the exception of version 11. This
version combines a pure gather kernel, with the single-sided Monaghan
artificial viscosity. This result is surprising in view of the
comparatively `normal' results for
versions 10, which differs in terms of the artificial viscosity, and
12 which differ in terms of the symmetrization.

\subsubsection{Summary}
All the versions of
SPH we have tested exhibit overcooling and this effect should be seen as
generic to the method itself. SPH will always experience difficulties
modelling arbitrarily steep density gradients.  The only implementation
that stands out as performing poorly is 11 which couples a
one-sided implementation of Monaghan 
artificial viscosity with the TC92 symmetrization procedure. When
the TC92 symmetrization is supplemented with kernel averaging the problem
is removed.

\subsection{Drag}
\label{drag}

There is concern that the over-merging problem encountered in N-body
simulations of clusters of galaxies is exacerbated in simulations which
use SPH (see Frenk {\it et al.} 1996). Excessive drag on small knots
of gas within a hot halo will cause the knots to spiral inward into
regions of stronger tidal forces where they may be disrupted (e.g., Moore
{\it et al.} 1996). In this section we model a cold dense clump moving
through a hot halo and investigate if the problem is sensitive to the
particular SPH implementation employed.

\subsubsection{Drag test model systems}
\label{drag.systems}

\begin{table}
\caption{The characteristics of the cold clumps and the hot media used in
the drag tests.}
\begin{tabular}{cccc}
                  & Slow cold clump & Fast cold clump \\
\hline
$\rho/\rho_c$     & $1000$          & $1000$          \\
$T$ ($\K$)        & $10^4$          & $10^4$          \\
$R$ ($\kpc$)      & $50$            & $50$            \\
$N$               & $100$           & $100$           \\
$m$ ($10^9\Msun$) & $1.7$           & $1.7$           \\
$V_o$ ($\kps$)    & $500$           & $1000$          \\
\hline
\hline
                  & Hot gas         & Very hot gas    \\
\hline
$\rho/\rho_c$     & $10$            & $10$	      \\
$T$ ($\K$)        & $10^7$          & $10^8$          \\
$N$               & $13000$         & $13000$         \\
$m$ ($10^9\Msun$) & $1.7$           & $1.7$	      \\
$V_s$ ($\kps$)    & $500$           & $1500$          \\
$R_J$ ($\Mpc$)    & $6$             & $18$	      \\
\hline
\end{tabular}
\medskip

Given are the overdensity, $\rho/\rho_c$ ($h_{100}=1$), the temperature,
$T$, the radius of the cold clump, $R$, the number of particles in the
medium, $N$, the mass resolution of the medium, $m$, the initial velocity
of the cold clump, $V_o$, the speed of sound in the hot medium, $V_s$, and
the Jeans length for the hot medium, $R_J$.  The simulation volume in all
cases is $(5 \Mpc)^3$.  The `fast cold clump' was used in the Mach 2 runs
in combination with the `hot gas'.  The Mach 1 runs used the `slow cold
clump' embedded in the `hot gas'.  The Mach 1/3 runs used the `slow cold
clump' in the `very hot gas'. 
\label{Eric.Tab.Drag.Init}
\end{table}

To cover a variety of infall speeds we examine the deceleration of a knot
of cold gas in three velocity regimes:
Mach 2, Mach 1, and Mach 1/3. The Mach 2 and Mach 1 tests differ in
terms of the speed of the cold knot (`fast' versus `slow') and not the
temperature of the hot gas. The Mach 1/3 test uses the same clump velocity 
as the Mach 1 test, but is performed in hotter (`very hot') gas.
Table~\ref{Eric.Tab.Drag.Init} gives the details of the cold clump and
hot gas phases. Clump characteristics are selected to loosely emulate a
poorly resolved galaxy with no dark matter, while the hot gas
media are typical of the intracluster medium. 

The hot gas was prepared from an initially random placement of
particles, and then allowed to relax to a stable state. The cold clump
was created by randomly placing particles within a sphere of radius
equal to the gravitational softening length. The cold clump was
allowed to relax in the same manner as the hot gas, before combining
the two systems.

The Jeans length, $R_J$, for the hot gas phases is sufficiently large to
ensure stability even in the presence of the perturbation from the cold
clump. Consequently, dynamical friction should not be important. This
conclusion was confirmed by passing a collisionless cold clump through the
hot medium -- it experienced negligible deceleration.
  
The box length, $5\Mpc$, was chosen so that the cold clump was well
separated from its images (arising from the periodic boundary
conditions employed) and would move across the box only once
without encountering its own wake. As in Section~\ref{cool}, an
appropriate value of the time-step normalization parameter, $\kappa$, was
found.  For these tests, a value of $\kappa=1.0$ is used.

\subsubsection{Expected deceleration}
An expected rate of deceleration may be approximated by considering a disc
sweeping through a hot medium, collecting all matter it encounters.  This
would represent a maximum expected rate of deceleration, ignoring
dynamical friction. The solution for the
velocity, $V$, of such a system is given by $V(t)=l/(t-t_l)$, where
$l$ is a characteristic length given by $l=M/2\pi R^2 \rho_g$ and
$t_l$ is a characteristic time-scale given by $t_l=t_o - l/V_o$.  Here, $M$
is the mass of the disc at the start, $R$ is the radius of the disc, and
$\rho_g$ is the density of the gas through which the disc is travelling. 
The clump starts with velocity $V_o$ at time $t_o$. For the tests that use
the slow clump, this estimate implies
the final velocity  should be $400 \kps$. For the
fast clump, the final clump velocity should be $670 \kps$.  These crude
estimates indicate that hydrodynamical forces should
indeed be important for the parameters being considered.

\subsubsection{Results of the SPH variants}
\label{drag.results}

\begin{table}
\caption{The relative final velocities of the cold clumps.}
\begin{tabular}{cccc}
Version & Mach 2     & Mach 1          & Mach 1/3 \\
\hline
 1 & $0.999 \pm 0.002$ & $1.004 \pm 0.009$ & $1.24 \pm 0.10$ \\
 2 & $0.991 \pm 0.006$ & $1.003 \pm 0.012$ & $1.29 \pm 0.08$ \\
 3 & $0.914 \pm 0.006$ & $1.009 \pm 0.008$ & $1.20 \pm 0.16$ \\
 4 & $1.018 \pm 0.005$ & $1.009 \pm 0.008$ & $0.78 \pm 0.07$ \\
 5 & $1.025 \pm 0.009$ & $1.004 \pm 0.009$ & $0.84 \pm 0.07$ \\
 6 & $0.984 \pm 0.008$ & $0.926 \pm 0.013$ & $0.67 \pm 0.05$ \\
 7 & $1.064 \pm 0.007$ & $1.060 \pm 0.004$ & $1.26 \pm 0.02$ \\
 8 & $1.045 \pm 0.007$ & $1.062 \pm 0.008$ & $1.34 \pm 0.01$ \\
 9 & $1.028 \pm 0.006$ & $0.979 \pm 0.006$ & $1.12 \pm 0.05$ \\
10 & $0.957 \pm 0.005$ & $0.951 \pm 0.007$ & $0.58 \pm 0.07$ \\
11 & $0.956 \pm 0.004$ & $0.955 \pm 0.006$ & $0.62 \pm 0.06$ \\
12 & $1.018 \pm 0.003$ & $1.038 \pm 0.009$ & $1.06 \pm 0.10$ \\
\hline
\end{tabular}
\medskip 

Given is the mean relative velocity of the cold clumps over the final
$0.5\Ga$ normalized by the mean velocity of all the cold clumps in that
velocity regime. 
\label{Eric.Tab.Drag.Results}
\end{table}

Four separate realizations of the same initial conditions were evolved
with the same version of the test code to look for variation due to
randomness in the initial conditions. There is variation on the order
of $10 \kps$ between the runs for the Mach 2 and Mach 1/3 scenarios,
$4 \kps$ for the Mach 1 scenario.

The cold clump size varies between implementations.  It is 2--3 times
larger for the $\nabla . \bmath{v}$ viscosity variants. However,
since the knot size is still much less than the smoothing radius for
these particles (at least a factor of three), the total clump size,
after smoothing, is approximately the same in all cases.

\begin{figure}
\epsfxsize=8.5cm
\centerline{\epsffile{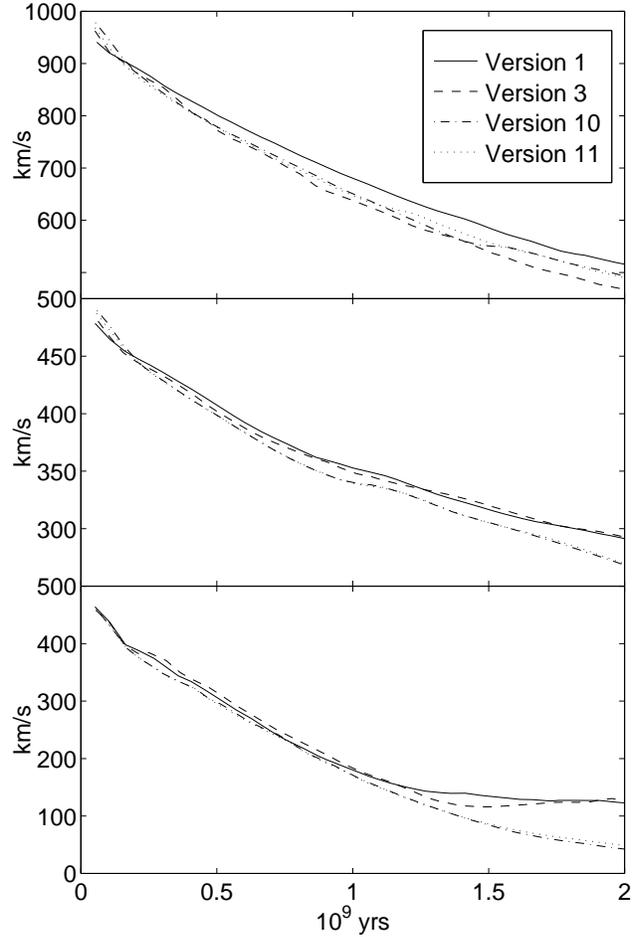}}
\caption{Variation of the cold-clump velocity with artificial
viscosity type (no shear-correction term is included).
In each panel the different lines distinguish different viscosity
implementations: standard \Hydra\ viscosity (TC92), solid;
localized TC92 viscosity, dashed; standard Monaghan viscosity,
dash-dot; single-sided Monaghan viscosity, dotted. 
The panels are, from top to bottom; Mach 2, Mach 1 and Mach 1/3.}
\label{Eric.Fig.Drag.Monaghan}
\end{figure}

Compared with $\nabla . \bmath{v}$ viscosity, Monaghan viscosity in
both the symmetric and single-sided forms leads to an increase in the
damping of the velocity of the clump when used with the TC92
symmetrization (Table~\ref{Eric.Tab.Drag.Results} and
\fig~\ref{Eric.Fig.Drag.Monaghan}). However, when 
Monaghan viscosity is used with TC92 symmetrization, supplemented by
kernel averaging, the deceleration becomes comparable to the 
$\nabla . \bmath{v}$ versions. The more localized estimate of the 
$\nabla . \bmath{v}$ viscosity does little except in the Mach 2 set of runs,
for which it increases the drag to match that of the Monaghan
viscosity. The inclusion of a shear-correction term reduces the drag
in the Mach 1/3 case as well as in the Mach 1 case when the clump
velocity has dropped below Mach 0.8.

\begin{figure}
\epsfxsize=8.5cm
\centerline{\epsffile{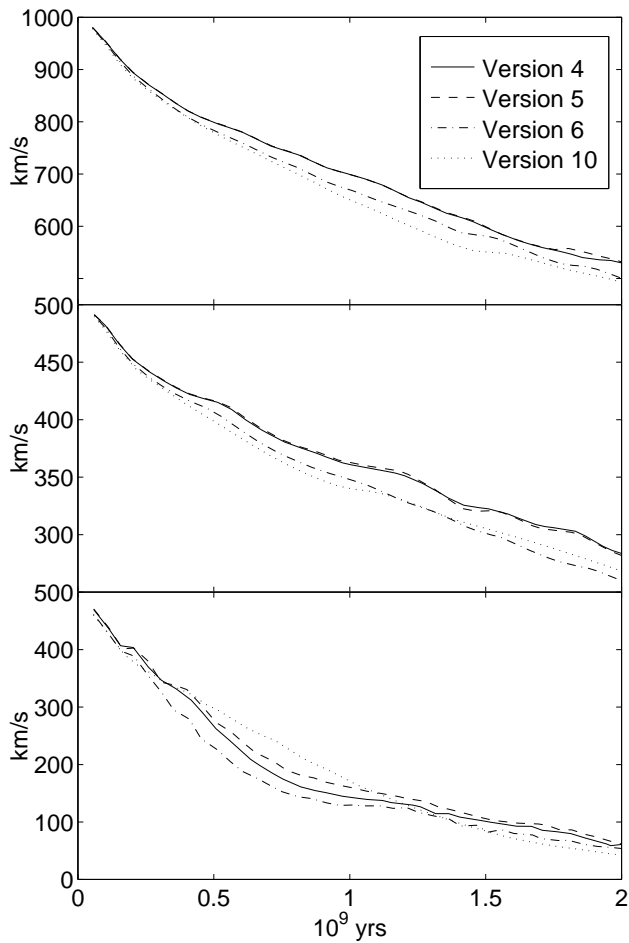}}
\caption{Variation of the cold-clump velocity with h-symmetrization.
The lines are: arithmetic average, solid; harmonic, dashed; kernel
averaging, dot-dash; TC92 symmetrization (version 10), dotted.
The panels are, from top to bottom; Mach 2, Mach 1 and Mach 1/3.}
\label{Eric.Fig.Drag.h-sym}
\end{figure}

Use of either the arithmetic or harmonic average for
$h_{ij}$ produces less drag then any other symmetrization method
(\fig~\ref{Eric.Fig.Drag.h-sym}) except the version with TC92
symmetrization combined with kernel averaging. On their own, kernel
averaging and the TC92 
symmetrization produce marginally higher deceleration,
particularly at supersonic speeds.

\subsubsection{Summary}
The tests favour (but cannot distinguish between) the harmonic and
arithmetic averages. The shear-correction term
lowers the drag at sub-sonic speeds.  The Monaghan viscosity coupled
with the TC92 symmetrization performs poorly.

\subsection{Cosmological simulation}\label{cosmo}
In this test we simulate a common astrophysical problem: the
formation of knots of cold, dense gas within a cosmological 
volume. For this problem we are interested in recovering 
accurate positions and masses for the objects. However, the 
resolution is such that no internal information (such as spiral
structure or radial density profiles) can be recovered.

\subsubsection{Initial Conditions}

The simulations presented here were of an $\Omega_0=1$, standard cold
dark matter universe with a box size of $10 h^{-1}\Mpc$. We take
$h=0.5$ throughout this section, equivalent to a Hubble constant of
$50\, \kmpspMpc$.  The baryon fraction, $\Omega_b$ was set from
nucleosynthesis constraints, $\Omega_bh^2 = 0.015$ \cite{cst} 
and we assume a constant gas metallicity of $0.5 Z_\odot$. Identical
initial conditions were used in all cases,
allowing a direct comparison to be made between the objects formed.

The initial fluctuation amplitude was set by requiring that the model
produce the same number-density of rich clusters as observed
today. To achieve this we take $\sigma_8 = 0.6$, the present-day
linear rms fluctuation on a scale of $8h^{-1}\Mpc$ (Eke,
Cole \& Frenk 1996, Vianna \& Liddle 1996).  
Each model began with $32^3$ dark matter
particles each of mass $1.58 \times 10^{10} \Msun$ and $32^3$ gas
particles each of mass $1.01
\times 10^9 \Msun$, smaller than the critical mass derived
by Steinmetz \& White (1997) required to prevent 2-body heating of the
gaseous component by the heavier dark matter particles. The
simulations were started at redshift 19. We employ a  
comoving Plummer softening of $10 h^{-1}\kpc$, which is typical
for modern cosmological simulations but still larger than required
to accurately simulate the dynamics of galaxies in dense environments.
This test case is identical to that extensively studied by Kay \etal
(1998) who used it to examine the effect of changing numerical and
physical parameters for a fixed SPH implementation.

\subsubsection{Extraction of glob properties}

\begin{figure}
 \centering
\psfig{file=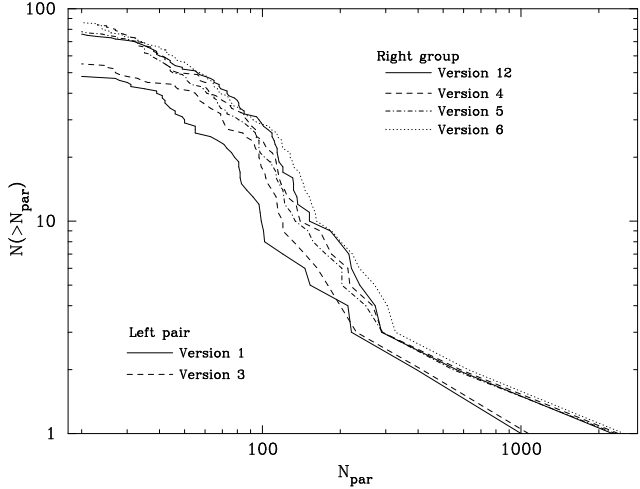,height=6.5cm}
\caption{The multiplicity functions for different versions. The six
implementations plotted span the range of outcomes as, in all cases, the
addition
of a shear-correction to the viscosity makes little difference.
Versions 10 and 11 produce very similar results to version 12,
whilst 1, 2 and 3 produce smaller objects than the other versions. 
}
\label{cosmo.mult}
\end{figure}

The gas is effectively in three disjoint phases. There is a cold, diffuse phase
which occupies the dark-matter voids and therefore most of the volume.
A hot phase occupies the dark-matter halos and at this
resolution is typically above $10^5\,$K. Finally, there is a cold,
dense phase 
consisting of tight knots of gas typically at densities several
thousand times the mean and at temperatures close to $10^4\,$K.
The relative proportions of the gas in each of these phases is
given in Table~\ref{cosmo.table}.
We follow Evrard, Summers \& Davis~\shortcite{esd} in defining a cooled
knot of particles
as a `glob' because the resolution is such that they can hardly be
termed a galaxy. The properties of the globs are calculated by first
extracting all the particles which are simultaneously below a
temperature of $10^5\,$K and at densities above 180 times the mean
and then running a friend-of-friends group-finder with a 
maximum linking length, $b$, of 0.08 times the mean interparticle
separation of the dark matter. In practice
the object set obtained is insensitive to the choice of $b$
because the globs are typically disjoint, tightly bound clumps. The
cumulative 
multiplicity function for the different implementations is shown in
\fig~\ref{cosmo.mult}. 

\begin{table}
 \caption{Properties of the cosmological test runs by version.}
 \begin{tabular}{cccccccc}   
  Version & $N_{steps}$ & Hours & $N_{>50}$ & $M_{big}$ & $f_{gal}$ & $f_{hot}$ & $f_{cold}$\\
\hline
 1 & 6060 & 34.7 & 30 & 1.01& 0.17 & 0.37 & 0.45 \\
 2 & 6348 & 35.0 & 34 & 1.07& 0.19 & 0.36 & 0.45 \\ 
 3 & 6532 & 43.7 & 41 & 1.08& 0.20 & 0.34 & 0.46 \\ 
 4 & 6984 & 65.1 & 49 & 2.41& 0.30 & 0.37 & 0.33 \\ 
 5 & 6113 & 60.5 & 48 & 2.34& 0.27 & 0.40 & 0.33 \\ 
 6 & 6769 & 57.4 & 56 & 2.53& 0.34 & 0.34 & 0.33 \\ 
 7 & 7562 & 63.4 & 47 & 2.43& 0.29 & 0.34 & 0.36 \\ 
 8 & 6782 & 93.6 & 47 & 2.38& 0.27 & 0.36 & 0.36 \\ 
 9 & 7688 &109.7 & 49 & 2.40& 0.32 & 0.37 & 0.31 \\ 
10 & 6858 & 60.5 & 50 & 2.69& 0.33 & 0.33 & 0.34 \\
11 & 6522 & 40.4 & 48 & 2.43& 0.31 & 0.35 & 0.34 \\
12 & 6205 & 38.5 & 51 & 2.28& 0.30 & 0.35 & 0.34 \\
\hline
 \end{tabular}
\medskip

Listed are the number of steps taken to reach $z=0$,
the number of hours required on a Sun Ultra II 300 workstation,
the number of groups of more than 50 cold particles
found at the endpoint, the mass of the largest clump (in units 
of $10^{12}\Msun$),
the fraction of the gas in galaxies, the fraction of gas above $10^5\,$K
and the fraction of the gas that remains diffuse and cold (all at $z=0$).
\label{cosmo.table}
\end{table}

\begin{figure}
 \centering
\psfig{file=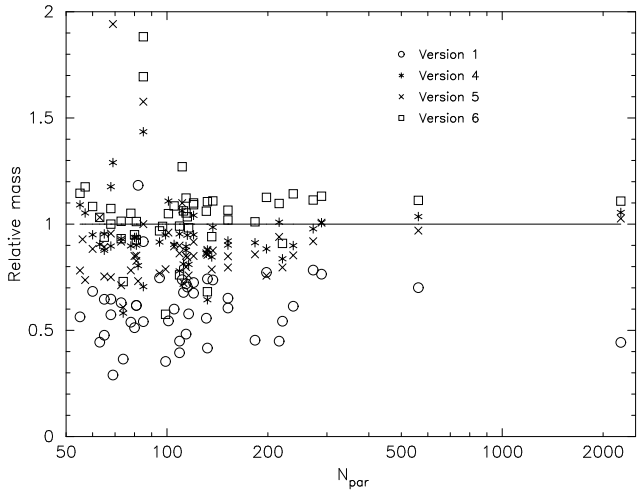,height=6.5cm}
\caption{Comparison of object masses in different implementations.  
The masses of objects found by the group-finder in version 12 of the
code are compared to the masses found for the corrsponding objects in
other versions.  Versions 1--3 of the code all produce objects of
about half the mass and many smaller objects are missing (because in
these runs they fall below the resolution limit of around 50
particles).  As for \fig~\ref{cosmo.mult} the addition of a
shear-correction makes little difference and versions 10 and 11 of the
code produce very similar masses to version 12.}
\label{cosmo.compare}
\end{figure}


\subsubsection{Results of cosmological test}

In all cases the largest object has `overcooled' in the sense
described in section~\ref{cool}. It is much too massive to be
expected in a simulation of this size and is only present because
gas within the hot halo has its cooling rate enhanced by the very
high-density gas contained in the globs.

A distinct difference can be seen in the morphologies of small objects
formed by versions 1--3 compared to those formed by 4--12. Versions
1--3 produce spherical objects since the $\nabla . \bmath{v}$
viscosity used does not damp random orbital motion within the softening
radius. Versions 4--12 produce disc-like objects as a result of the
effective dissipation provided by the pairwise trigger and
conservation of angular momentum. Both spherical and disc objects are of size
approximately equal to the softening length. We do not expect merging
to play a significant role in this simulation due to the low
particle-number in the majority of globs.

The major discriminant between the versions is the different artificial
viscosities. As shown in section~\ref{shock}, versions 1, 2 and 3 produce
broader shock fronts because the viscosity employed is a locally averaged
quantity, whereas for all the other versions the viscosity is calculated
on a pairwise basis. The pairwise viscosity shock-heats the gas more
efficiently and leads to a larger fraction of hot gas
in versions 4--12, (see Table~\ref{cosmo.table}). 

%
The fraction of matter present in globs and the number of groups with
$N>50$ is clearly lower for versions 1--3 than for other versions. The
lower mass-fraction in globs is a combination of both the smaller number
of groups 
found above the threshold and versions 1--3 producing lighter
objects. As was demonstrated in 
section~\ref{evrard}, the $\nabla . \bmath{v}$ viscosity produces a
shallow collapse, with much less dissipation. When collapse occurs in an
object that has approximately $N_{smooth}$ particles, there will be
virtually no shock heating and the gas particles will free-stream within
the shallow potential well. In simulations with cooling the $\nabla .
\bmath{v}$ viscosity provides a marginally higher pressure support than
the Monaghan viscosity (see section~\ref{rotcloud}) which can be sufficient
to prevent collapse of surrounding material. At low resolution this
results in an object not achieving $N>50$, whilst at higher resolution the
object has lower mass. 

Of the Monaghan variants 4--6, version 5 has the lowest fraction of matter
in the glob phase and the highest fraction of hot gas. Fig. 
\ref{cosmo.compare} shows that it also produces systematically lighter
objects. Version 6 has the highest fraction of matter in the glob phase,
the lowest fraction of hot gas and tends to produce the heaviest objects. 
These results are due to the symmetrization scheme causing the artificial
viscosity to produce different amounts of dissipation. Similarly, for
versions 10--12 the fraction of hot gas can be traced to the amount of
dissipation. These results are consistent with those in
section~\ref{rotcloud}, where they are discussed in detail. The 
trend for Monaghan-type viscosities is distinct: versions that produce
more dissipation form lighter objects as the hot halo gas is heated to
higher temperatures where the cooling time is longer. 

In section~\ref{shock} it was shown that the shear-correction term is less
able to capture shocks and consequently produces lower shock heating. For
the $h$-averaging implementations (4, 5), the hot gas fraction is reduced
upon adding shear correction, which agrees with the shock tube
result. For the kernel-averaged version this is not the case -- the hot
gas fraction increases.  This result is probably not significant; in
section~\ref{rotcloud} versions 4--6 all show reduced dissipation upon
including the shear-correction term. 

\subsubsection{Summary} 
Any of the versions discussed in this paper could be used effectively for
this problem. Differences in the amount of gas in each of the hot, cold
and glob phases are produced, which can be explained in terms of the
amount of dissipation produced by each scheme. The $\nabla .
\bmath{v}$-based viscosities produce objects of spherical morphology,
while the Monaghan variants produce objects with disc morphology.

\subsection{Rotating cloud collapse}\label{rotcloud}
 \begin{table}
 \caption{Results of rotating cloud collapse test.}
 \begin{tabular}{@{}cccccc}   
 Ver & $N_{step}$ & $\Delta E/E (\times 10^{-3})$ & $\Delta L/L
(\times10^{-4})$ & $Q_{peak}$ & 
$f_{peak}$ \\
\hline
 1 & 1461 & $2.8$ & $+0.23$ & $0.001$ & $0.095$ \\
 2 & 1627 & $1.7$ & $+0.98$ & $0.001$ & $0.118$ \\
 3 & 1588 & $1.7$ & $+3.8$  & $0.024$ & $0.212$ \\
 4 & 1806 & $2.8$ & $+9.9$  & $0.023$ & $0.159$ \\
 5 & 1703 & $1.8$ & $+9.2$  & $0.049$ & $0.215$ \\
 6 & 1834 & $1.8$ & $+10.$  & $0.009$ & $0.124$ \\
 7 & 2087 & $4.4$ & $-6.2$  & $0.015$ & $0.128$ \\
 8 & 2002 & $4.4$ & $-6.9$  & $0.030$ & $0.164$ \\
 9 & 2074 & $4.8$ & $-5.7$  & $0.004$ & $0.095$ \\
10 & 1789 & $2.2$ & $+10.$  & $0.002$ & $0.074$ \\
11 & 1748 & $3.1$ & $+7.4$  & $0.023$ & $0.143$ \\
12 & 1705 & $2.9$ & $+6.6$  & $0.036$ & $0.175$ \\
\hline
 \end{tabular}
\medskip

With the exception of the peak variables values are given at t=256. 
$\Delta L/L$, the fractional change in angular momentum, is measured to be
positive for a loss of momentum and is expected to be zero. $Q_{peak}$ is
the maximum value of the thermal energy (as a fraction of the initial
total mechanical energy) and $f_{peak}$ is the peak fraction of gas
shocked to high temperature. 
 \label{galsum}
\end{table}

A standard test problem for galaxy-formation codes is presented in Navarro
\& White \shortcite{nw}. In this test a cloud of dark matter and gas is
set in solid-body rotation. Gravitational collapse combined with radiative
cooling leads to a cool, centrifugally supported gaseous disc.

\subsubsection{Initial conditions}
The initial radius of the gas cloud is chosen to be 100 kpc, and the total
mass (dark matter and gas) is $10^{12}\,\Msun$. The spin parameter,
$\lambda$, is set to, \begin{equation} \lambda={|\bmath{L}| |E|^{1/2}
\over GM^{5/2}} \simeq 0.1, \end{equation} where $\bmath{L}$ is the
angular momentum, $E$ the binding energy, and $M$ the mass. The baryon
fraction is set to $10\%$, and the cooling function for
primordial-abundance gas is interpolated from Sutherland \& Dopita
\shortcite{sd}. 

\begin{figure*}
\vspace{225mm}
\begin{minipage}{170mm}
\includegraphics{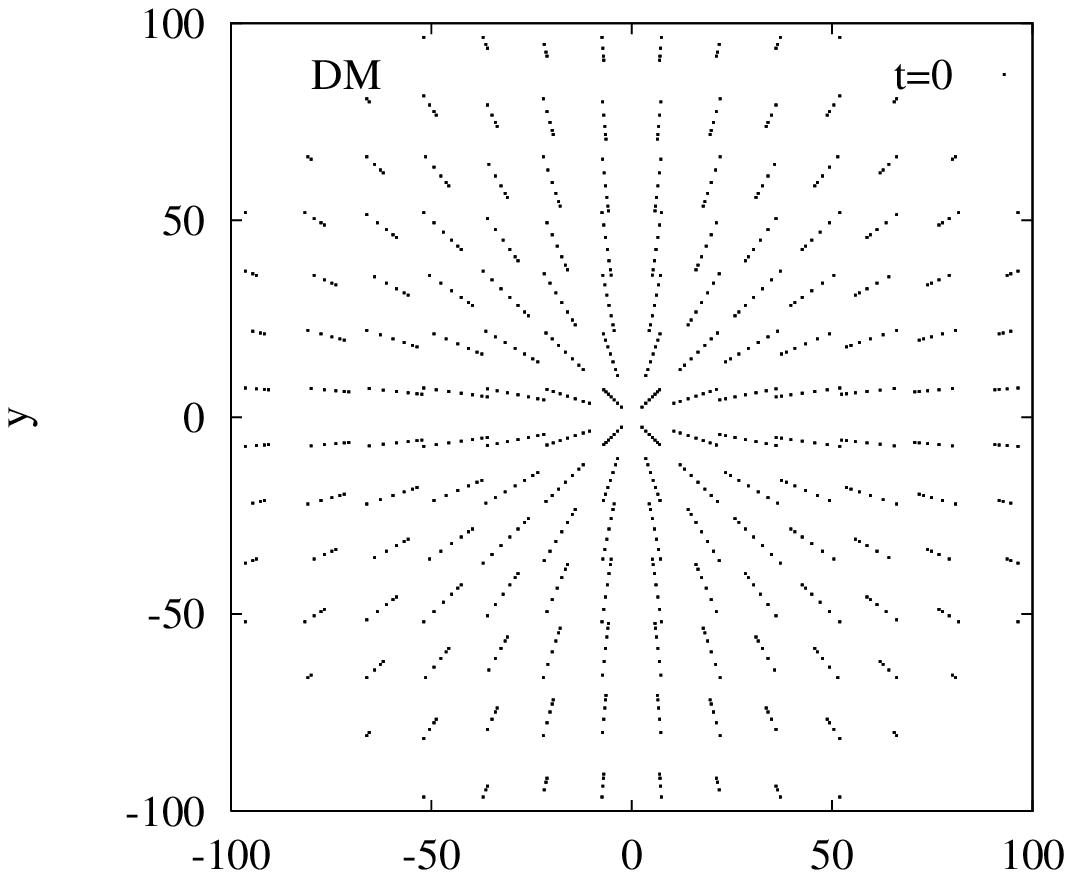}
\includegraphics{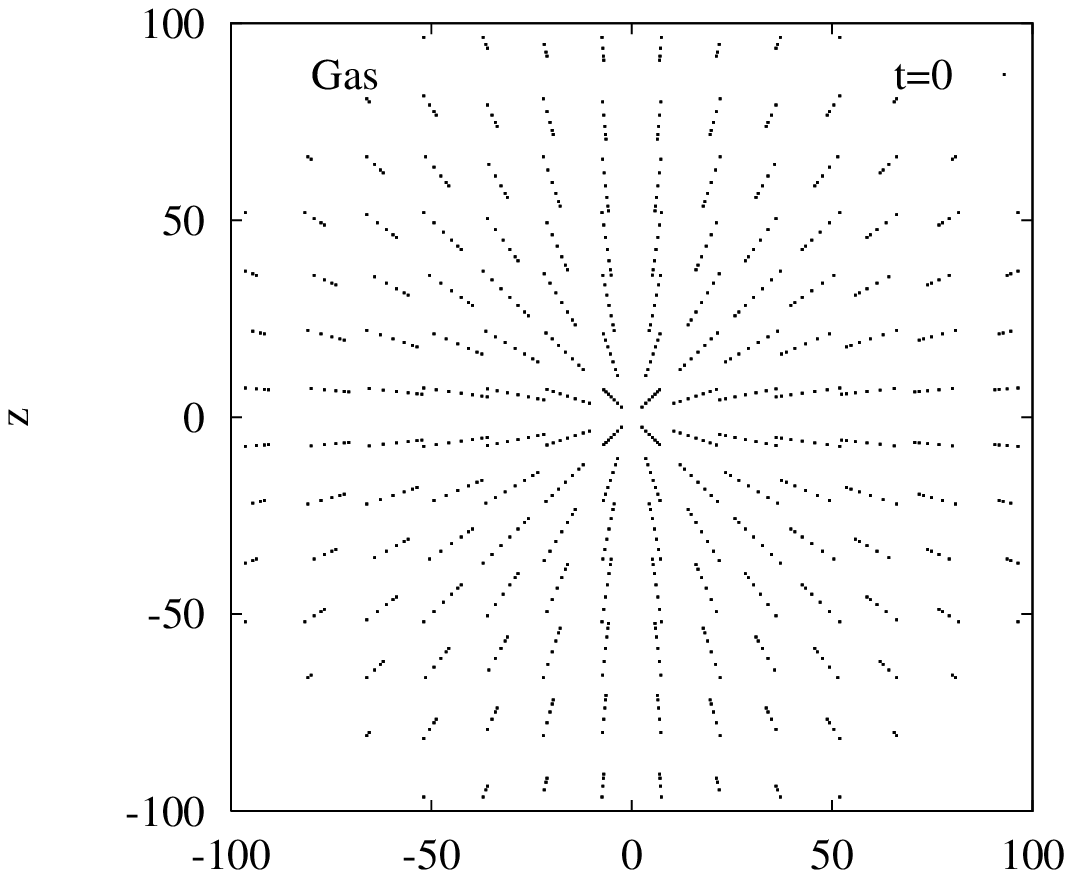}
\includegraphics{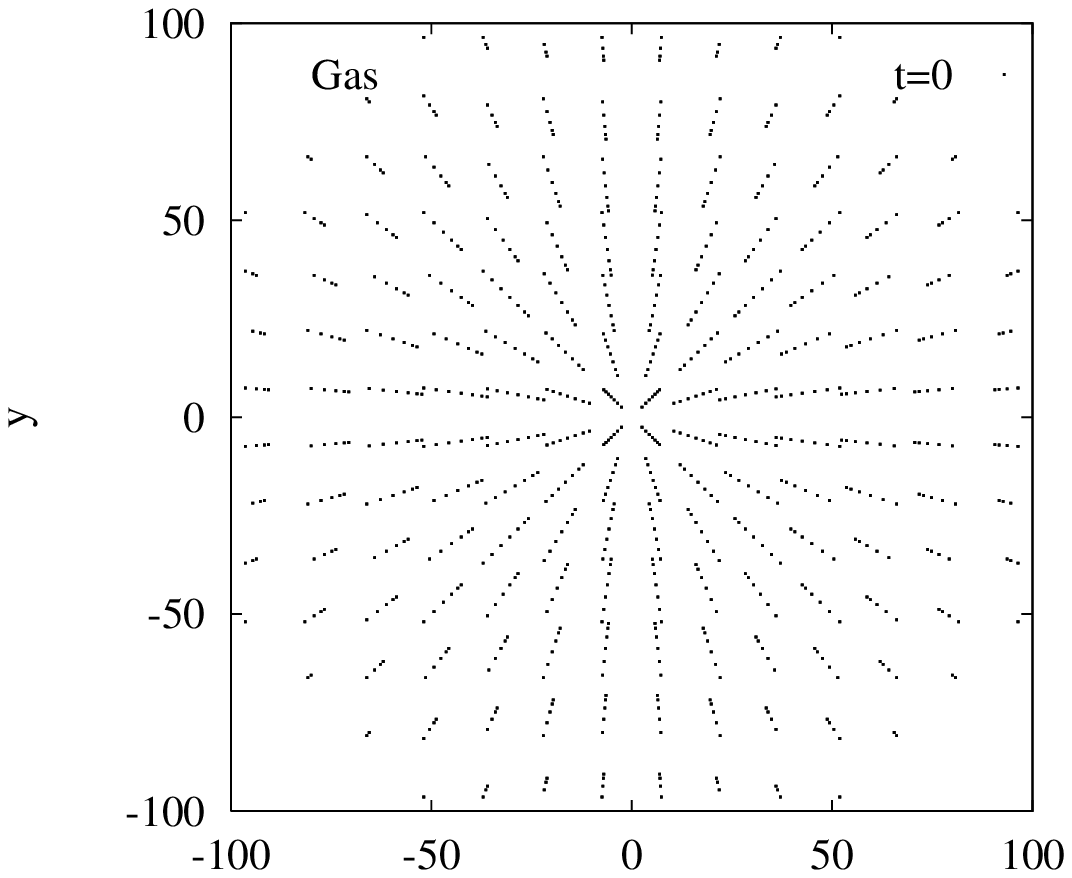}
\includegraphics{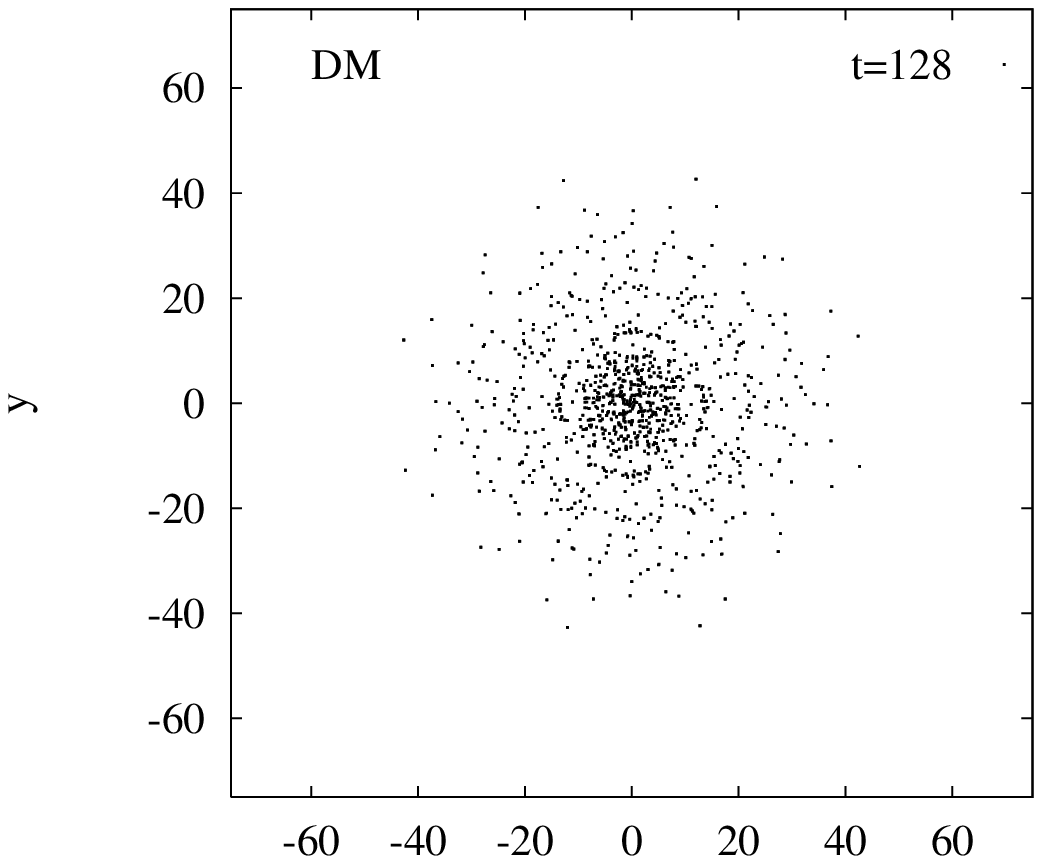}
\includegraphics{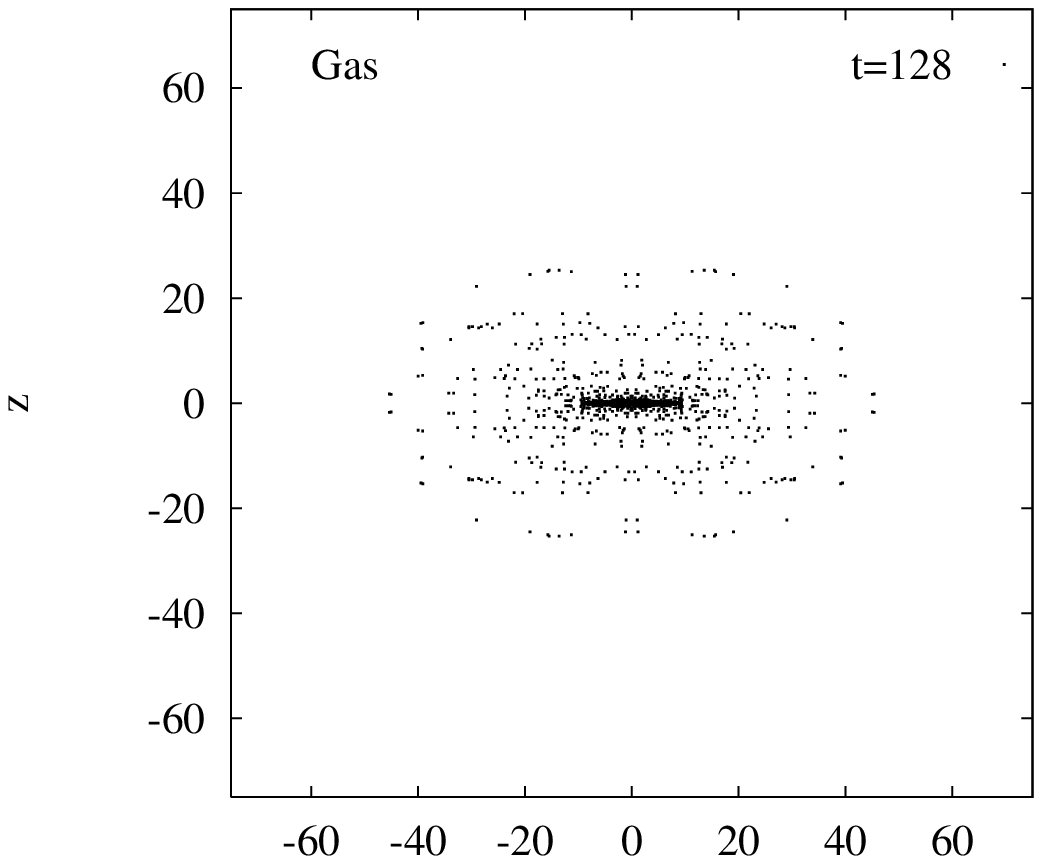}
\includegraphics{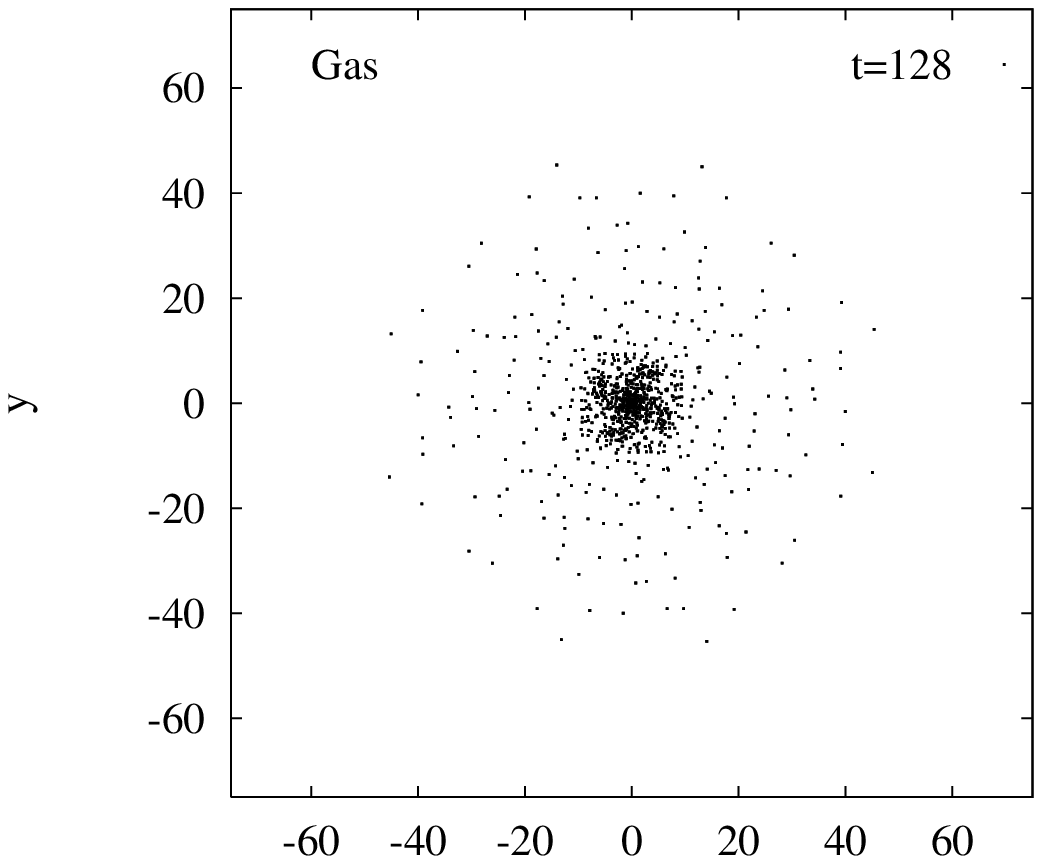}
\includegraphics{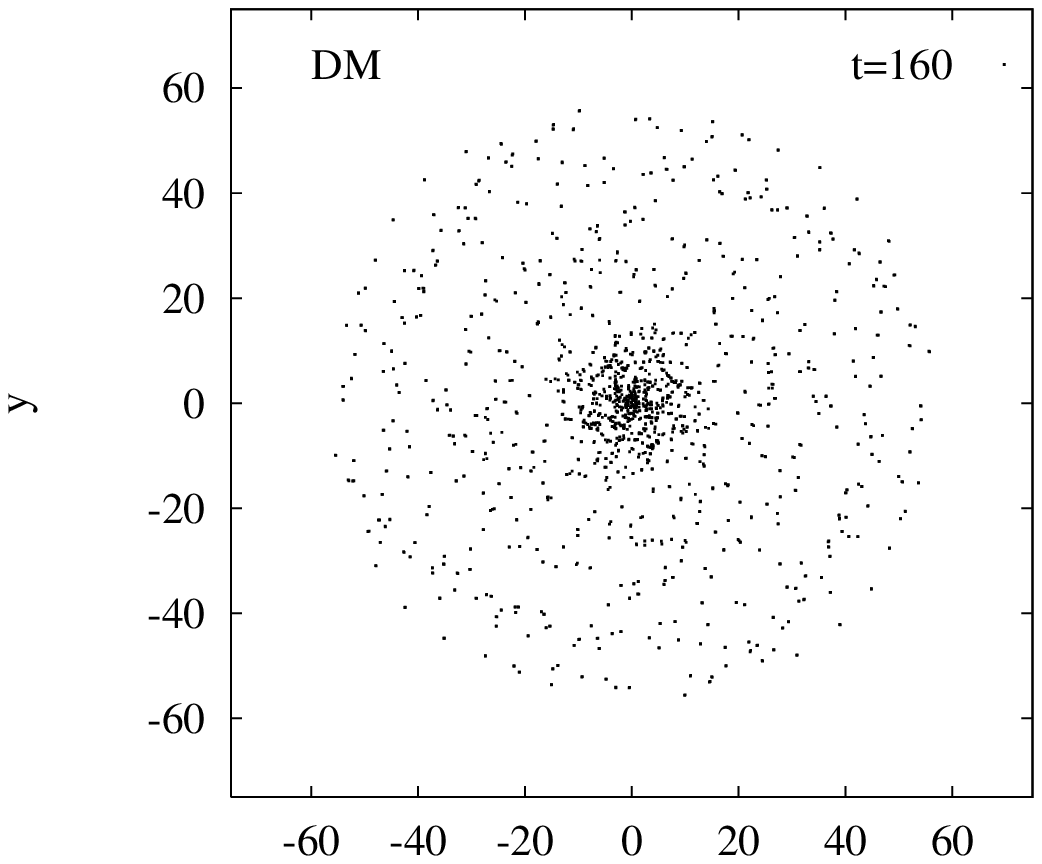}
\includegraphics{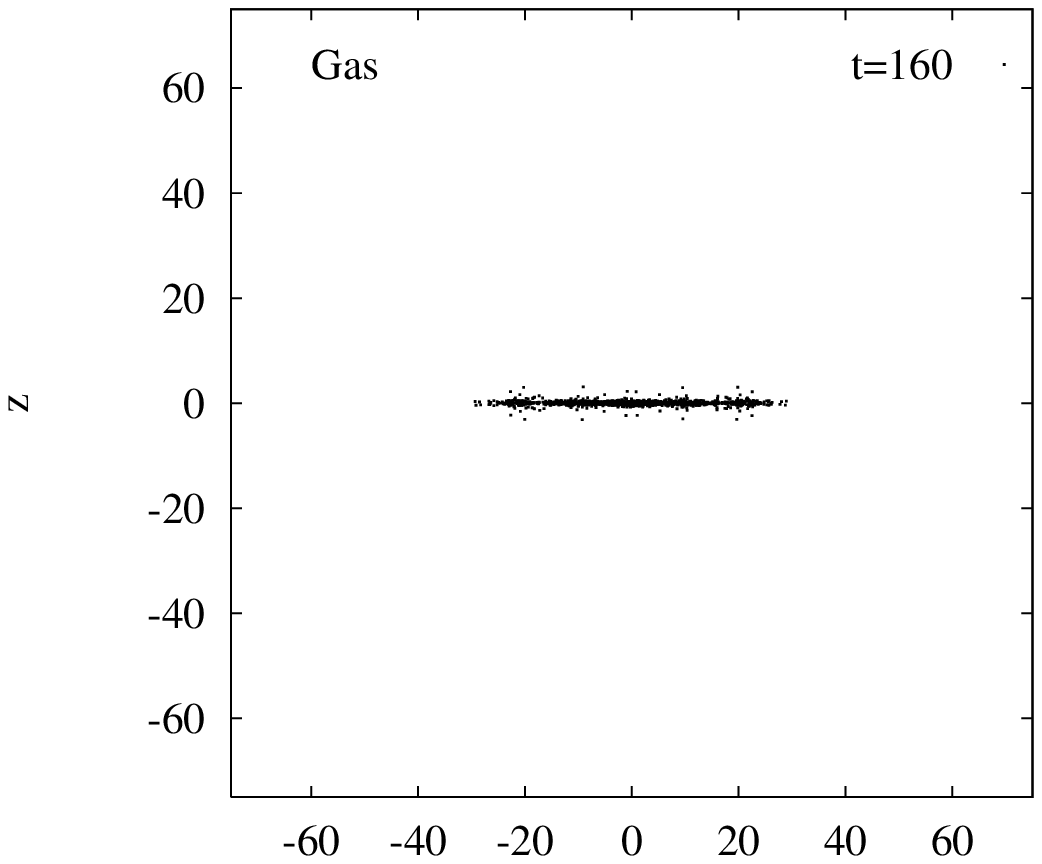}
\includegraphics{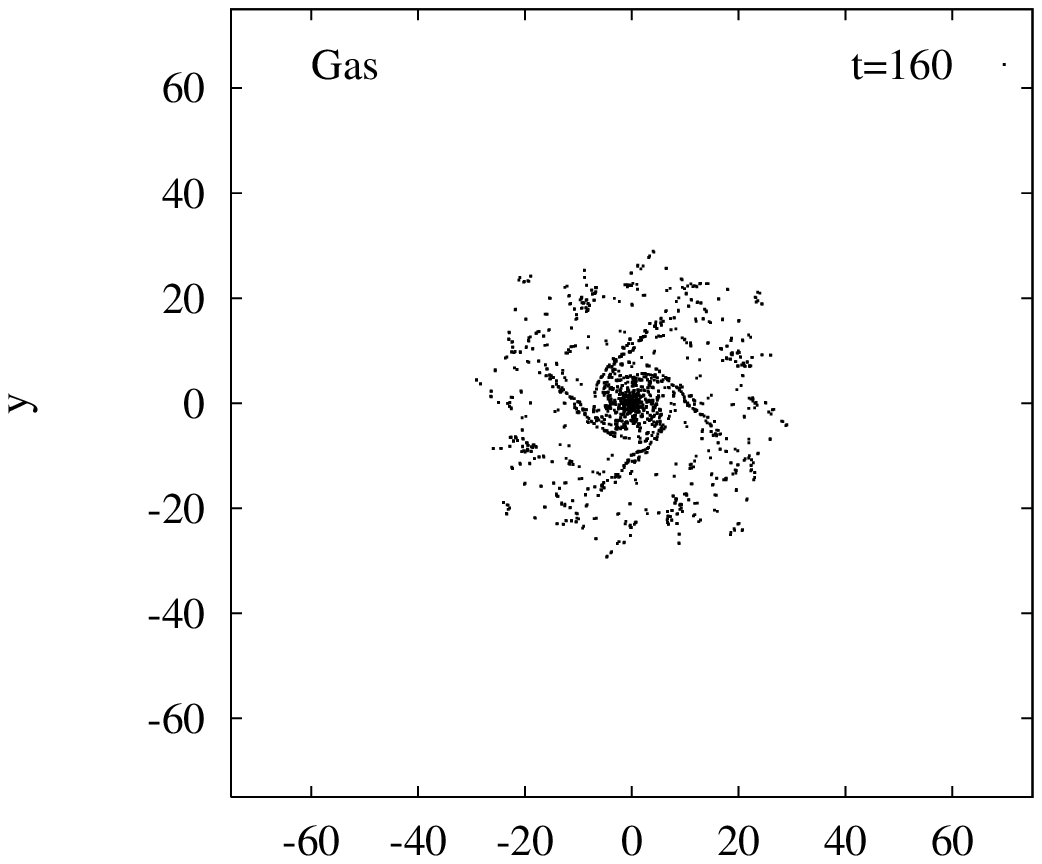}
\includegraphics{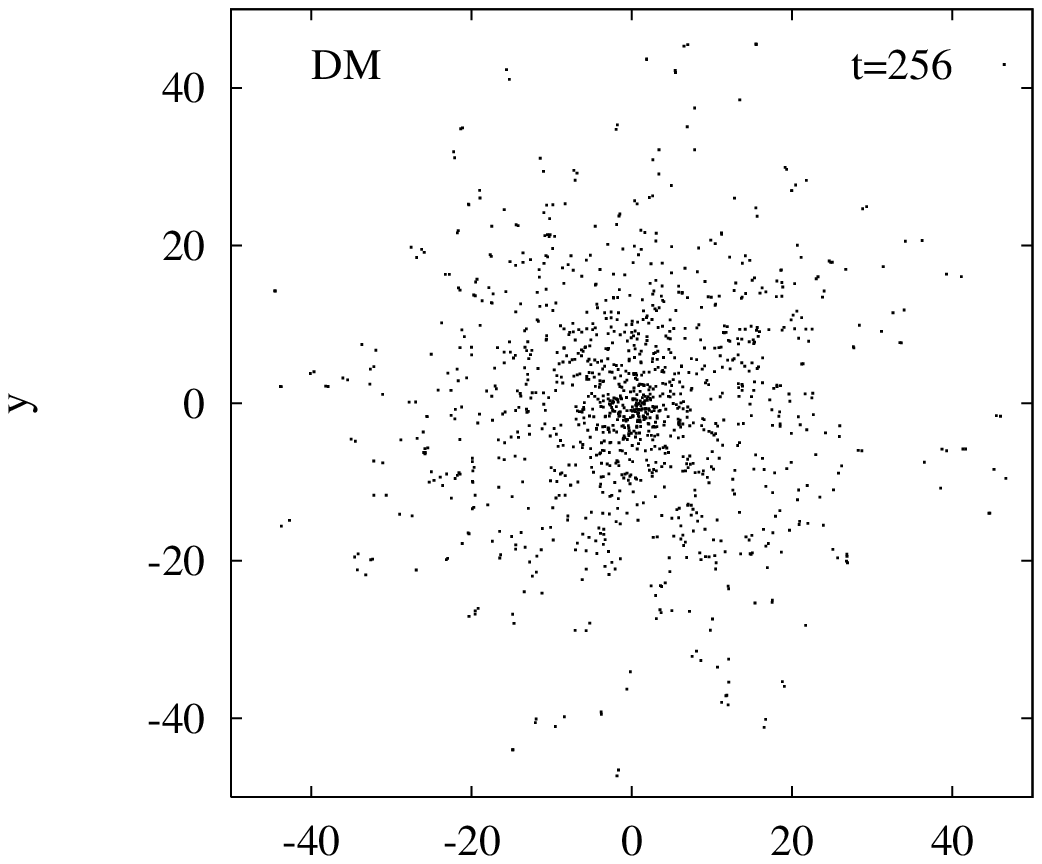}
\includegraphics{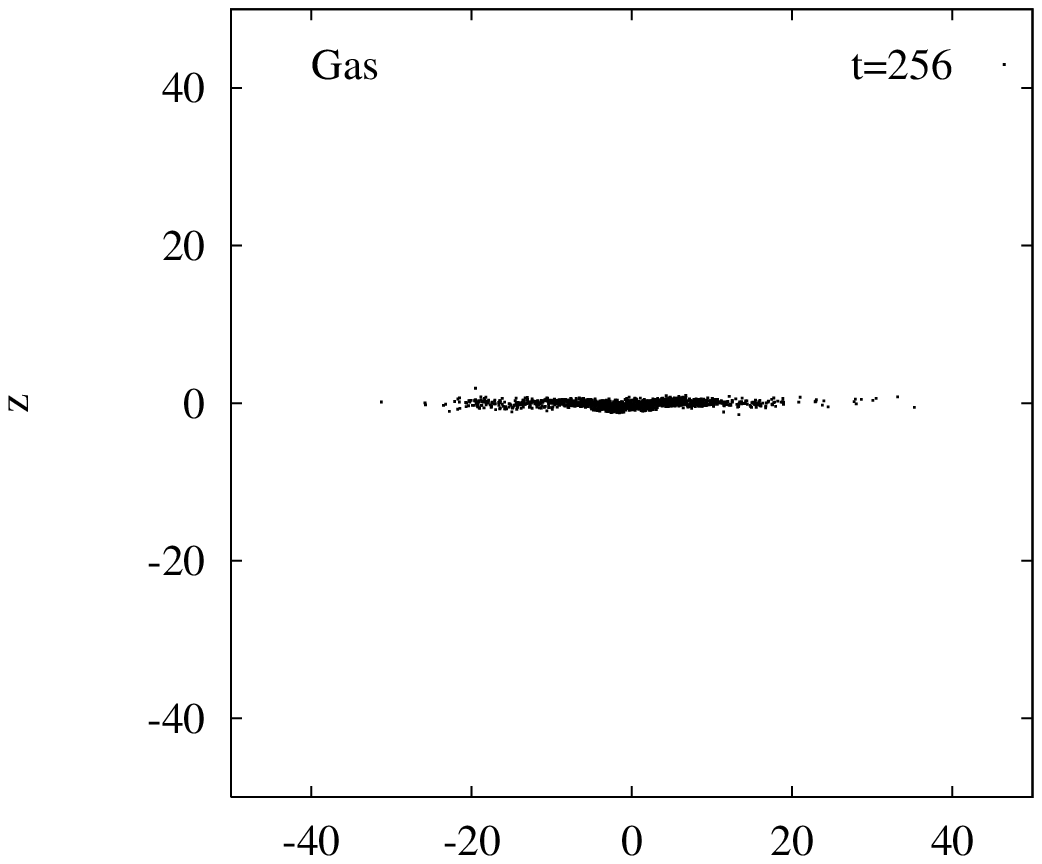}
\includegraphics{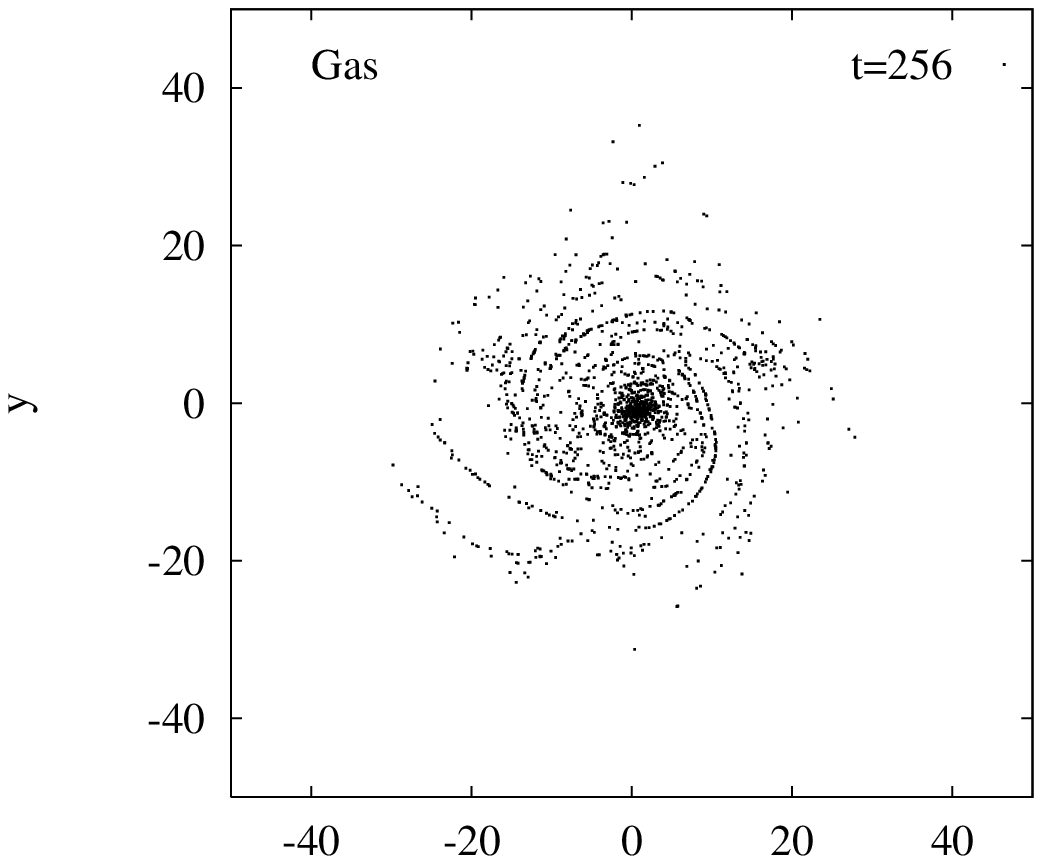}
\includegraphics{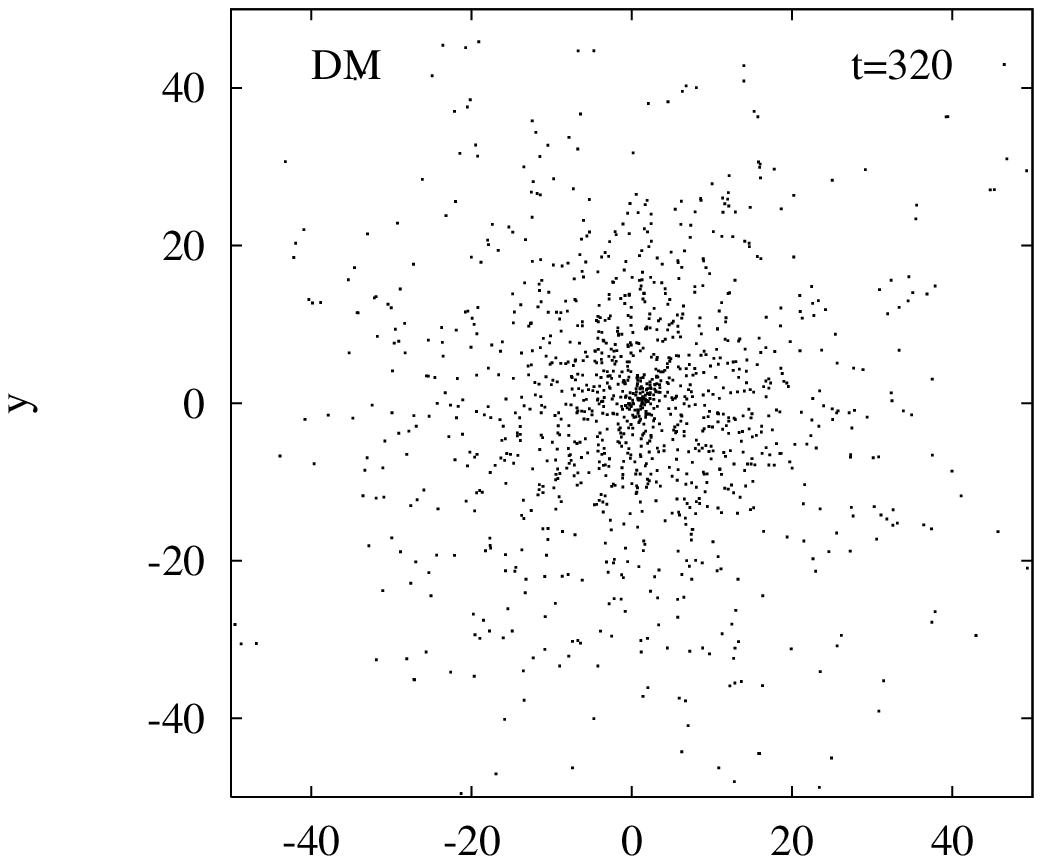}
\includegraphics{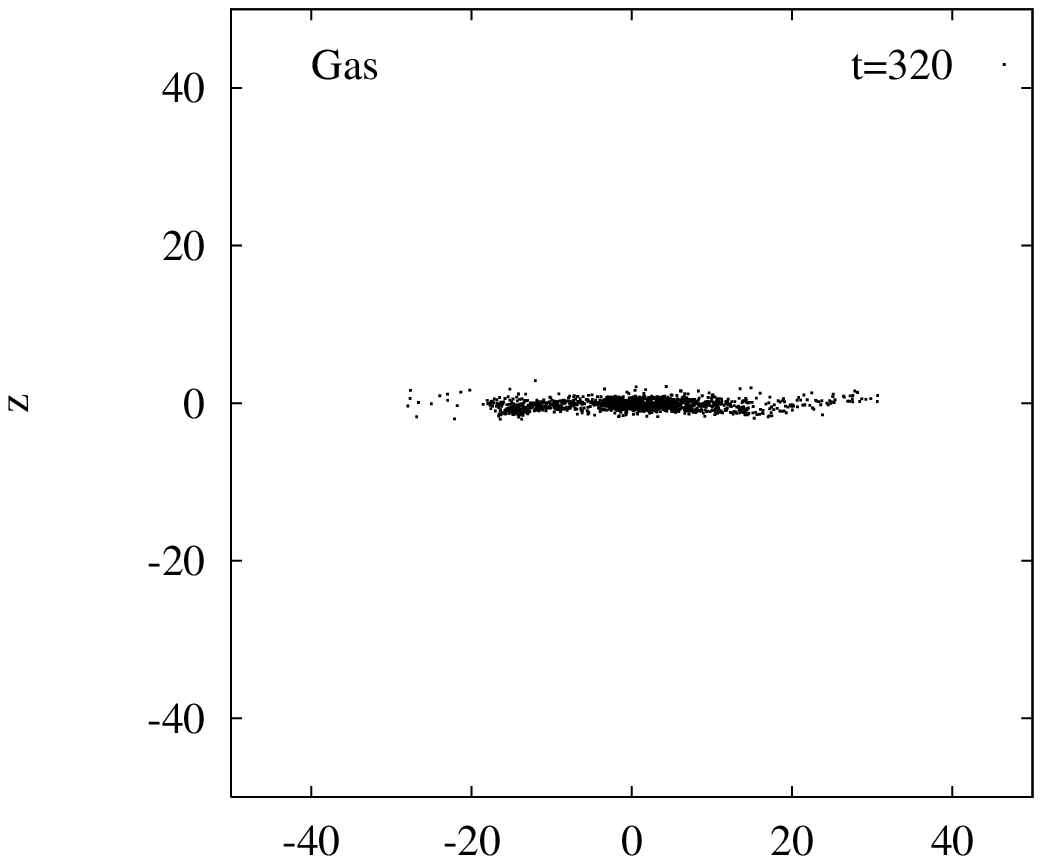}
\includegraphics{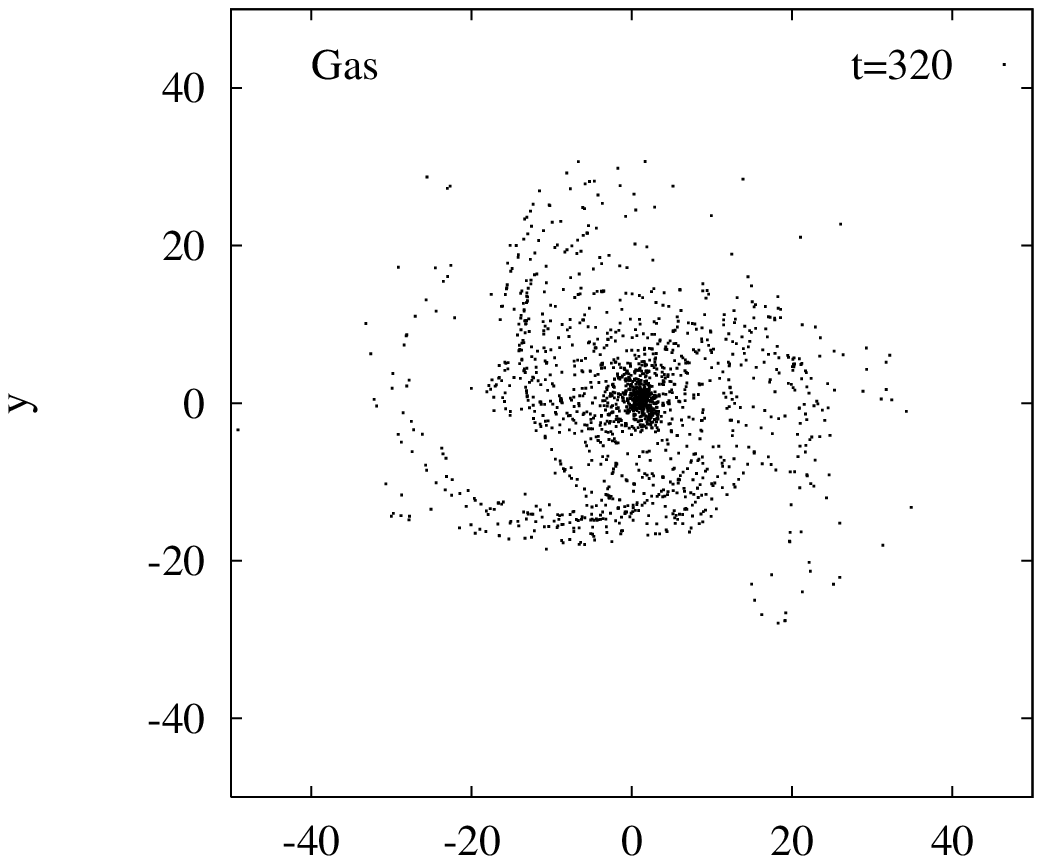}
\caption{Evolution of gas and dark matter in $2\times 1736 $ particle
collapse. The results for version 10 are plotted, which produces little
shocked gas during collapse. The morphological evolution of the system
agrees well with previous work, with minor differences being attributable
to differing initial conditions. The results presented here preserve
symmetry above and below the equatorial plane for longer than seen in
other work, this may be a consequence of the excellent momentum
conservation exhibited by grid based gravity solvers and our smoothing
over all particles within $2h_{min}$.}
\label{1736zpro}
\end{minipage}
\end{figure*}
 
For our tests we consider simulations with $N=2\times1736$ particles. The
gravitational softening length is set at 2 kpc for both dark matter and gas
particles. This is different from previous authors who have set the dark-matter
softening to be 5 kpc and the gas softening to 2 kpc. As a result we form
a smaller central dark-matter core. Times are quoted in the units
of Navarro \& White~\shortcite{nw} ($4.7\times 10^6\,$yr).
One rotation period at the half-mass radius corresponds to 
approximately 170 timesteps.
 
The resulting evolution of the system is shown in \fig~\ref{1736zpro},
and test results summarized in Table~\ref{galsum}. 
Radiative cooling during the collapse causes the gas to form a flat disc.
The dark matter virialises quickly after collapse, leaving a tight core.
Because of the large amount of angular momentum in the initial
conditions a `ring' of dark matter is thrown off. Swing amplification
causes transitory spiral features early in the
evolution which are later replaced by spiral structure that
persists for a 
number of rotations. If shocked gas is developed during the collapse it
forms a halo around the disc.

\begin{figure}
\vspace{50mm}
\includegraphics{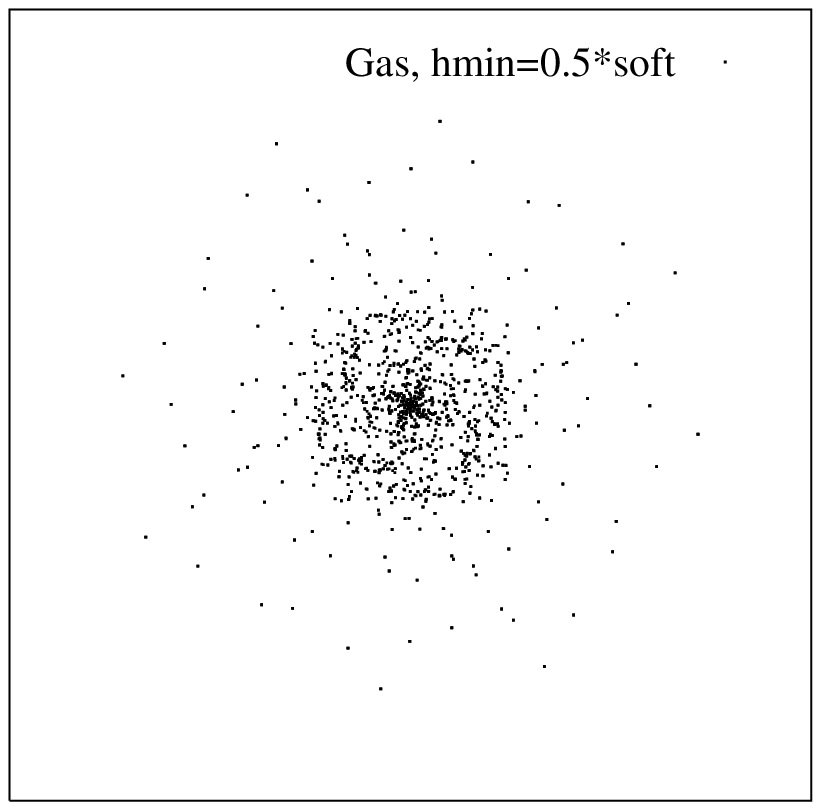}
\includegraphics{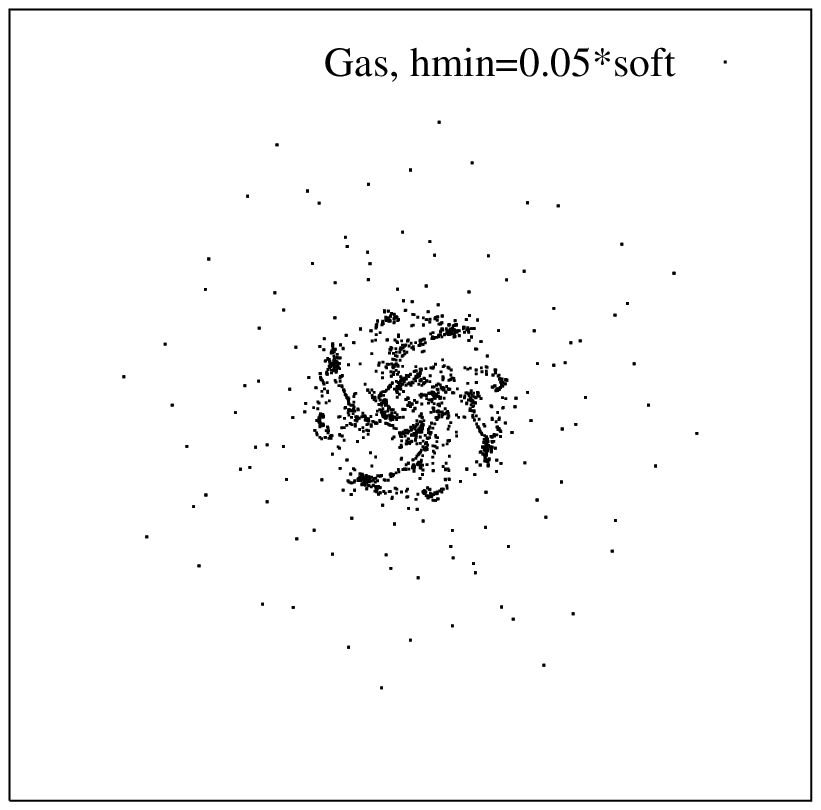}

 \caption{Comparison of morphology for $N=1736$ run under varying
$h_{min}$, at $t=128$. Each panel is 100 kpc across, and gravitational
softening was set at 2 kpc.  Implementation 12 was used to run the
simulations.  The right-hand panel clearly shows more structure on scales
near and below the gravitational softening length.} 
\label{hcomp}
\end{figure}
 
\subsubsection{Non-implementation-specific results}

We have found marginally different results for our codes when
compared with
other work.  This is due to two factors. Firstly, using a 2 kpc softening for
the dark-matter particles has a significant effect on the final
morphology. The 2-body interaction between gas and dark matter is much
stronger than would be expected if the dark matter had a longer softening
length. Secondly, most of the particles in the disc have an $h$ value
close to $h_{min}$ which in turn sets a significant limit on the minimum
mass of a clump that may be resolved. We have run a simulation with
$h_{min}=0.05\epsilon$ to see the effect of this. \fig~\ref{hcomp} shows
a comparison of the simulation run with the smaller $h_{min}$ to the
standard $h_{min}=0.5\epsilon$ simulation. Far more structure is evident
on scales close to the gravitational softening length, which must be viewed as
being unphysical since at this scale the gravitational forces are
severely softened.

\begin{figure*}
\vspace{225mm}
\begin{minipage}{170mm}
\includegraphics{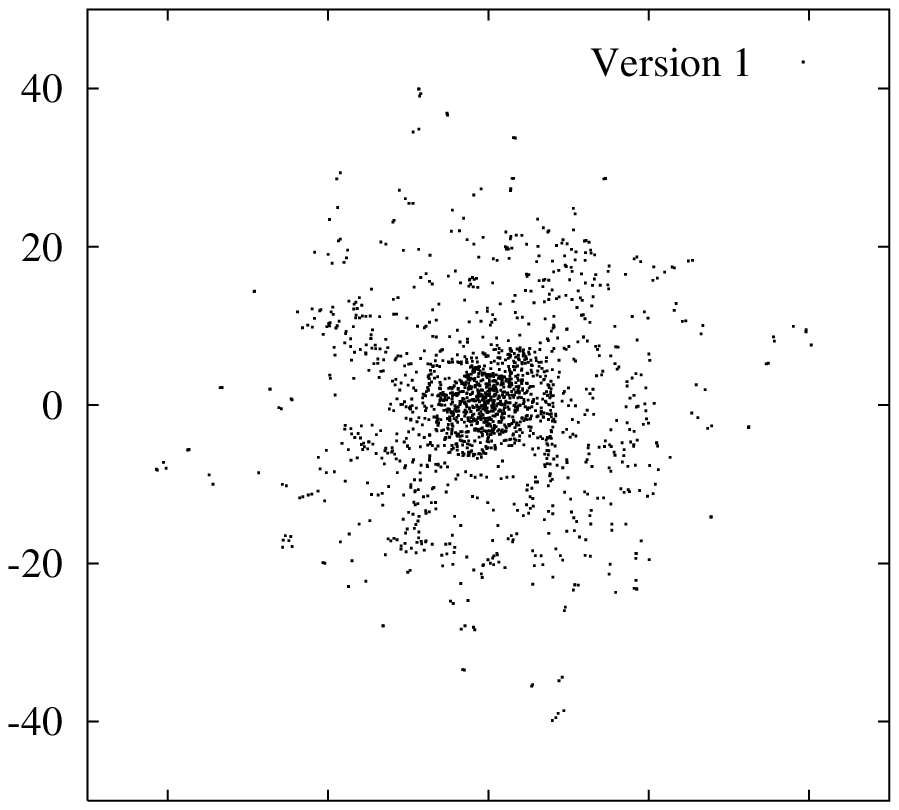}
\includegraphics{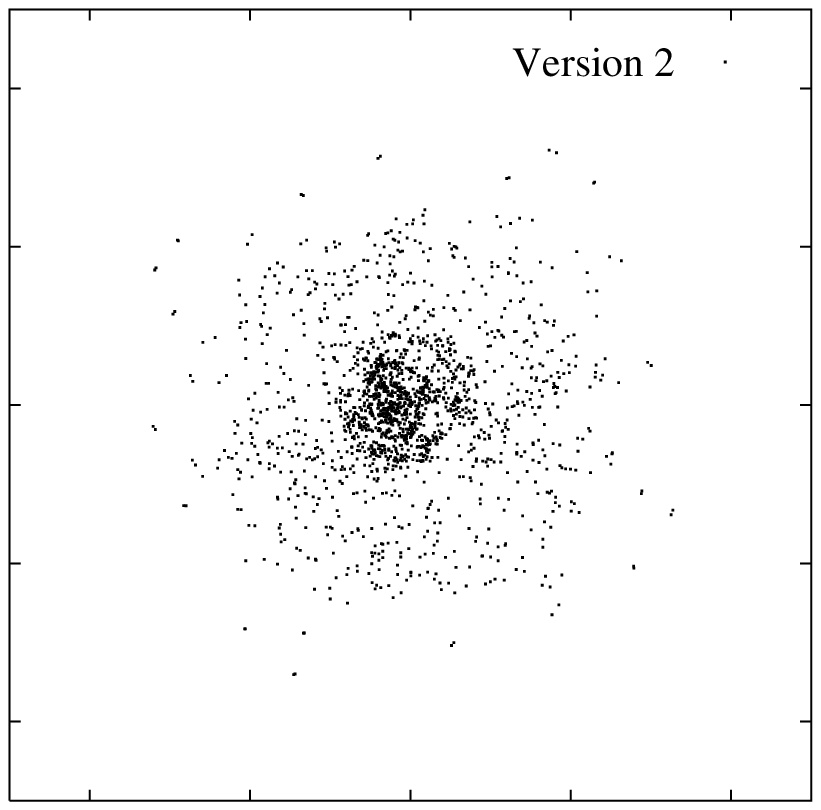}
\includegraphics{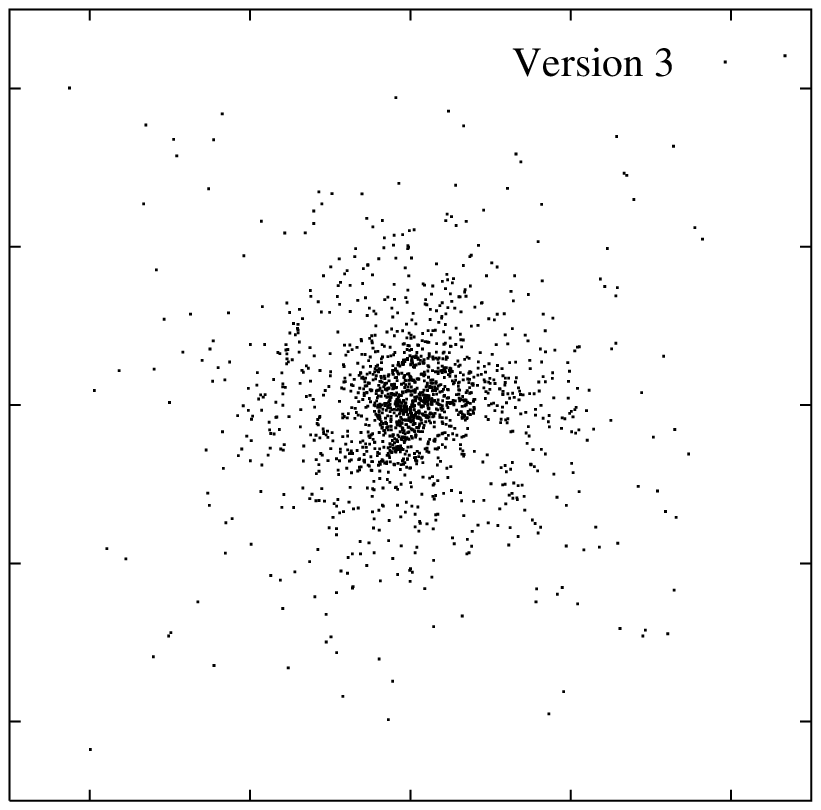}
\includegraphics{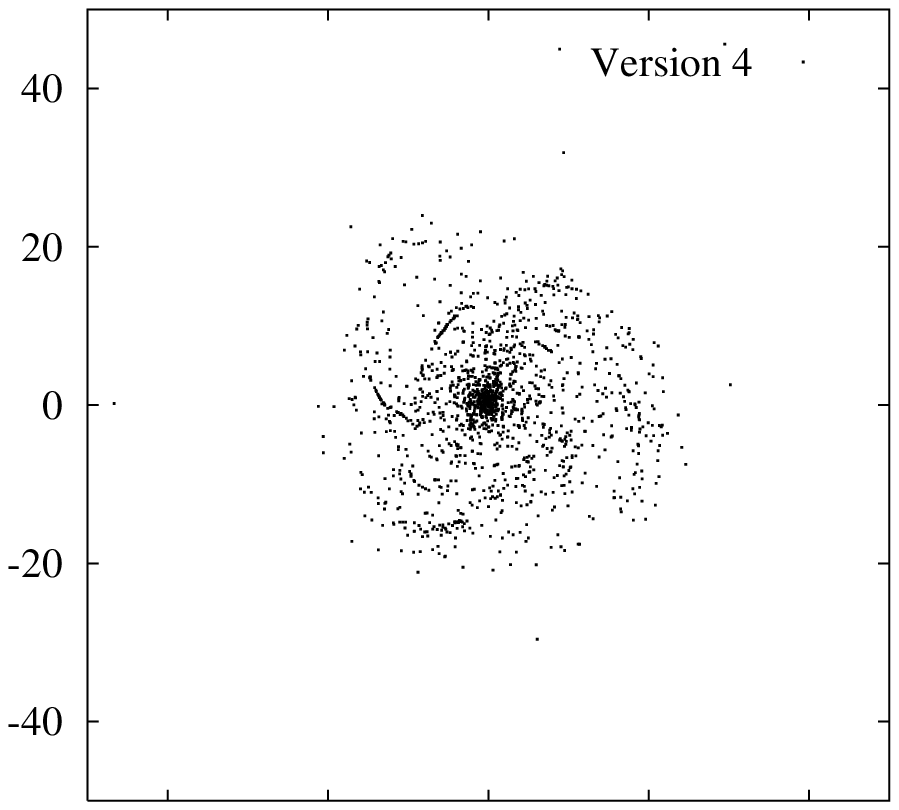}
\includegraphics{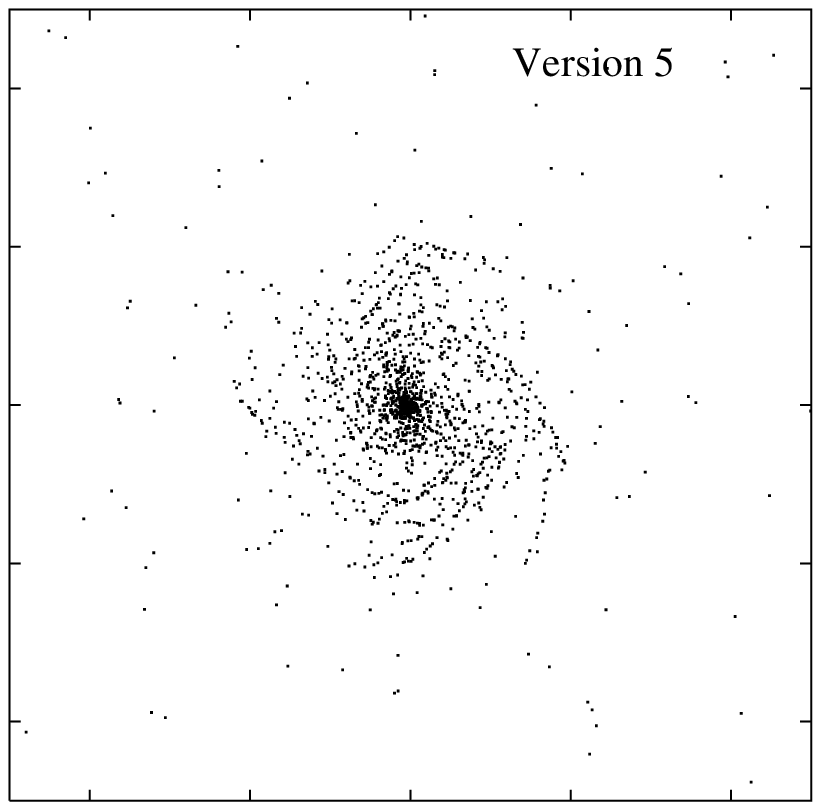} 
\includegraphics{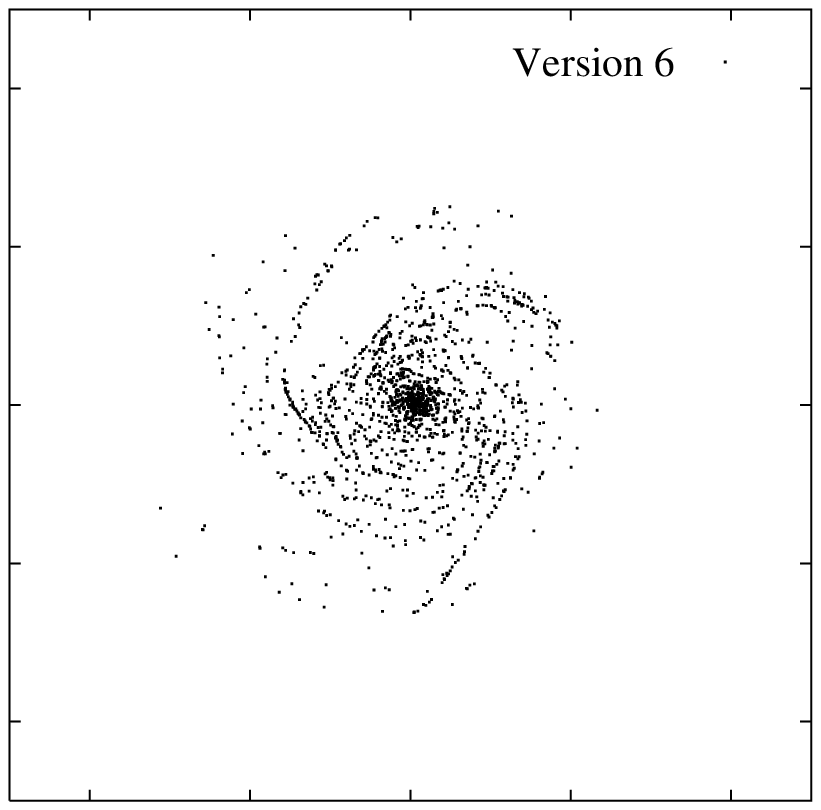}
\includegraphics{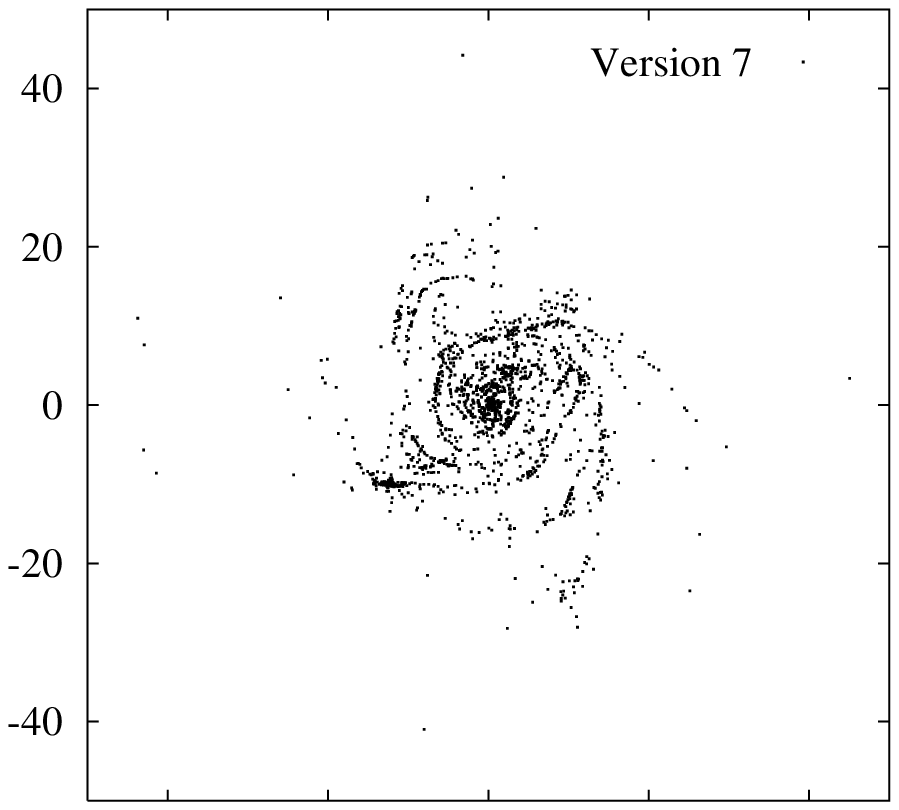}  
\includegraphics{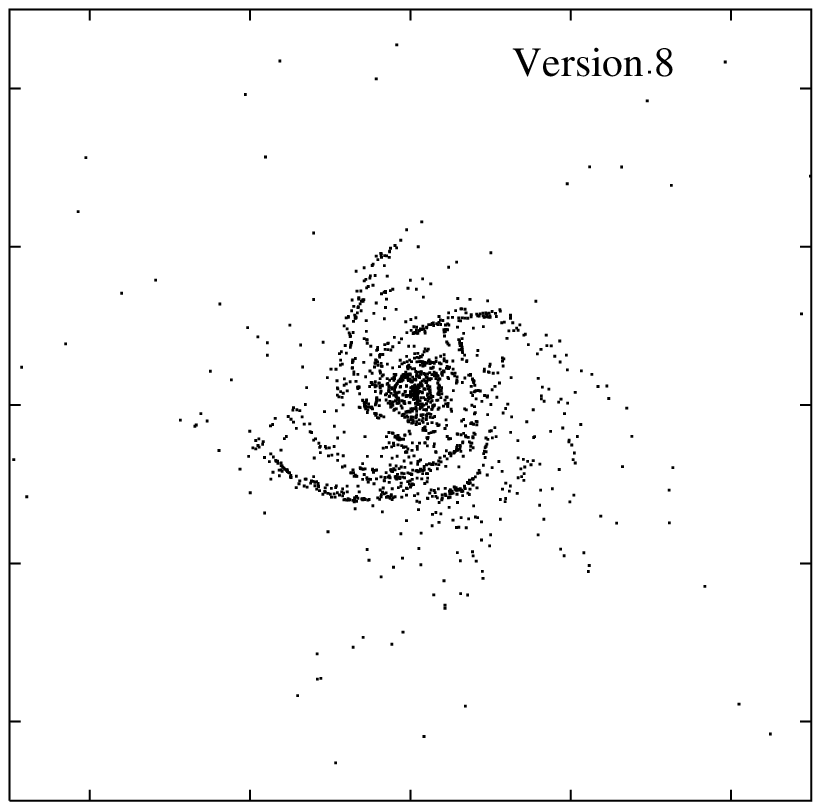}
\includegraphics{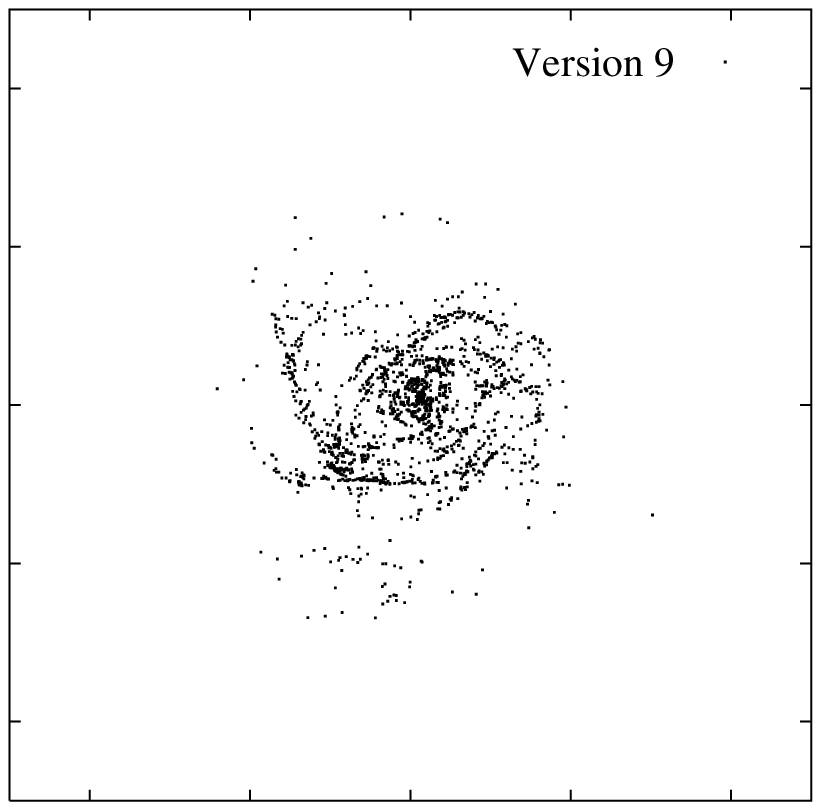}
\includegraphics{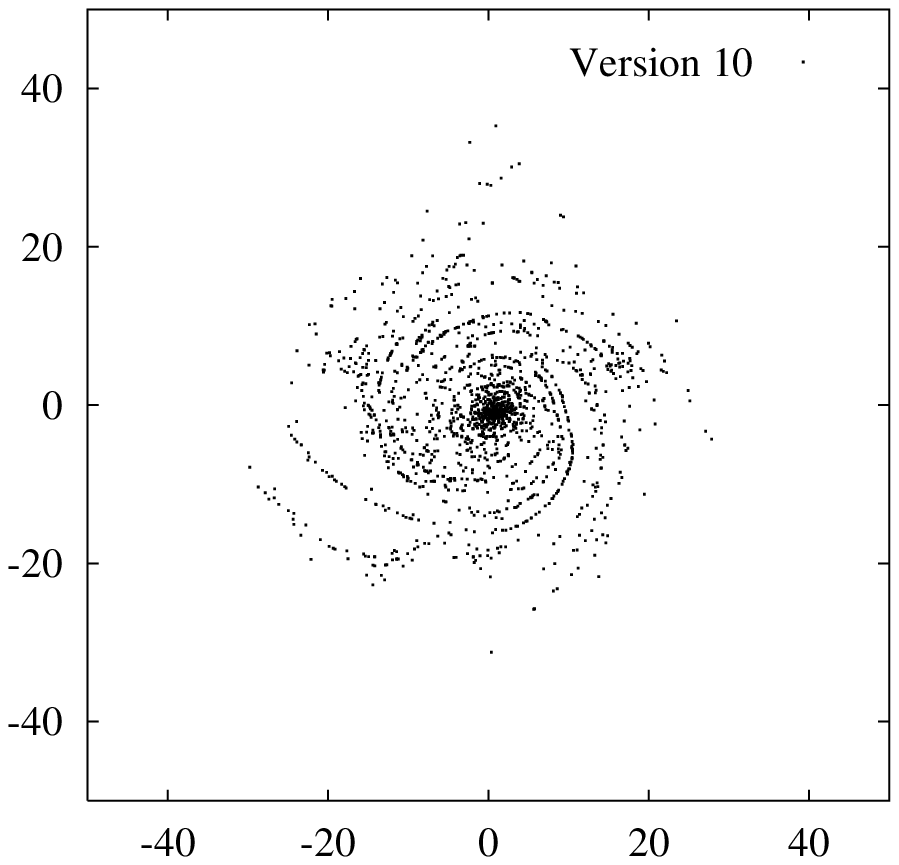} 
\includegraphics{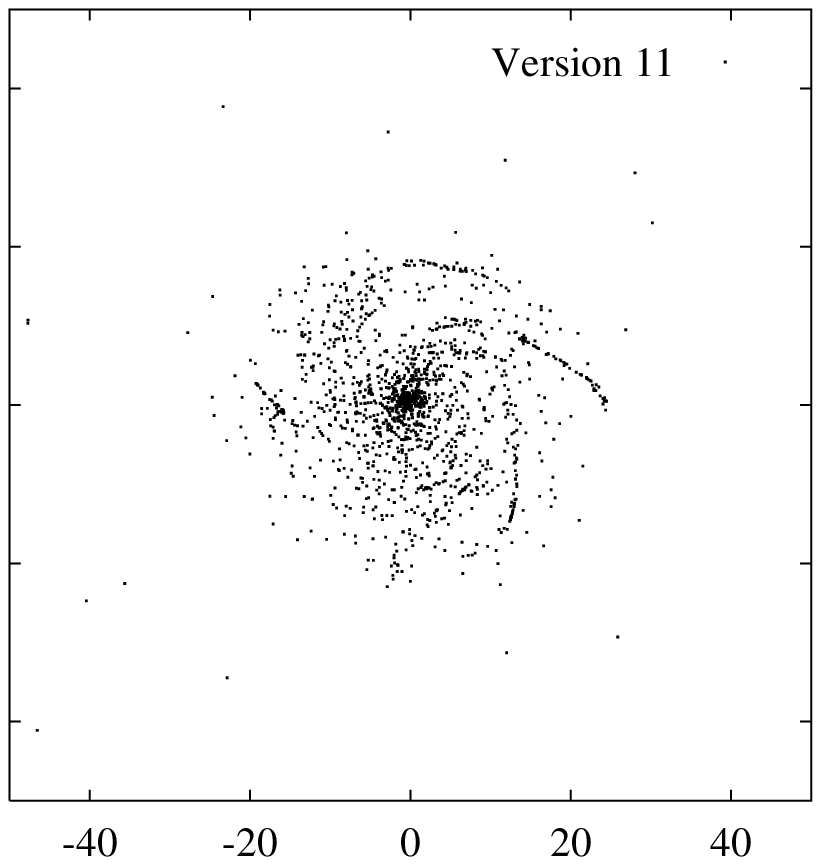} 
\includegraphics{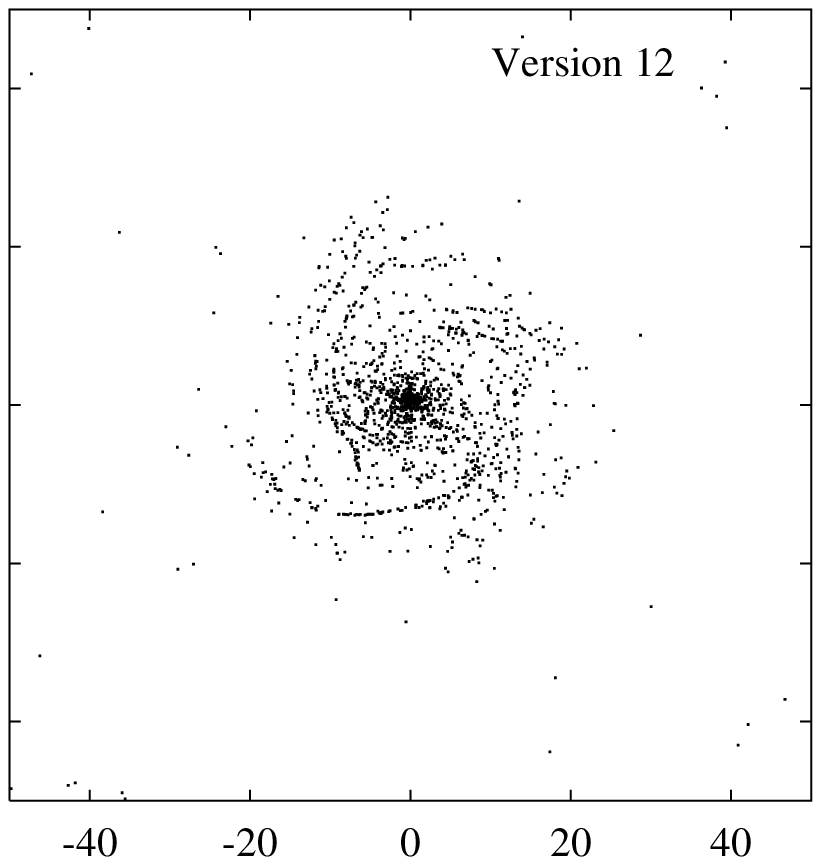} 
 \caption{Comparison of gas morphology for $2\times1736$ particle collapse. Time
is t=256 and axis scales are in kpc. 
Clearly different implementations exhibit different spiral
structures, indicating that at this resolution the structures are poorly
defined.} 
\label{1736comp} 
\end{minipage}
\end{figure*}

\begin{figure}
\vspace{58mm}
\includegraphics{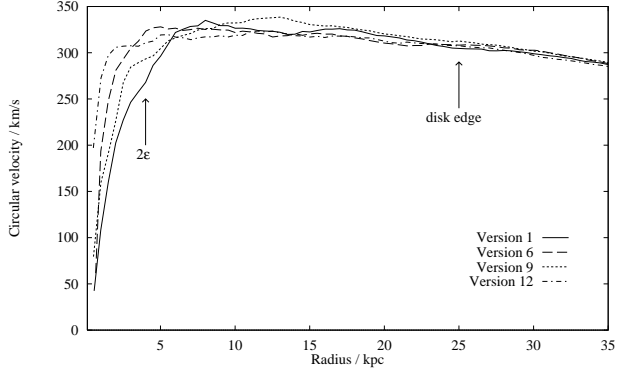}
 \caption{Rotation curves for four different implementations. The shear
correction in version 9 produces a higher rotational velocity at half the disc
radius. This is because the outward transport of angular momentum is reduced,
thereby reducing the inward movement of the half mass radius.} 
\label{rotcur}
\end{figure}

\begin{figure}
\vspace{60mm} 
\includegraphics{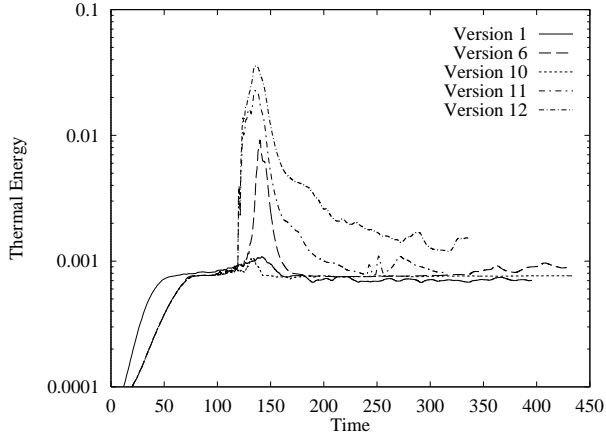}  
 \caption{Thermal energy, divided by the total initial mechanical energy,
for 5
different implementations. A marginal
change in the artificial viscosity can produce a 25-fold change in
the peak total thermal energy. This is indicative of the test sitting at the
edge of the Rees-Ostriker (1977) criterion.} 
\label{rcctherm}
\end{figure}

\begin{figure}
\vspace{60mm}
\includegraphics{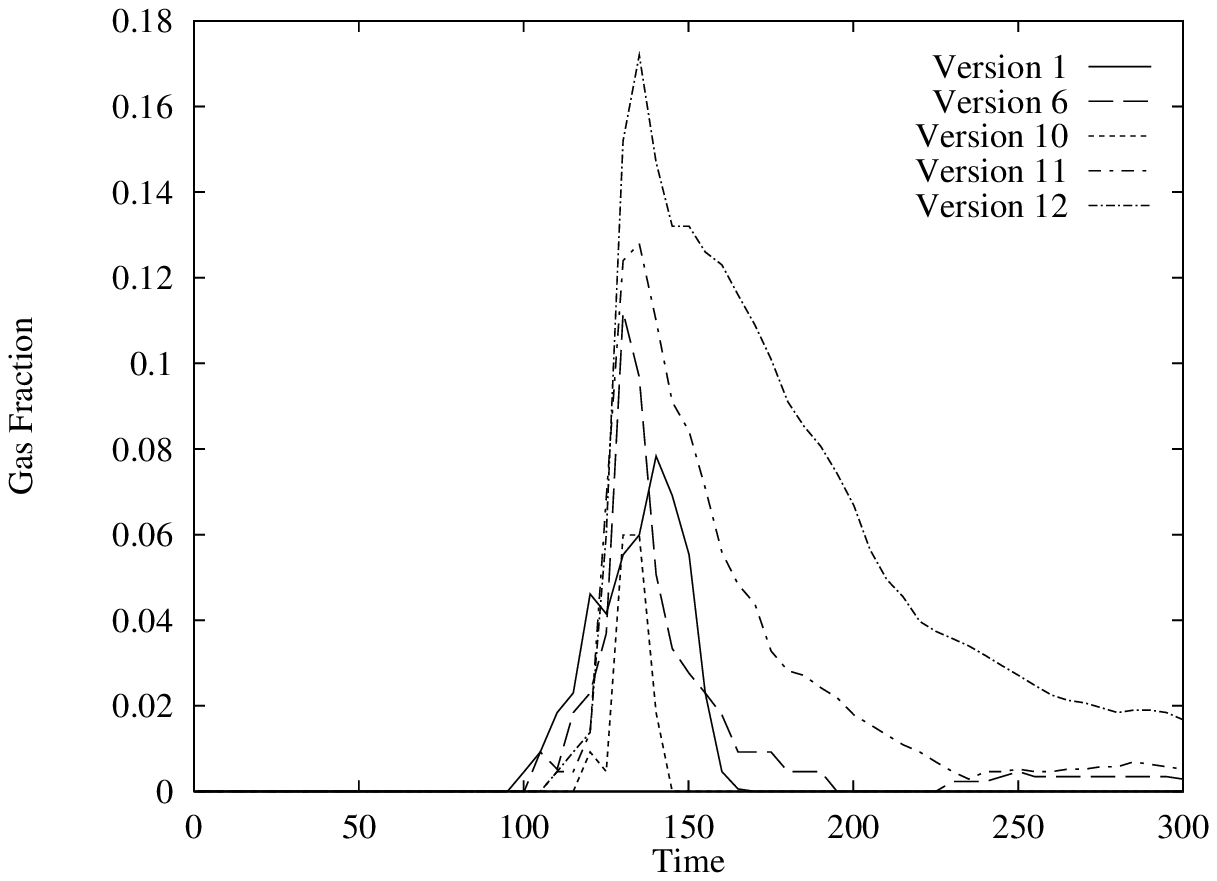}
 \caption{Fraction of gas mass above 30,000 K. This gives an approximate
measure of the amount of gas which is shocked. The amount of shocked gas is
extremely sensitive to the artificial viscosity implementation. Compare
versions 10 and 11 which differ only by the $\tilde{\rho}_{ij}$
substitution, the
peak
amount of shocked gas is different by a factor of two.}
\label{rccshock}
\end{figure}

Since the circular velocity is calculated from $[GM(<r)/r]^{1/2}$ and
the dark matter has the dominant mass contribution we expect little
difference among the rotation curves for the different
implementations.  In \fig~\ref{rotcur} we plot the rotation curves for
four different implementations. Apart from a visibly lower central
mass concentration for version 1 there is comparatively little
difference.

\subsubsection{Implementation-specific results}
\begin{figure}
\vspace{50mm}
\includegraphics{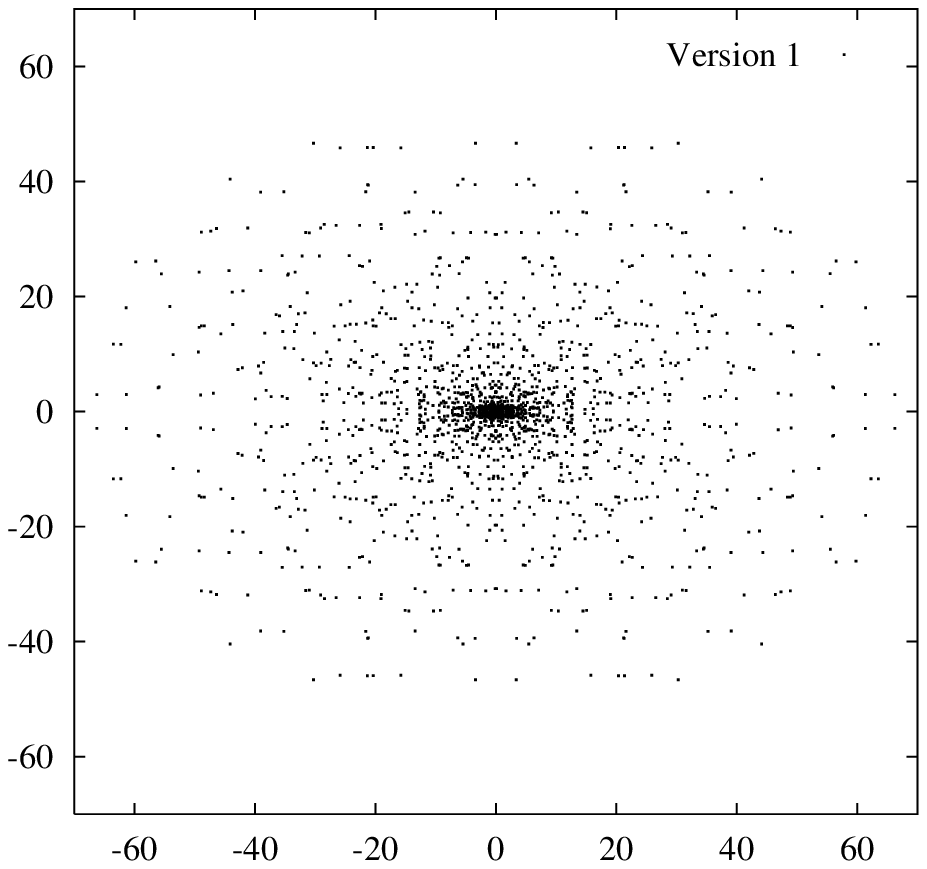}
\includegraphics{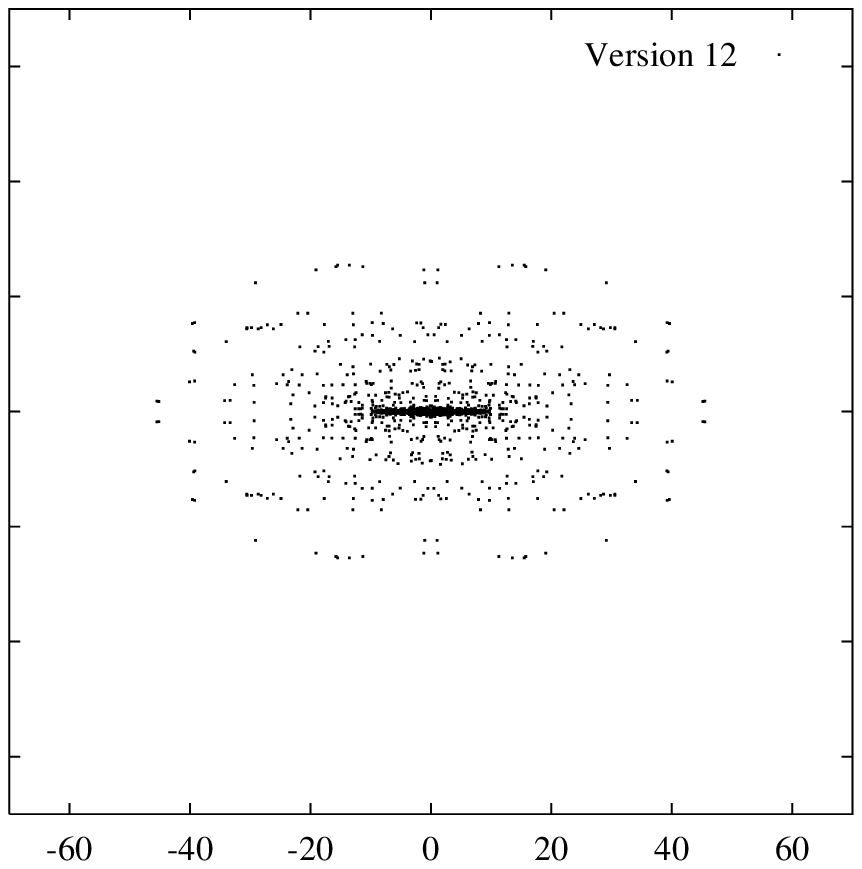}
\caption{Comparison of gas halo size at $t=128$ for versions 1 and 12.
Axis scales are given in kpc. The gas halo for version 1 is clearly
larger.
}
\label{halo.comp}
\end{figure}

Before the disc has formed (prior to t=128) versions 1--3 have an
extended gas halo compared with the remaining versions. The halo for
version 1 is as much as 40\% larger than those for versions 4--12.  In
\fig~\ref{halo.comp} we compare the gas structure of version 1 to that
of version 12.  The source of the extended halo is the $\nabla
. \bmath{v}$ artificial viscosity, which acts to increase the local
pressure. This is seen in the early rise in the thermal energy for
version 1 in \fig~\ref{rcctherm}. The more local estimate used in version
3 produces less pressure support and a smaller halo.
The Monaghan viscosity does not
provide pressure support as the pairwise $\bmath{r}.\bmath{v}$ term is
very small.  The different artificial viscosities also lead to
different disc morphologies. The $\nabla .
\bmath{v}$ viscosity fails to damp collapse along the z-axis
sufficiently and allows far more interpenetration than the Monaghan
viscosity leading to thicker discs in versions 1--3. 
 
The angular momentum losses in Table~\ref{galsum} show a noticeable trend.
For most codes $\Delta L/L$ is small and positive (by definition
indicating
a loss of angular momentum). However the shear-corrected Monaghan variants
show an {\em{increase}} in the angular momentum of the system. 
However, since the magnitude of the angular momentum is approximately the
same as that of the other codes, we do not place strong significance on this
result. Version 2 also has the shear-correction term, but we
attribute the similar performance to version 1 as being due to
the low amount of dissipation produced by the $\nabla . \bmath{v}$
viscosity.

Examining the thermal energy during collapse yields very interesting
results. \fig~\ref{rcctherm} shows a plot of the thermal energy of the
cloud versus time. As a fraction of the total energy the
thermal energy makes a small contribution because the baryon
fraction is only 10\%. However, the relative differences in thermal energy
between versions can be significant. This situation is analogous to the
differences seen in the kinetic energy in the Evrard collapse test (see
section \ref{evrard}). For this test it is
important to note that the differences in the thermal energy arise from
the amount of shocked gas present in the simulation, which is
determined by the artificial viscosity. This is demonstrated in the
comparison plot of the fraction of gas above 30000 K, shown in 
\fig~\ref{rccshock}. A small change in the artificial viscosity can have a very
significant change in the amount of shocked gas. If one considers the
effect of changing the $h$-averaging scheme from the arithmetic mean to
the harmonic mean, then the artificial viscosity will be higher in most
situations. This is because the term used to prevent divergences,
$0.01 \bar{h}_{ij}^2$ in the denominator of equation~\ref{muij}, is now
{\em{always}} smaller, and hence
can lead to larger values of $\Pi_{ij}$. This explains why version 5 has
so much shocked gas. Similarly, the $\rho_{ij}$ replacement in versions 11
and 12 leads to a larger $\Pi_{ij}$ as the $(h_i/h_j)^3\rho_i$ replacement
systematically tends to underestimate 
the value of $\rho_j$, and hence we
get more shocked gas. We note, however, that the effect is only noticeable in
cases of extreme density contrast, e.g., halo particles just above a cold
gaseous disc.

It is also evident that a change in the symmetrization procedure can have
a significant effect, codes 4 and 6 have differing amounts of shocked gas.
This fact suggests that the halo gas in this collapse problem must sit at
the edge of the Rees-Ostriker \cite{ro} cooling criterion, namely that the
free fall time is approximately equal to the gas cooling time. We have
checked this by running simulations with masses a factor of five higher
and two lower. The lower mass system produces less (12\% by mass, compared
to 18\%) shocked gas, whilst the high mass system produces a very large
(75\%) amount of shocked gas. In view of these results and that the
thermal energy is a very small fraction of the total, we do not place
strong emphasis on the differences in shocked gas between the versions.
Comparison of the results of Serna \etal \shortcite{se} and those of
Navarro \& White \shortcite{nw} confirms this conclusion as the amount of
shocked gas in the former differs visibly from the latter. 

\subsubsection{Summary}
Setting aside the differences in the amount of shocked gas among
implementations, there is comparatively little variation among the
final results. There are differences in morphology: versions 1, 2 and
3 have a thicker disc and, during the initial collapse, show a more
extended gas halo. Both effects can be traced to the $\nabla
. \bmath{v}$ viscosity. The potential and kinetic energies during
collapse, however, are all similar, due to the dominance of the dark
matter.

\subsection{Disc stability}\label{disk}
Disc stability is a critical issue in galaxy formation. It is now widely
known (Balsara \shortcite{ba}, Navarro \& Steinmetz \shortcite{sn}, for
example) that the
standard Monaghan viscosity introduces spurious angular momentum
transport (as opposed to the physical transfer of angular momentum that
occurs in differentially rotating discs). This spurious transfer can have
a significant effect of the development of
a small $N$ object, as the angular momentum may be transported to the outer
edge of the disc in a few rotations \cite{sn}. 

\begin{figure*}
\vspace{220mm}
\begin{minipage}{170mm}
\includegraphics{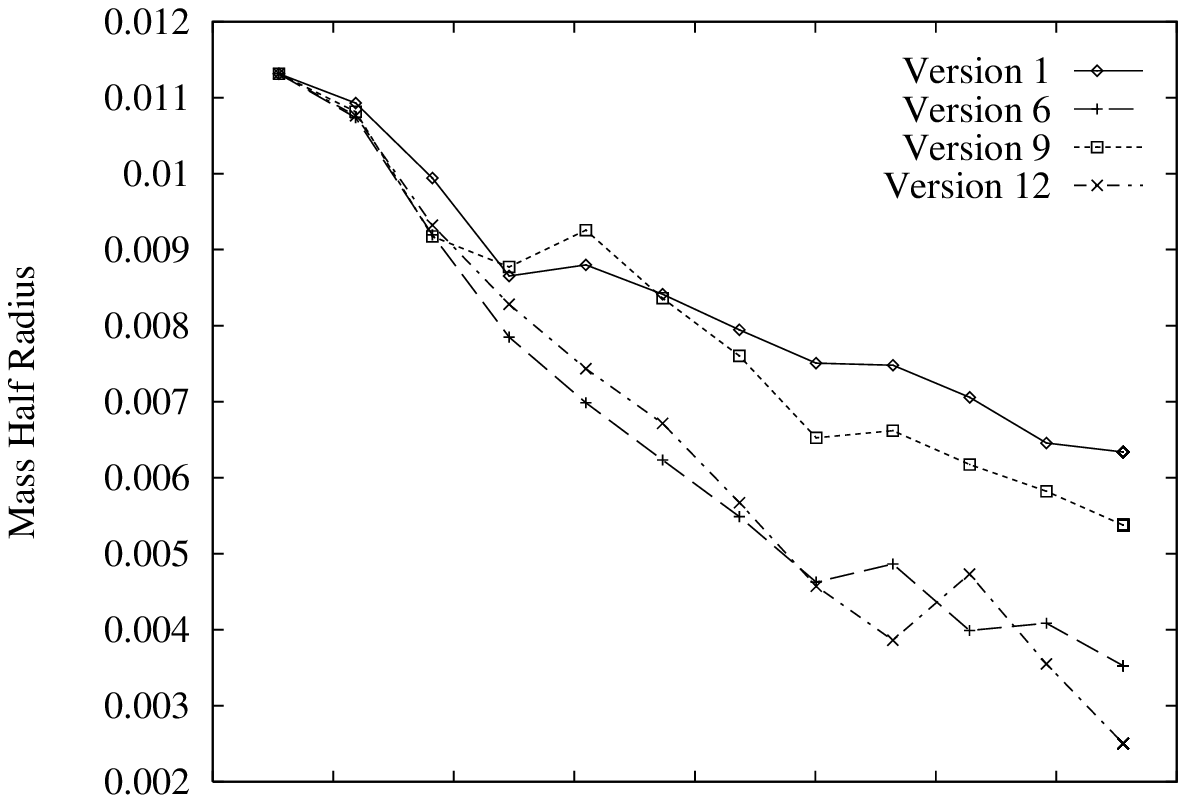}
\includegraphics{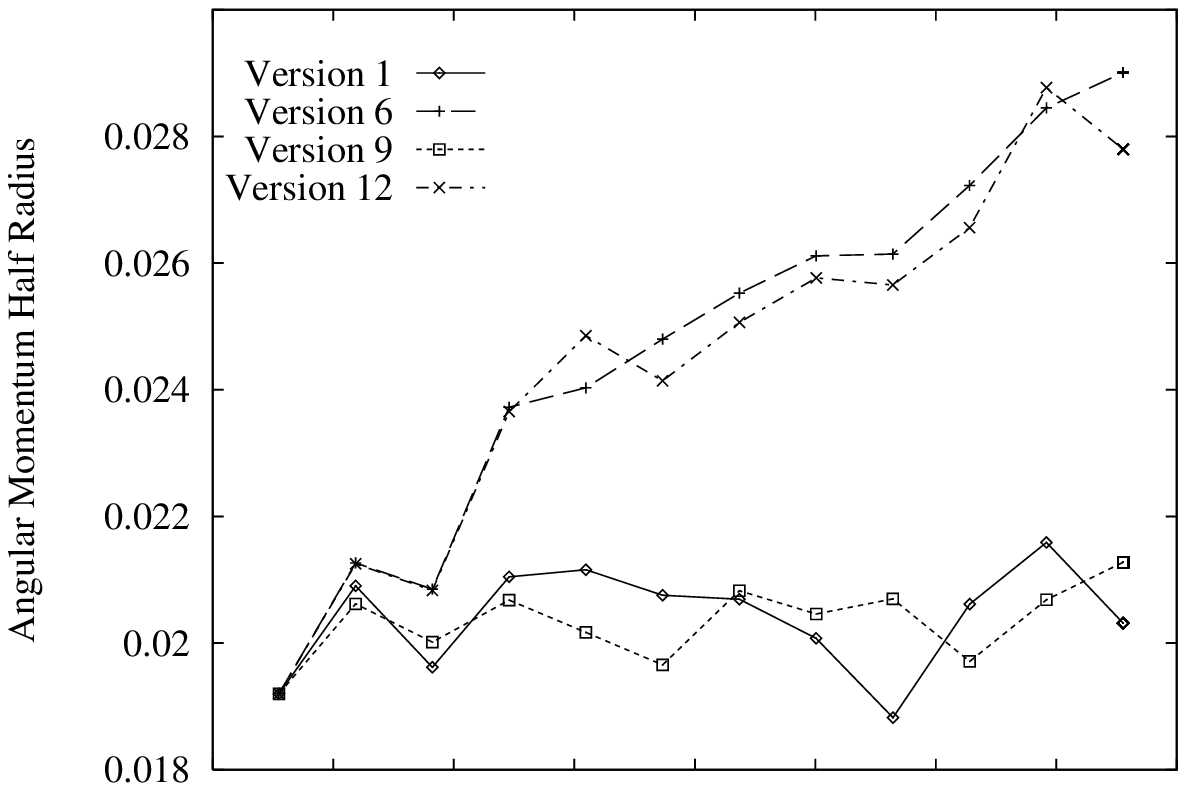}
\includegraphics{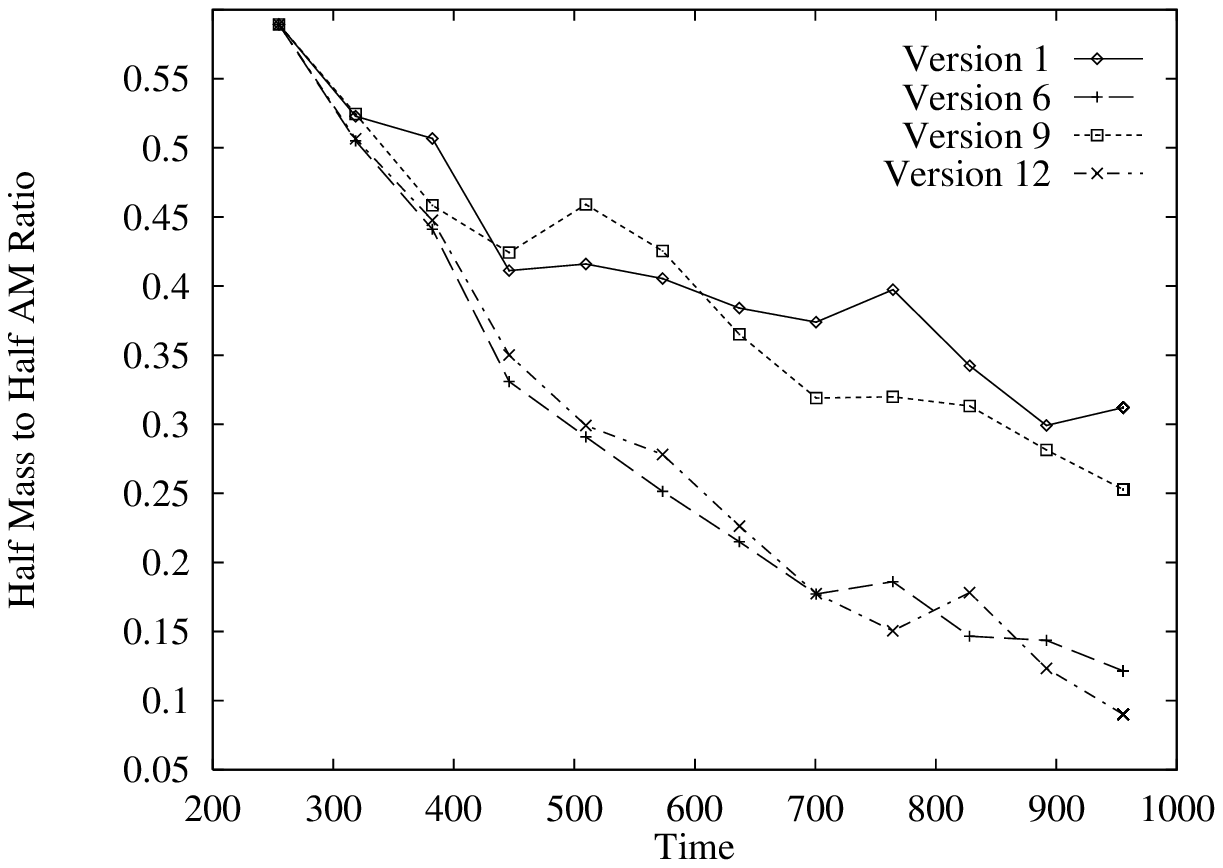}
\caption{The half-mass radius, half-AM radius
and the ratio of the two. 
Versions 1 and 9 are best, producing
a negligible increase in the angular momentum half-mass radius and a
comparatively slow reduction in the half-mass radius. The Monaghan
variants that are not shear-corrected perform worst, transporting the
angular momentum rapidly, resulting in a fast decrease of the half-mass
radius.
Radii are plotted in code units (1 unit equals 800 kpc).} 
\label{disk_stab} 
\end{minipage} 
\end{figure*}  

In this section we compare the growth of the half-angular-momentum 
radius (half-AM radius) of the disc, the half-mass radius 
and ratio of the two to determine which SPH
implementation is least susceptible to this problem. A similar
investigation was first performed by \cite{sn}.
For initial conditions we use
the disc formed by version 9 during the rotating cloud collapse (see
section~\ref{rotcloud}). 
We cut out the central 100 kpc diameter region and evolved this 
for a sufficient time for relaxation of the system as well as for transport
of angular momentum. We did not run a higher resolution test as
increasing the resolution leads to a disc that is unstable to
perturbations \cite{nw}. The limit at which this occurs is set by the
Toomre stability criterion \cite{at}, $Q$, where
\begin{equation}
Q={\sigma_r \upsilon \over 3.36 G \Sigma},
\end{equation}
$\sigma_r$ is the velocity dispersion, $\upsilon$ the epicycle frequency
and $\Sigma$ the surface density. 

\subsubsection{General evolution properties}
Given the comparatively quiet disc environment the morphological evolution
of the different versions were comparatively similar. The only noticeable
difference could be seen in the disc thickness, which for the $\nabla .
\bmath{v}$
viscosity variants (1, 2 and 3), was much larger than that of the rest
of the versions.  This effect was also observed in section
\ref{rotcloud}, although here the gas has `diffused' away from an
initially thin disc.

\subsubsection{Implementation-specific results}
The evolution of the  half-AM radii are plotted in 
\fig~\ref{disk_stab}. 
The figure indicates that the codes fall into two groups, with one
group suffering a stronger decay of the half-mass radius and an
associated growth of the half-AM radius. The other group, which shows
less decay, consists of versions 1--3, 7--9. The artificial
viscosities for this group are the $\nabla . \bmath{v}$ variants
(1--3) and the shear-corrected Monaghan version (7--9). For this group
the half-AM radius does not change significantly during the simulation
which corresponds to approximately 30 rotations and 5000
time-steps. The half-mass radius decays by approximately 50\%, leading
to a similar reduction in the half-mass to half-AM radii
ratio.

Within this first group of codes there is a sub-division determined by
the disc thickness at the end of the simulation. The
$\nabla. \bmath{v}$ 
variants all have a thick disc, due to the failure of the
algorithm to adequately damp convergent motions in the
z-direction. For the shear-corrected Monaghan variants this is not a
concern.

The
second group, comprising versions 4--6, 10--12 (all Monaghan
viscosity variants), shows
significant outward transport of the angular momentum, and by the
end of the simulation the half radius has increased by approximately
50\%. There is also a larger decay in the half-mass radius, it being
50\% greater than the decay seen for the other group of codes. These
two results contribute to make the half-mass to half-AM radii
ratio decay to only 25\% of its initial value at the beginning of the 
simulation. 
This result confirms that seen in Navarro \& Steinmetz~(1997) where it
was shown that the shear correction significantly reduces angular momentum
transport in disc simulations. Our results show a faster increase
in transport, but this is probably due to the simulations being
dissimilar; Navarro \& Steinmetz place a disc in a predefined halo,
and then evolve the combined system.

\subsubsection{Summary}
Whilst preserving the half-mass to half-AM radii ratio as well as the
shear-corrected scheme, the $\nabla.\bmath{v}$ viscosity is not a
significant improvement. The large increase in disc thickness and loss
of definition more than outweighs the improvement in the decay of the
half-mass radius. Both the Monaghan and shear-corrected artificial
viscosities maintain the disc structure. On the basis of these tests
any of the shear-corrected implementations (7--9) is to be preferred.

\section{Conclusions}\label{conc}
\begin{table*}
 \begin{tabular}{@{}ccccccc}
   & Shock capturing & Cooling & AM loss & Morphology &
Drag & Execution \\
Version & 3.1, 3.2 & 3.3 & 3.7 & 3.6 (3.5, 3.7) & 3.4 & time\\
\hline
 1  & $\times$ & $\surd$  & $\surd$  & $\times$ & $\surd$  & 1.1  \\
 2  & $\times$ & $\surd-$  & $\surd$  & $\times$ & $\surd$  & 1.1  \\
 3  & $\times$ & $\surd$  & $\surd$  & $\times$ & $\surd$  & 1.1  \\
 4  & $\surd$  & $\surd$  & $\times$ & $\surd$  & $\surd$ & 2.4   \\
 5  & $\surd$  & $\surd$  & $\times$ & $\surd$  & $\surd$ & 2.4   \\
 6  & $\surd$  & $\surd$  & $\times$ & $\surd$  & $\surd-$  & 2.4  \\
 7  & $\surd-$ & $\surd$  & $\surd$  & $\surd$  & $\surd+$ & 2.4   \\
 8  & $\surd-$ & $\surd$  & $\surd$  & $\surd$  & $\surd+$ & 2.4   \\
 9  & $\surd-$ & $\surd$  & $\surd$  & $\surd$  & $\surd$  & 2.4  \\
 10 & $\surd$  & $\surd-$ & $\times$ & $\surd$  & $\times$ & 1.2  \\
 11 & $\surd$  & $\times$ & $\times$ & $\surd$  & $\times$ & 1.1   \\          
 12 & $\surd$ & $\surd-$  & $\times$ & $\surd$  & $\surd$ & 1.1  \\
\hline
\end{tabular}
 \caption{Qualitative summary of the strengths and weaknesses of each
implementation. We categorize each version using a $\surd$ to indicate
preferable performance
and an $\times$ to indicate inferior performance.
+ and - signs differentiate between similarly grouped implementations.
Assessments in the morphology column refer to the success with which the
implementation can produce and maintain thin discs.
Execution times represent the best performance to be
expected from each algorithm and are relative to the \Hydra\ code without
the new $h$-update algorithm. In high-resolution simulations
($2\times128^3$ particles, for example) versions 4--9
may well be significantly slower. Section numbers from which these   
conclusions are
drawn are shown at the top of the columns.
}
\label{conc.table}
\end{table*} 
We have presented a series of tests designed to determine the differences
in performance of various SPH implementations in 
scenarios common in simulations of cosmological hierarchical clustering.
Special attention was paid to how the codes perform
for small-$N$ problems. A summary of our findings is presented in
Table~\ref{conc.table}. 

Our principle conclusions follow.
\begin{enumerate}
\item We
recommend schemes that use the Monaghan viscosity supplemented with the
shear-correction. Of those methods that do not use the shear-correction,
version 12 is preferable because of its high speed and accuracy (and
see vii below). 

\item Several implementations introduce programming difficulties, such
as changing the 
neighbour search from a gather process to a hybrid gather-scatter. 
This is especially 
problematic in the adaptive refinements where different symmetrization
schemes lead to different choices for which particles to place in a
refinement. The TC92 symmetrization is by far the easiest to program
in this respect.

\item The choice of artificial viscosity is the primary factor in
determining code performance. The equation of motion and particle
symmetrization schemes have only a secondary role, albeit significant for
small-$N$. In particular, the artificial viscosity used in the current  
implementation of \Hydra, and variants of it, produce a large amount of
scatter in local variables, such as the velocity field and temperature,
and also lead to less thermalisation. These
characteristics indicate that the relative performance of an artificial
viscosity is determined by its effective resolution. Viscosities which use
an estimate of $\nabla.\bmath{v}$ to determine whether the viscosity
should be applied will always be less able to capture strong flow
convergence and shocks than those which use the pairwise
$\bmath{r}.\bmath{v}$ trigger. 

\item Instabilities inherent in simple smoothing-length update algorithms
can be removed. By using a `neighbour-counting kernel' and a weighted
update average, stability can be increased
without requiring an expensive exact calculation of the correct $h$ to
yield a constant number of neighbours.

\item We strongly agree with the conclusion presented in SM93 that to
accurately calculate local physical variables in dynamically evolving
systems, at least
$10^4$ particles are required. The difference in morphologies observed in
the rotating cloud collapse clearly indicates that the belief SPH can
accurately predict galaxy morphologies with as few as 1000 particles is 
overly optimistic. This implies that in studies of galaxy formation,
where the    
internal dynamics define morphology, an object may only be considered
well resolved if it contains a minimum of $10^4$ particles.

\item The introduction of a shear-corrected viscosity leads to reduced
shock capturing, although the effect is small and primarily visible in
the local velocity field. We have also confirmed the results of other
authors, namely that the shear-corrected viscosity does indeed reduce
viscous transport of angular momentum.

\item It is possible to implement a scheme (11 and 12) which uses the
Monaghan artificial viscosity, that does not
require the precomputation of all density values before solving the
hydrodynamic equation of motion. Further, the resulting implementation
conserves momentum and energy to an extremely high accuracy. This
implementation
removes the need to compute the list of neighbours twice or alternatively
the need to store it for every particle. Hence it is significantly faster
than other schemes. Additionally, although not done in this work, the
shear-correction can be incorporated with little extra effort.  

\end{enumerate}

\section*{Acknowledgments}
RJT thanks Professor Don Page for financial support during the writing of
this paper. The authors thank NATO for providing NATO Collaborative
Research Grant CRG 970081 which facilitated their interaction.

\label{lastpage}

\end{document}